\pdfoutput=1
\newcommand*{\ATLASLATEXPATH}{latex/}
\documentclass[cernpreprint,texlive=2016,txfonts,USenglish]{\ATLASLATEXPATH atlasdoc}
\makeatletter
\DeclareOldFontCommand{\rm}{\normalfont\rmfamily}{\mathrm}
\DeclareOldFontCommand{\sf}{\normalfont\sffamily}{\mathsf}
\DeclareOldFontCommand{\tt}{\normalfont\ttfamily}{\mathtt}
\DeclareOldFontCommand{\bf}{\normalfont\bfseries}{\mathbf}
\DeclareOldFontCommand{\it}{\normalfont\itshape}{\mathit}
\DeclareOldFontCommand{\sl}{\normalfont\slshape}{\@nomath\sl}
\DeclareOldFontCommand{\sc}{\normalfont\scshape}{\@nomath\sc}
\makeatother

\usepackage[subfigure=true,block=none]{\ATLASLATEXPATH atlaspackage}
\usepackage{\ATLASLATEXPATH atlasbiblatex}
\usepackage{\ATLASLATEXPATH atlascontribute}
\usepackage{\ATLASLATEXPATH atlasphysics}
\graphicspath{{logos/}}
\usepackage{TOPQ-2016-09-PAPER-defs}

\usepackage{color, colortbl}
\definecolor{Gray}{gray}{0.9}
\usepackage{rotating}
\usepackage{pdflscape}
\usepackage{multirow}
\usepackage{placeins}
\usepackage{nicefrac}
\usepackage{tikz}
\usepackage{array}

\AtlasTitle{Measurements of $t\bar{t}$ differential cross-sections of highly boosted top quarks decaying to all-hadronic 
final states in $pp$ collisions at $\sqrt{s}=13\,$ \TeV\ using the ATLAS detector}

\author{The ATLAS Collaboration}

\makeatletter
\renewcommand\AB@affilsepx{, \protect\Affilfont}
\makeatother
\renewcommand\Authands{, } 
\renewcommand\Affilfont{\itshape\small} 

\AtlasRefCode{ATLAS-TOPQ-2016-09}

\AtlasNote{ATLAS-TOPQ-2016-09}

\PreprintIdNumber{CERN-EP-2017-226}

\AtlasJournal{Phys.\ Rev.\ D.}
\AtlasAbstract{%
Measurements are made of differential cross-sections of highly boosted pair-produced top quarks as a function of 
top-quark and $t\bar{t}$~system 
kinematic observables using proton--proton collisions at a center-of-mass energy of $\sqrt{s} = 13$~\TeV. 
The data set corresponds to an integrated luminosity of $36.1$ fb$^{-1}$, recorded in 2015 and 2016 with the ATLAS detector at the CERN Large Hadron Collider.
Events with two large-radius jets in the final state, one with transverse momentum $p_{\rm T} > 500$~\GeV\ and a 
second with $p_{\rm T}>350$~\GeV, are used for the measurement. 
The top-quark candidates are separated from the multijet background using jet substructure information and 
association with a $b$-tagged jet.
The measured spectra are corrected for detector effects to a particle-level fiducial phase space and a parton-level limited
phase space, and are compared to several Monte Carlo simulations by means of calculated $\chi^2$\ values. 
The cross-section for $t\bar{t}$ production in the fiducial phase-space region is
$292 \pm 7 \ \rm{(stat)}  \pm 76 \rm{(syst)}$~fb, to be compared to the theoretical prediction 
of $384 \pm 36$~fb.
}

\AtlasCoverSupportingNote{ATL-COM-PHYS-2016-1696}{https://cds.cern.ch/record/2234845}
\AtlasCoverSupportingNote{HepData materials}{https://cds.cern.ch/record/2281524}
\AtlasCoverCommentsDeadline{27th of October 2017}

\AtlasCoverAnalysisTeam{
Peter Berta,
Ye Chen,
Kyle Cormier,
Riccardo Di Sipio,
Federica Fabbri,
Petr Ja\v{c}ka,
Ji\v{r}\'i Kvita,
Rupert Leitner,
Roman Lys\'{a}k,
Shima Shimizu Noda,
Jan Palicka,
Marino Romano,
Pekka Sinervo,
Francesco Spano,
Yuji Yamazaki
}

\AtlasCoverEdBoardMember{Arnulf Quadt~(chair), Michael Begel, Mark Oreglia}

\AtlasCoverEgroupEditors{atlas-TOPQ-2016-09-editors@cern.ch}

\AtlasCoverEgroupEdBoard{atlas-TOPQ-2016-09-editorial-board@cern.ch}

\hypersetup{pdftitle={ATLAS document},pdfauthor={The ATLAS Collaboration}}

\begin{document}

\maketitle

\section{Introduction}
\label{sec:intro}

The large top-quark pair-production cross-section at the Large Hadron Collider (LHC) allows detailed 
studies of the characteristics of the production of top--antitop (\ttbar) quark pairs, providing an opportunity to 
further test the Standard Model (SM). 
Focusing on highly boosted final states probes the QCD \ttbar\ production processes in the \TeV\ scale range, 
a kinematic region where theoretical calculations based on the SM still present large 
uncertainties \cite{Czakon:2017wor,NNLO_calc,nnloMtt}. 
High-precision measurements, especially in kinematic regions that have not been explored extensively, are necessary 
to better constrain the models currently in use.
Furthermore, effects beyond the SM can appear as modifications of $\ttbar{}$ differential 
distributions with respect to the SM predictions~\cite{Hill:1993hs,Frederix:2009,top_bkg_interference} that may not be 
detected with an inclusive cross-section measurement. 

In the SM, the top quark decays almost exclusively to a~\Wboson{} boson and a~$b$-quark. 
The signature of a~\ttbar\ final state is therefore determined by the \Wboson{} boson decay modes. 
The ATLAS~\cite{atlasDiff1,atlasDiff2,atlasDiff3,atlasDiff4,atlasDiff5,atlasDiff6,atlasDiff7,atlasDiff8}
and  CMS~\cite{cmsDiff1,cmsDiff2,cmsDiff3,cmsDiff4,cmsDiff5} Collaborations have published measurements of 
the $\ttbar{}$ differential cross-sections at center-of-mass energies of $\sqrt{s}=7$ \TeV{}, 
$\sqrt{s}=8$ \TeV{} and $\sqrt{s}=13$ \TeV{} in $pp$ collisions using final states containing leptons. 
The analysis presented here makes use of the all-hadronic \ttbar{} decay mode, where 
only top-quark candidates with high transverse momentum (\pt{}) are selected.
This highly boosted topology is easier to reconstruct than other final-state configurations 
as the top-quark decay products are collimated into a 
large-radius jet by the Lorentz boost of the top quarks. 
This analysis is performed on events with the leading top-quark jet having $\pTtone > 500$~\GeV\ and the 
second-leading top-quark jet having $\pTttwo > 350$~\GeV. 
These jets are reconstructed from calorimeter energy deposits and tagged as top-quark candidates to separate 
the \ttbar\ final state from background sources.
The event selection and background estimation follows the approach used in Ref. \cite{TOPQ-2012-15}, but with 
updated tagging methods and data-driven multijet background estimates.

These measurements are based on data collected by the ATLAS detector in 2015 and 2016 from $pp$ collisions at 
$\sqrt{s}=13\,\TeV$, corresponding to an integrated luminosity of \lumitot.  
Measurements are made of the \ttbar\ differential cross-sections by unfolding the detector-level distributions 
to a particle-level fiducial phase-space region.
The goal of unfolding to a particle-level fiducial phase space and of using variables directly related to 
detector observables is to allow precision tests of QCD by avoiding model-dependent extrapolation 
of the measurements to a phase-space region outside the detector acceptance. 
Measurements of parton-level differential cross-sections are also presented, where the detector-level 
distributions are unfolded to the top quark at the parton-level in a limited phase-space region.
These allow comparisons to the higher-order calculations that are currently restricted to stable top quarks 
\cite{Czakon:2017wor,NNLO_calc,nnloMtt}. 

These differential cross-sections are similar to those studied in dijet measurements at large jet 
transverse momentum~\cite{ATLAS:dijet8TeV, PhysRevLett.81.2642}\ and are sensitive to effects of 
initial- and final-state radiation (ISR and FSR), to different parton distribution functions (PDF) and to different schemes for 
matching matrix-element calculations to parton shower models.

Measurements are made of the differential cross-sections for the leading and second-leading top quarks as a function of
$\pTtone$\ and $\pTttwo$, as well as the rapidities of the top quarks.
The rapidities of the leading and second-leading top quarks in the laboratory frame are denoted by $\ytone$ and $\yttwo$, 
respectively, 
while their rapidities in the $\ttbar{}$ center-of-mass frame are $\ystar = \nicefrac{1}{2}\left(\ytone - \yttwo\right)$ and $-\ystar$. 
These allow the construction of the variable $\chittbar=\exp{2|\ystar|}$, which is of particular interest as 
many processes not included in the Standard Model are predicted to peak at low values of \chittbar{}~\cite{EXOT-2016-21}.
The longitudinal motion of the \ttbar~ system in the laboratory frame is described by the rapidity 
boost $\boostttbar=\nicefrac{1}{2} \left( \ytone + \yttwo \right) $\ and is sensitive to PDFs.
Measurements are also made of the differential cross-sections as a function of
the invariant mass, \pT\ and rapidity of the \ttbar\ system;
the absolute value of the azimuthal angle between the two top quarks, \deltaPhittbar;
the absolute value of the out-of-plane momentum, \absPoutttbar\ (i.e., the projection of the three-momentum of one 
of the top-quark jets onto the direction perpendicular to a plane defined by the other top quark and the beam axis ($z$) 
in the laboratory frame \cite{PhysRevLett.81.2642}); the cosine of the production angle in the Collins--Soper\footnote{The Collins--Soper frame is the rest frame of the \ttbar{} pair, wherein the two top quarks have equal and opposite momenta; 
thus, each makes the same angle $\theta^{\star}$ with the beam direction.} reference frame, \cosThetaStar; and
the scalar sum of the transverse momenta of the two top quarks, \HTttbar\ \cite{Denner:2010jp,Bevilacqua:2010qb}.
Some of the variables (e.g. \deltaPhittbar\ and \absPoutttbar) are more sensitive to additional radiation in the main 
scattering process, and thus are more sensitive to effects beyond leading order (LO) in the matrix elements.
All of these variables are sensitive to the kinematics of the \ttbar\ production process.

The paper is organized as follows. Section ~\ref{sec:detector} briefly describes the ATLAS detector, while Sec.~\ref{sec:datasets} describes the data and simulation samples used in the measurements. 
The reconstruction of physics objects and the event selection is explained in Sec.~\ref{sec:selection}\ and the background estimates 
are discussed in Sec.~\ref{sec:backgrounds}.
The procedure for unfolding to particle level and parton level are described in Sec.~\ref{sec:unfolding}. 
The systematic uncertainties affecting the measurements are summarized in Sec.~\ref{sec:uncertainties}.
The results of the measurements are presented in Sec.~\ref{sec:measurements} and comparisons of these results with theoretical predictions are made in Sec.~\ref{sec:comparisons}. 
A summary is presented in Sec.~\ref{sec:conclusion}.

\section{ATLAS detector}
\label{sec:detector}
The ATLAS experiment \cite{PERF-2007-01}\ at the LHC uses a multi-purpose detector with a forward-backward 
symmetric cylindrical geometry and near 4$\pi$\ coverage in solid angle.\footnote{
ATLAS uses a right-handed coordinate system with its origin at the nominal interaction point (IP) 
in the center of the detector and the $z$-axis along the beam pipe. 
The $x$-axis points from the IP to the center of the LHC ring and the $y$-axis points upward. 
Cylindrical coordinates $(r,\phi)$ are used in the transverse plane, $\phi$ being the azimuthal angle around the $z$-axis. 
The pseudorapidity is defined in terms of the polar angle $\theta$ as $\eta=-\ln\tan(\theta/2)$.
}\ It consists of an inner tracking detector 
surrounded by a superconducting solenoid magnet creating a 2~T axial magnetic field, electromagnetic and
hadronic calorimeters, and a muon spectrometer. 

The inner tracking detector covers the 
pseudorapidity range $|\eta| <  2.5$.
Consisting of silicon pixel, silicon microstrip and transition radiation tracking detectors, the inner 
tracking detector allows highly efficient reconstruction of 
the trajectories of the charged particles produced in the $pp$\ interactions.
An additional silicon pixel layer, the insertable B-layer, 
was added between 3 and 4~cm from the beam line to improve
$b$-hadron tagging \cite{ATLAS-TDR-19}.
Lead/liquid-argon (LAr) sampling calorimeters provide electromagnetic (EM) energy measurements with
high granularity and shower-depth  segmentation. 
A hadronic (steel/scintillator-tile) calorimeter covers the
central pseudorapidity range ($|\eta| < 1.7$). 
The endcap and forward regions are instrumented with LAr
calorimeters for EM and hadronic energy measurements up to $|\eta| = 4.9$.
The muon spectrometer
is located outside of the calorimeter systems and is based on
three large air-core toroid superconducting magnets with eight
coils each. 
It includes a system of precision tracking chambers and detectors with sufficient timing resolution to enable triggering of events. 

A two-level trigger system is used to select events~\cite{TRIG-2016-01}. 
The first-level hardware-based trigger 
uses a subset of the detector information to reduce the rate of accepted events to a design maximum of 100 kHz.
This is followed by a software-based trigger with a maximum average accepted event rate of 1 kHz.
\section{Data sets and Monte Carlo event generation}
\label{sec:datasets}
The data used for this analysis were recorded with the ATLAS detector at a center-of-mass energy of 13~\TeV\ in 2015 and 2016 and correspond to an integrated luminosity of \lumitot. Only the data-taking periods in which all the subdetectors were operational are considered. 

The events for this analysis were collected using an inclusive anti-$k_{t}$\ jet trigger with radius parameter $R=1.0$\ 
and nominal \pT\ thresholds of 360~\GeV\ and 420~\GeV\ for the 2015 and 2016 data-taking periods, respectively.  
These triggers were fully efficient for jets with $\pT > 480$~\GeV~\cite{TRIG-2016-01}.

The signal and several background processes are modeled using Monte Carlo (MC) event generators. 
Multiple overlaid proton--proton collisions (pileup) are simulated with 
the soft QCD processes of \PYTHIA~8.186 \cite{Sjostrand:2007gs} using a set of tuned 
parameters called the A2 tune~\cite{ATL-PHYS-PUB-2011-014} and the 
MSTW2008LO \cite{MSTW}\ PDF set.
The detector response is simulated using the \GEANT{4}\ framework~\cite{GEANT4,SOFT-2010-01}. 
The data and MC events are reconstructed with the same software algorithms.

Several next-to-leading-order (NLO) MC calculations of the \ttbar\
process are used in the analysis,
and to compare with the measured differential cross-sections. The \textsc{Powheg-Box}~v2\ \cite{POWHEGBOX}, 
\MGMCatNLO{}\ \cite{MCATNLO}\ and \SHERPA\ \cite{Gleisberg:2008ta}\ Monte Carlo event generators encode different 
approaches to the matrix element calculation and different matching 
schemes between the NLO QCD matrix-element calculation and the parton shower algorithm. 
A more detailed explanation of the differences among these event generators can be found in Ref. \cite{Hoeche:2011fd}.
 
The nominal sample uses
the \textsc{Powheg-Box}~v2 \cite{POWHEGBOX} event generator employing the NNPDF30 PDF set interfaced with 
the \PYTHIA{8}\ parton shower and 
hadronization model (hereafter also referred to as PWG+PY8).  
The \POWHEG{}\ $h_{\rm damp}$ parameter, which controls the \pT\ of the first additional emission beyond the Born 
configuration, is set to $1.5$ times the top-quark mass \cite{ATL-PHYS-PUB-2016-020}. 
The main effect of this is to regulate the high-\pT\ emission against which the \ttbar\ system recoils.
To enhance the production of top quarks in the high-\pT\ region, the \POWHEG{}\ parameter \bornsuppfact\ is set to 
$p_{{\rm T,supp}}=500$~\GeV\ \cite{POWHEGBOX,POWHEGdijet}.
The \PYTHIA{8}\ parameters are chosen for good agreement with ATLAS
Run-1 data by employing the A14 tune~\cite{ATL-PHYS-PUB-2014-021} with the NNPDF23LO PDF set \cite{Ball:2012cx}.

Two alternative \POWHEG{}+\PYTHIA{8}\ samples with systematic variations of the \POWHEG{}\ and \PYTHIA{8}\ parameters
probe the effects of the experimental tuning of the MC event generators. 
One sample, which primarily increases the amount of initial- and final-state radiation, uses $h_{\rm damp}=3\mt$, the factorization
and renormalization scale reduced by a factor of 2 and the A14 Var3c Up tune variation~\cite{ATL-PHYS-PUB-2014-021}.
The second sample, which decreases the amount of initial- and final-state radiation, uses $h_{\rm damp} = 1.5\mt$, the factorization and renormalization scale increased by a factor of 2 and the A14 Var3c Down tune variation \cite{ATL-PHYS-PUB-2014-021}.

An alternative matrix element calculation and matching with the parton shower is realized 
with the \MGMCatNLO{}\ event generator (hereafter referred to as \MGAMCatNLO) \cite{MCATNLO}\ interfaced 
with the \PYTHIA{8} parton shower and hadronization model using the same tune as the nominal sample. 
This sample requires the leading top quark in each event to have $\pT > 300$~\GeV\ to ensure that
the high-\pT\ region is adequately populated.
The effects of using alternative parton shower and hadronization models is probed by interfacing the nominal \POWHEG{}\ setup 
with the \HerwigSeven\ parton shower and hadronization model
\cite{HerwigSeven} employing the H7UE tune (hereafter also referred to
as PWG+H7).   
Another calculation using the 
\SHERPA~v2.2.1 event generator \cite{Gleisberg:2008ta} with the default \textsc{Sherpa}\ parton shower and hadronization model
merges the NLO \ttbar\ matrix element with matrix element calculations including up to four additional jets using the \textsc{MEPS@NLO}\ setup \cite{Hoeche:2012yf}.

The $Wt$\ single-top-quark processes are modeled using 
the \textsc{Powheg-Box}~v2 event generator with the 
CT10 PDF set \cite{CT10}. 
For the single-top-quark process, the top quarks are decayed 
using \textsc{MadSpin}~\cite{Artoisenet:2012st}. 
The parton shower, fragmentation and the underlying event for these processes are simulated using 
the \PYTHIA~6.428 event generator \cite{Sjostrand:2006za} with the CTEQ6L1 PDF sets and the corresponding Perugia 2012 tune (P2012)~\cite{Skands:2010ak}. 
Electroweak $t$- and $s$-channel single-top-quark events are not explicitly modeled because of the small cross-section 
of these processes and the low jet multiplicity in the final state. 
Their contribution is accounted for in the data-driven background estimate.

The associated production of \ttbar\ pairs with \Wboson, \Zzero\ and Higgs bosons is modeled using 
the \MGAMCatNLO\ event generator \cite{MCATNLO}\ coupled 
to the \PYTHIA{8}\ parton shower and hadronization model using the same PDF sets and tunes as the \ttbar{} sample.

The top-quark mass is set to $\mt=172.5$~\GeV\ for all samples and the renormalization and factorization scales are 
set to $\mu_{\rm R/F} = \sqrt{\mt^2 + \frac{1}{2}(\pT(t)^2 + \pT(\bar{t})^2)}$ for all \ttbar\ samples except where explicity noted above.
 The EvtGen v1.2.0 program~\cite{EvtGen} is used for modeling the properties of the bottom and charm hadron decays for 
all event generator setups other than for the \SHERPA\ sample. 

The $t\bar{t}$ samples are normalized using the next-to-next-to-leading-order cross-section plus 
next-to-next-to-leading-logarithm corrections (NNLO+NNLL) 
$\sigma_{t\bar{t}} = 832^{+46}_{-51}$~pb~\cite{Czakon:2013goa}, where the uncertainties reflect the effect of scale
and PDF variations. 
The single-top-quark cross-section is normalized to the NLO 
predictions~\cite{Kidonakis:2013}. 
The associated production of \ttbar\ pairs with \Wboson, \Zzero\ and Higgs bosons are normalized to 
0.603~pb, 0.586~pb and 0.231~pb, respectively, as predicted by the \MGAMCatNLO\ event generator.
\section{Event reconstruction and selection}
\label{sec:selection}

This analysis makes use of jets, electrons and muons as well as event-based measures formed from their
combinations. 
The event reconstruction and selection are summarized in the following subsections.

\subsection{Event reconstruction}
\label{sec:objects_selection}

Electron candidates are identified from high-quality inner detector tracks matched to 
calorimeter deposits consistent with an electromagnetic shower.  
The calorimeter deposits have to form a cluster with $E_{\rm T} >$ 25~\GeV, $|\eta| < 2.47$\ and 
be outside the transition region $1.37 \leq |\eta| \leq 1.52$\ 
between the barrel and endcap calorimeters. 
A  likelihood-based requirement is used to suppress misidentified jets 
(hereafter referred to as fakes), 
and calorimeter- and track-based isolation requirements are
imposed~\cite{PERF-2016-01, ATLAS-CONF-2016-024}.  
Overall, these criteria result in electron identification efficiencies of $\sim90$\%\ for 
electrons with $\pt > 25$~\GeV\ and 96\%\ for electrons with $\pt > 60$~\GeV.

Muon candidates are reconstructed using high-quality inner detector tracks combined with tracks reconstructed in the muon 
spectrometer.
Only muon candidates with $\pt > 25$~\GeV\ and $|\eta| < 2.5$\ are considered.
Isolation criteria similar to those used for electrons are used~\cite{PERF-2015-10}. 
To reduce the impact of non-prompt leptons, muons within 
$\Delta R =\sqrt{(\Delta\eta)^2+(\Delta\phi)^2} = 0.4$\ of a jet are removed. 

The \antikt\ algorithm implemented in the FastJet package \cite{akt1,Cacciari:2011ma}\ is used to define two types of jets for this analysis: \smallR\ jets with a radius 
parameter of $R = 0.4$\ and \largeR\ jets with $R = 1.0$.
These are reconstructed independently of each other from topological clusters in the
calorimeter.
The clusters used as input to the \largeR\ jet reconstruction are calibrated using the local calibration method described in 
Ref.~\cite{PERF-2011-03}.
The \smallR\ jet energy scale is obtained by using an energy- and
$\eta$-dependent calibration scheme resulting from simulation and {\it in situ}\
corrections based on data~\cite{jer_2, jes:2013,Aad:2012vm,PERF-2016-04}.
Only  \smallR\ jets that have $|\eta| <$ 2.5 and $\pt > 25$~\GeV\ are considered. 
To reduce pileup effects, an algorithm that determines whether the primary vertex is the origin of the 
charged-particle tracks associated with a jet candidate is used to
reject jets coming from other interactions~\cite{PERF-2014-03}.
This is done only for jet candidates with $\pt<50$~\GeV\ and $|\eta|<$ 2.4. 
The \smallR\ jet closest to an electron candidate is removed if they are separated by no more than $\Delta R = 0.2$.
Small-$R$\ jets containing $b$-hadrons are identified ($b$-tagged) using a multivariate discriminant that 
combines information about secondary vertices and impact parameters.  
The \smallR\ jets are considered $b$-tagged if the value of the discriminant is larger than a threshold that 
provides 70\%\ efficiency. 
The corresponding rejection factors for gluon/light-quark jets and charm-quark jets are 
approximately 125 and 4.5, respectively \cite{PERF-2012-04,ATL-PHYS-PUB-2016-012}.

The \largeR\ jet energy scale is derived by using energy-
and $\eta$-dependent calibration factors derived from
simulation and {\it in situ}\ measurements~\cite{jer_2,jes:2013,atlasJetSubstructure}.
The \largeR\ jet candidates are required to have $|\eta| < 2.0$\
and $\pt >300$~\GeV. 
A trimming algorithm~\cite{trimming}\ with parameters $R_{\rm sub} = 0.2$ and $f_{\rm cut} = 0.05$\ is 
applied to suppress gluon radiation and further mitigate pileup effects. 
A top-tagging algorithm~\cite{ATL-PHYS-PUB-2015-053} is applied that 
consists of \pT-dependent requirements on two variables: the jet mass $m_{J}$, measured from clusters in the calorimeter, 
and the $N$-subjettiness ratio $\tau_{32}$~\cite{thaler_2011,thaler_2012}. 
The $N$-subjettiness variable $\tau_{N}$ expresses how well a jet can be described as containing $N$ or fewer subjets. 
The ratio $\tau_{32}=\tau_{3}/\tau_{2}$ allows discrimination between jets containing a three-prong structure and jets containing a two-prong structure.
The \pT-dependent requirements provide a 50\%\ top-quark tagging efficiency independent of \pT, with 
a light-quark and gluon jet rejection factor
of $\sim 17$\ at $\pT=500$~\GeV\ and decreasing with increasing \pT\ to $\sim 10$\ at $\pT=1$~\TeV. 
This combination of variables used with trimmed \largeR\ jets provides the necessary rejection for this analysis, and is
insensitive to the effects of pileup.

\subsection{Event selection}
\label{sec:events_selection}

The event selection identifies fully hadronic \ttbar\ events where both top quarks have high \pT.
Each event is required to have a primary vertex with five or more associated tracks with $\pT > 0.4$~\GeV.
In order to reject top-quark events where a top quark has decayed semileptonically, the 
events are required to contain no reconstructed
electron or muon candidate.
To identify the fully hadronic decay topology,  events must have at least two \largeR\ jets 
with $\pt > 350$ \GeV, $|\eta| < 2.0$ and $ |m_{J} - \mt| < 50$ \GeV, 
where the top-quark mass \mt\ is set to $172.5$~\GeV. 
The leading jet is required to have  $\pt > 500$ \GeV\
and the event must contain at least two \smallR\ jets with $\pt > 25$ \GeV\  and $|\eta| < 2.5$. 
This preselection results in an event sample of 22.7 million events.

To reject multijet background events, the two highest \pt\ \largeR\ jets
must satisfy the top-tagging criteria described in Sec.~\ref{sec:objects_selection}.  
Furthermore, both top-tagged \largeR\ jets are required to have an associated \smallR\ $b$-tagged jet.
This association, hence referred to as $b$-matching, is made by requiring $\Delta R$ < 1.0 between the \smallR\ and 
\largeR\ jets.
These two highest \pT\ \largeR\ jets are the leading and second-leading top-quark candidate jets (or ``top-quark jets'' in what follows).
The candidate \ttbar\ final state is defined as the sum of the four-momenta of the two \largeR\ top-quark jets.

This selection defines the signal region, which has 3541 events.

\section{Background estimation}
\label{sec:backgrounds}

There are two categories of background sources: those involving one or more top quarks in the final state
and those sources where no top quark is involved.
The background processes involving top quarks are estimated using MC calculations.
The largest background source is events where the two leading jets both arise from gluons or
$u$, $d$, $s$, $c$, or $b$~quarks (which are referred to as ``multijet'' events).
Monte Carlo predictions of multijet events have large uncertainties coming  from the 
relatively poorly understood higher-order contributions that 
produce a pair of massive jets \cite{STDM-2013-11,STDM-2011-38}.
To avoid these large uncertainties the multijet background is determined using a data-driven technique. 
A similar method was used in previous work \cite{TOPQ-2012-15}.

A  \POWHEG{}+\PythiaEight{}\ \ttbar\ sample is used to estimate the number of \ttbar\ events in the 
sample that arise from at least one top quark decaying semileptonically.
This includes contributions from decays resulting in $\tau$\ leptons, as no attempt is made to 
identify $\tau$\ lepton candidates and reject them.
The rate is estimated to be only $\sim 4$\% in the signal region, primarily due to the top-tagging requirements.  
However, this category of \ttbar\ events contributes to control and validation regions where the 
top-tagging and/or $b$-tagging requirements are relaxed.
Thus, this MC prediction is used to estimate this contamination.
Single-top-quark production in the $Wt$-channel makes a small contribution to the signal sample, which is estimated using the MC predictions described earlier.  
The $t$-channel single-top-quark process is not included, but is partially accounted for in the multijet background estimate.

\begin{table}[b!]
\begin{center}
\caption{Region labels and expected proportion of \ttbar{} events used for the data-driven 
background prediction of multijet events.
A top-quark tagged jet is defined by the tagging algorithm described in the text, and denoted ``1t'' in the table, while
a jet that is not top-tagged is labeled ``0t''.
A $b$-match is defined as $\Delta R(J,b) <$ 1.0, where $J$ represents a \largeR~ jet 
and ``$b$'' represents a $b$-tagged jet.
The labels ``1b'' and ``0b'' represent \largeR\ jets that either have or not have a $b$-match.
Regions $K$, $L$, $N$ and $M$ have an expected contribution from sources involving one or more top quarks of at least 
15\%\ of the observed yield. 
In other regions, the expected contribution from signal and backgrounds involving top quarks is 
less than 15\%\ of the observed event rate.
}
\vspace{0.4cm}
\small{
\newcommand\T{\rule{0pt}{2.6ex}}
\begin{tabular}{c|c|c|c|c|c|}
\cline{2-6}
\multirow{4}{*}{\rotatebox{90}{2nd large-$R$ jet}}&1t1b & 
\cellcolor{green}J (7.6\%) & \cellcolor{cyan}K (21\%) & 
\cellcolor{cyan}L (42\%) & \cellcolor{red}S \T \\ \cline{2-6}
&  0t1b & \cellcolor{green}B (2.2\%) & \cellcolor{green}D (5.8\%) & 
\cellcolor{green}H (13\%) & \cellcolor{cyan}N (47\%) \T \\ \cline{2-6}
&  1t0b & \cellcolor{green}E (0.7\%) & \cellcolor{green}F (2.4\%) & 
\cellcolor{green}G (6.4\%) & \cellcolor{cyan}M (30\%) \T \\ \cline{2-6}
&  0t0b & \cellcolor{green}A (0.2\%) & \cellcolor{green}C (0.8\%) & 
\cellcolor{green}I (2.2\%) & \cellcolor{green}O (11\%) \T \\ \cline{2-6}
& &0t0b&1t0b&0t1b&1t1b \T \\ \cline{2-6}
\multicolumn{1}{c}{}& \multicolumn{5}{c}{Leading \largeR\ jet} \T \\
\end{tabular}
}
\label{tab:ABCD:16regions}
\end{center}
\end{table}

The data-driven multijet background estimate is performed using a set of control regions.
Sixteen separate regions are defined by classifying each event in the preselection
sample according to whether the leading and second-leading jets are top-tagged or $b$-tagged. 
Table~\ref{tab:ABCD:16regions}\ shows the 16 regions that are defined in this way, and illustrates
the proportion of expected \ttbar\ events in each region relative to the observed rate. 
Region S is the signal region, while the regions with no $b$-tags (A, C, E and F) and the regions
with one $b$-tag and no top-tags (B and I) are dominated by multijet backgrounds.

After subtracting the estimated contributions of the \ttbar\ signal and of the other background sources to each of the control
regions, the number of events in region J divided by the number of events in region A gives an estimate of the 
ratio of the number of multijet events in region S to the number of multijet
events in region O.  

Thus one can use these relationships to estimate the multijet background rate in region S, i.e. $S=O\times J/A$,
where $O$, $J$\ and $A$\ are the number of observed events in each region, while $S$\ is the estimate of
the multijet background in region S.

This ``ABCD'' estimate assumes that the mistagging rate of the leading jet does not depend on how the 
second-leading jet is tagged. 
This assumption is avoided by measuring the correlations in
background-dominated regions, e.g. comparing the ratio of the numbers
of events in regions F and E (giving the leading jet top-tagging rate when the second-leading jet is top-tagged) with 
the ratios of events in regions C and A (giving the leading jet top-tagging rate when the second leading jet is 
not top-tagged).  
This results in a refined data-driven estimate of the size of the multijet background given by
\begin{eqnarray}
 S &=& \dfrac{J \times O}{A}\cdot\dfrac{D\times A}{B\times C}\cdot\dfrac{G\times A}{E\times I}\cdot\dfrac{F\times A}{E\times C}\cdot\dfrac{H\times A}{B\times I}, \nonumber \\
&=& 
\dfrac{ J\times O \times H \times F \times D\times G\times A^{3}}{\left( B\times E\times C\times I \right)^{2}},
\label{eq:ABCD16}
\end{eqnarray}
where the region name is the number of observed events in that region. 
The measured correlations in the tagging of background jets result in
an increase of $(12\pm3)$\%\ in the background estimate compared with the estimate assuming that the tagging rates are independent.  
This estimate is also valid when a variable characterizing the kinematics of the events in all the regions is 
further restricted to range between specific values. 
This provides a bin-by-bin data-driven background estimate with
uncertainties that come from the number of events in 
the regions used in Eq.~(\ref{eq:ABCD16}).

Regions L and N are estimated to consist of approximately  equal
numbers of $t\bar{t}$ signal events and multijet background events.
They are used as validation regions to verify that the 
signal and background estimates are robust. In these cases, the multijet background is estimated using different combinations of control regions, namely $N = H\times D/B$ and $L = H\times G/I$.

The number of multijet events in the signal region is calculated by applying Eq.~(\ref{eq:ABCD16}) to the 
number of events in the control regions.  
This results in an estimate of $810\pm 50$\ multijet events in the signal region, where the uncertainty
takes into account the statistical uncertainties as well as the systematic uncertainties in the \ttbar\ 
signal subtraction.

\begin{figure*}[htbp]
\centering
\subfigure[]{ \includegraphics[width=0.45\textwidth]{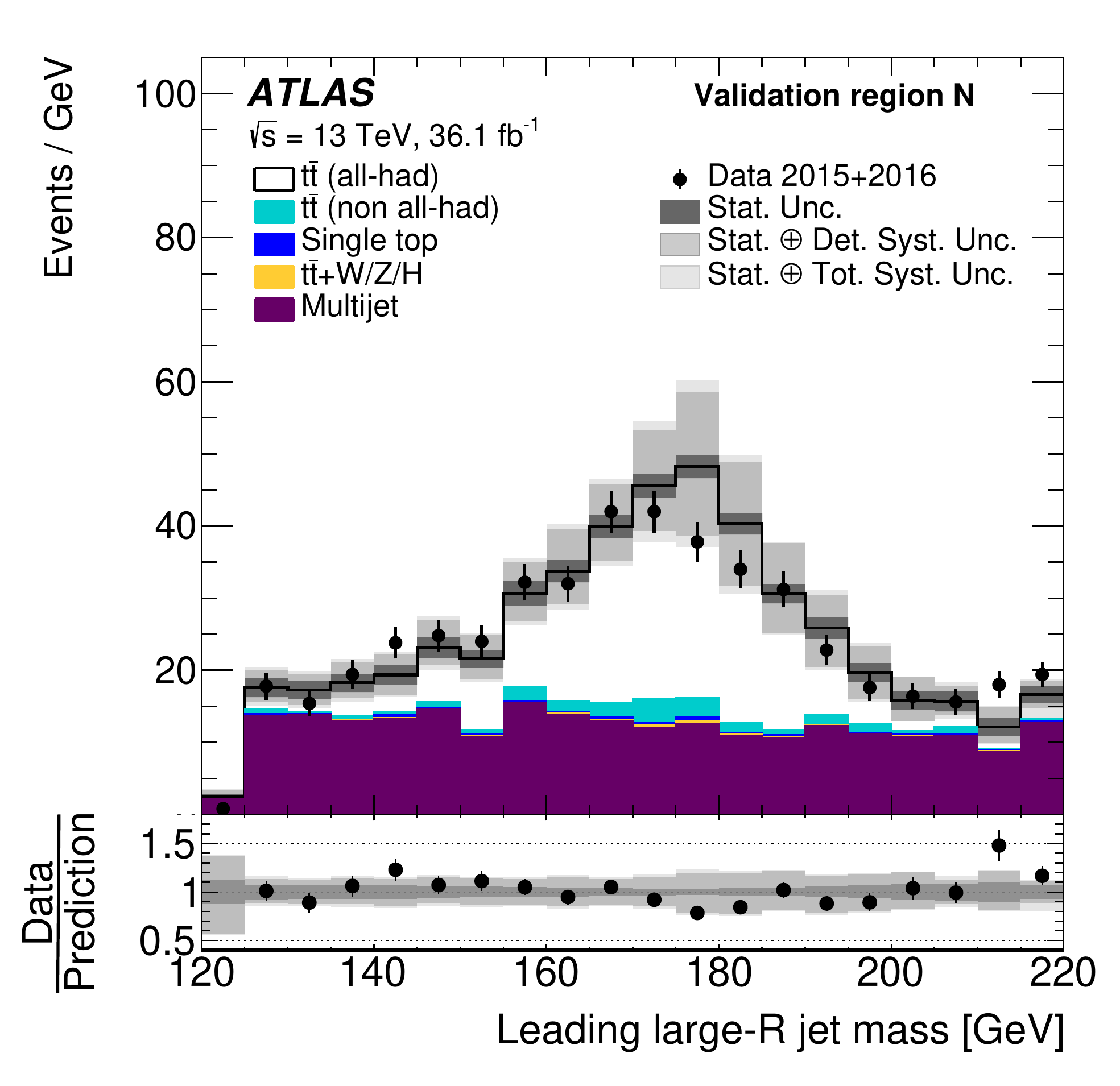}\label{fig:VR:t1_M}}
\subfigure[]{ \includegraphics[width=0.45\textwidth]{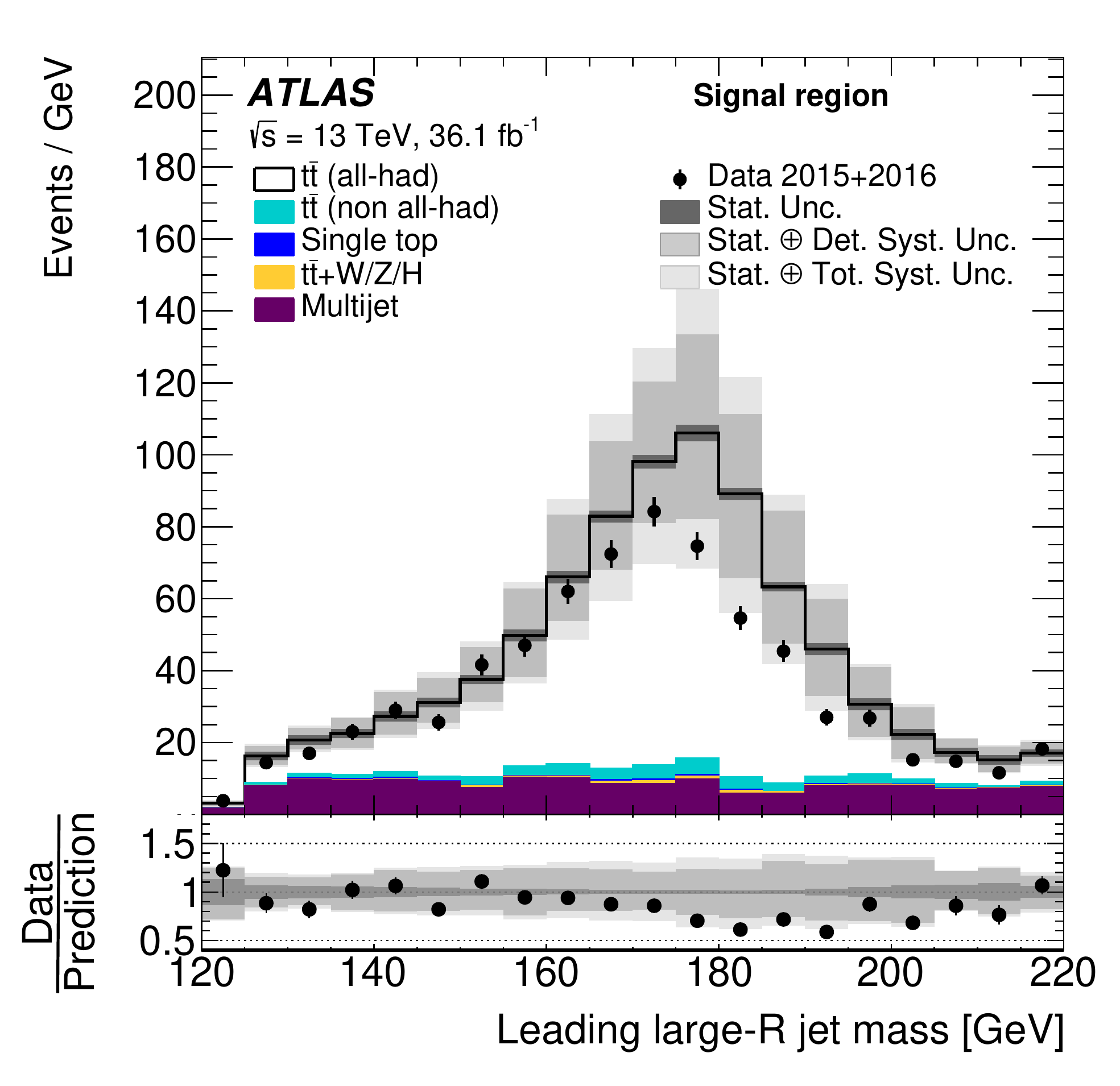}\label{fig:SR:t1_M}}
\subfigure[]{ \includegraphics[width=0.45\textwidth]{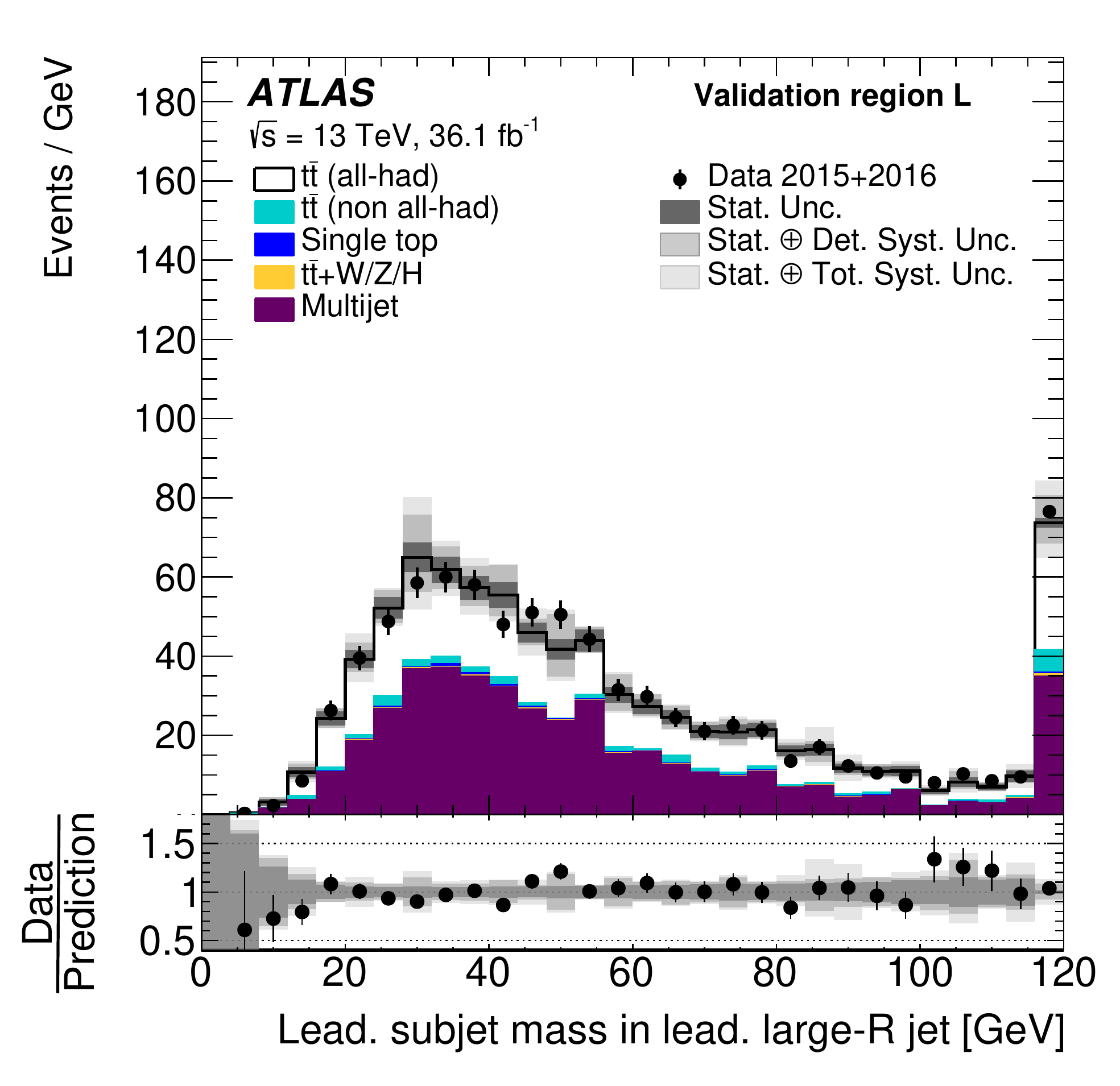}\label{fig:VR:t1_sjM}}
\subfigure[]{ \includegraphics[width=0.45\textwidth]{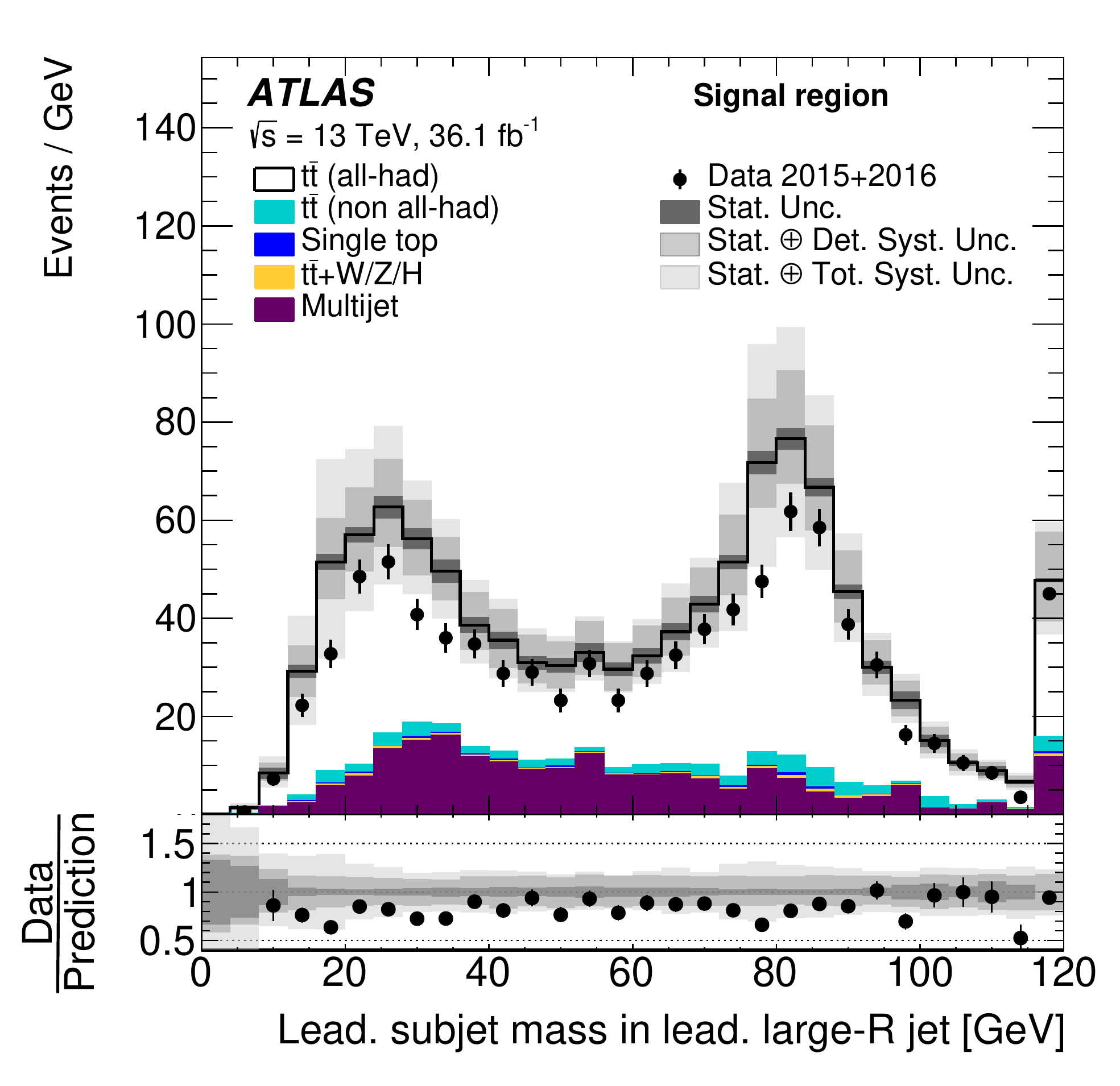}\label{fig:SR:t1_sjM}}
\caption{\small{Kinematic distributions of top-quark candidate jets in the signal region S and in the two validation regions N and L.  
The leading \largeR\ jet mass distributions for the events in the validation region N 
and the signal region S
are shown in \subref{fig:VR:t1_M}\ and \subref{fig:SR:t1_M}, respectively.  
The mass distribution of the leading small-$R$ subjet in the leading \largeR\ jet for events in the validation 
region L and in the signal region are 
shown in \subref{fig:VR:t1_sjM}\ and \subref{fig:SR:t1_sjM}, respectively.
The signal prediction (open histogram) is based on the \POWHEG{}+\PythiaEight{} event generator normalized to the NNLO+NLL cross-section.
The background is the sum of the data-driven multijet estimate (dark histogram) and the MC-based expectation for the contributions of non-all-hadronic \ttbar\ and single-top-quark processes. Events beyond the $x$-axis range are included in the last bin.  
The gray area indicates the combined statistical and systematic uncertainties, 
including \ttbar\ modeling uncertainties.
}}
\label{fig:VRSR:t1_M}
\end{figure*}

\begin{figure*}[htbp]
\centering
\subfigure[]{ \includegraphics[width=0.45\textwidth]{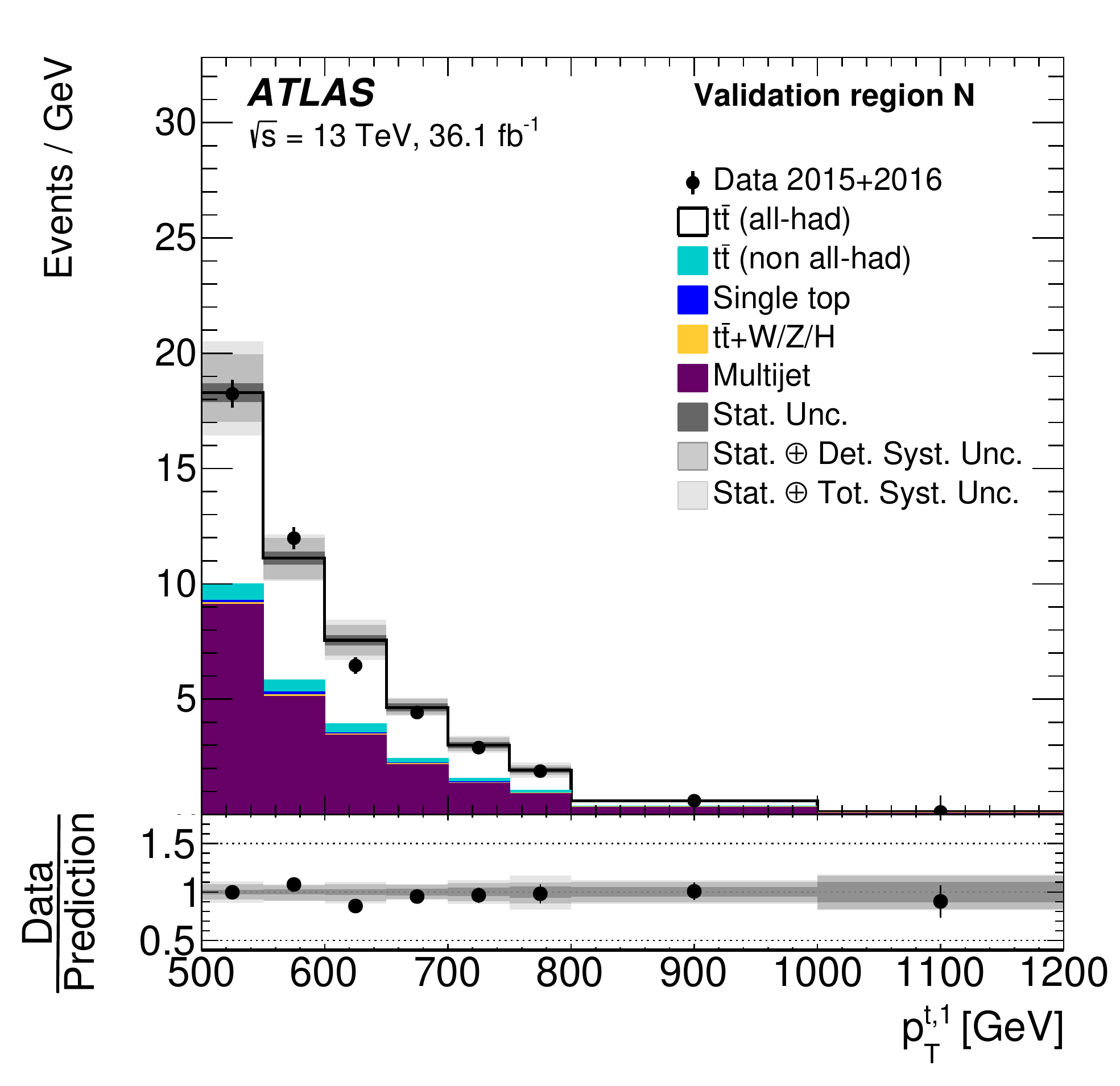}\label{fig:CP:VR:t1_pt}}
\subfigure[]{ \includegraphics[width=0.45\textwidth]{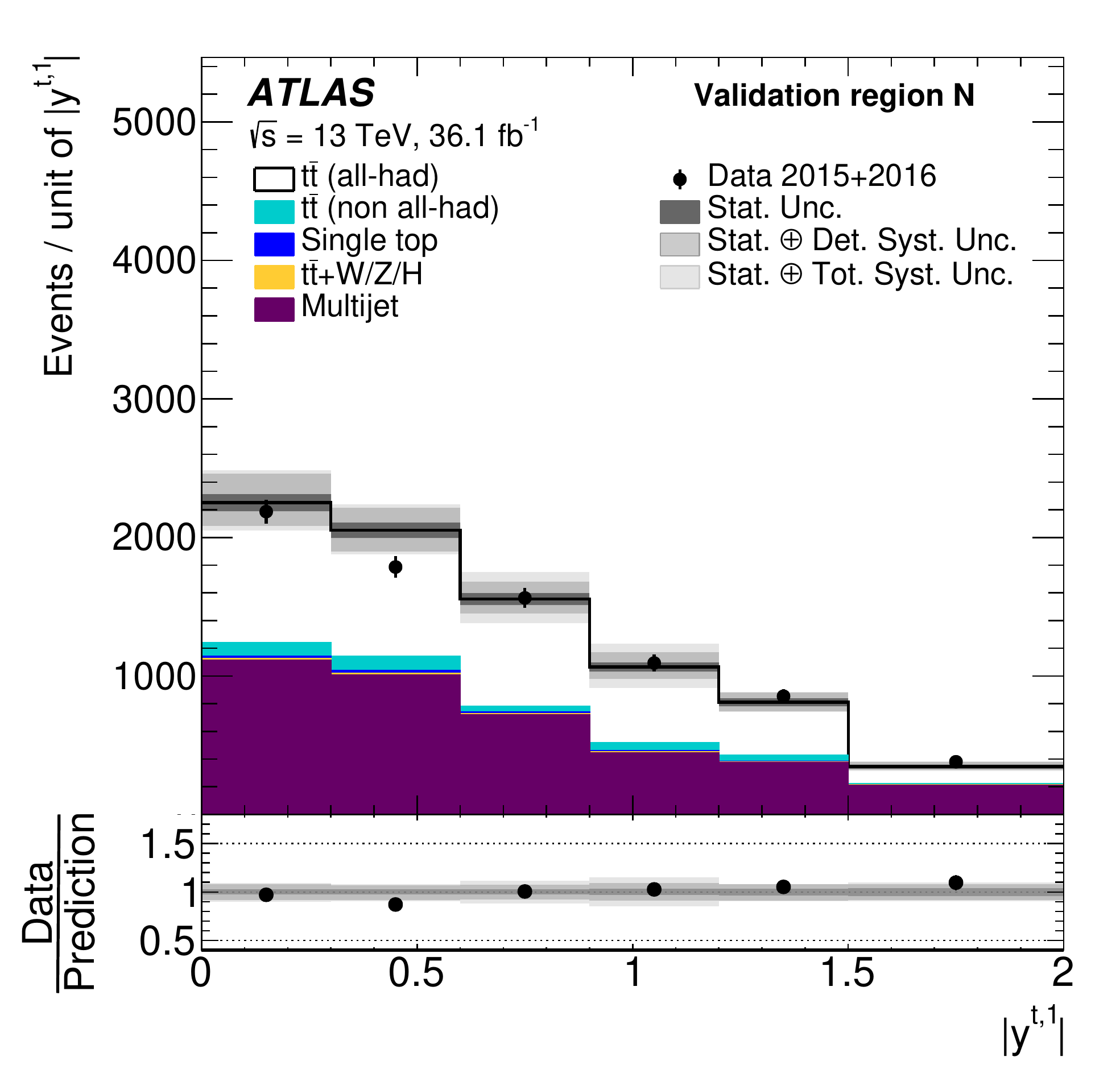}\label{fig:CP:VR:t1_y}}
\subfigure[]{ \includegraphics[width=0.45\textwidth]{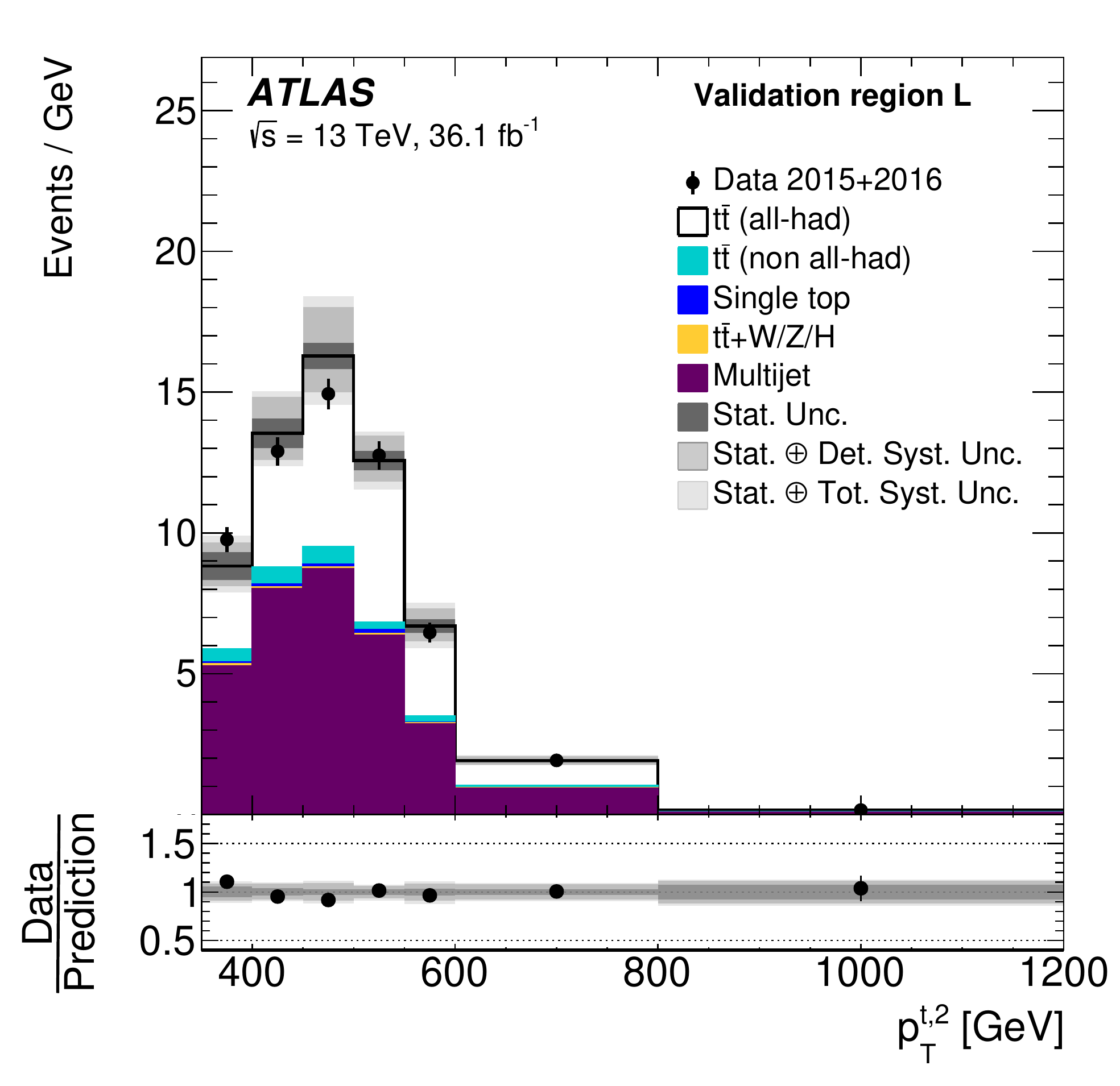}\label{fig:CP:SVR:t2_pt}}
\subfigure[]{ \includegraphics[width=0.45\textwidth]{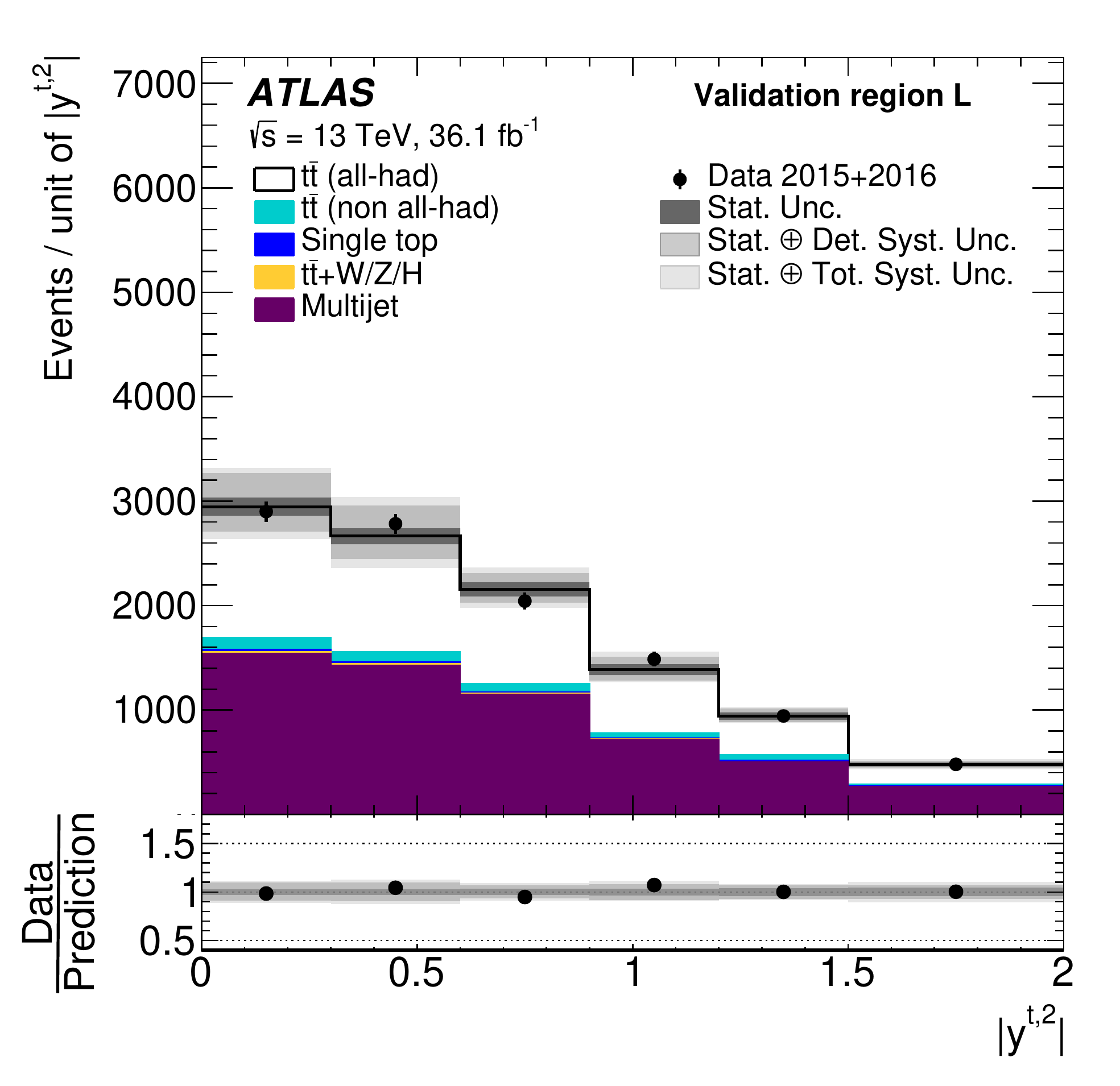}\label{fig:CP:SVR:t2_y}}
\caption{\small{Kinematic distributions of top-quark candidate jets in the two validation regions $N$ and $L$: 
\subref{fig:CP:VR:t1_pt} transverse momentum and \subref{fig:CP:VR:t1_y} absolute value of the rapidity of the leading \largeR\ jet, 
\subref{fig:CP:SVR:t2_pt} transverse momentum and \subref{fig:CP:SVR:t2_y} absolute value of the rapidity of the second-leading \largeR\ jet. 
The signal prediction (open histogram) is based on the \POWHEG{}+\PythiaEight{} event generator normalized to the NNLO+NLL cross-section. The background is the sum of the data-driven multijet estimate (dark histogram) and the MC-based expectation for the contributions of non-all-hadronic \ttbar\ and single-top-quark processes. Events beyond the $x$-axis range are included in the last bin.
The gray area indicates the combined statistical and systematic uncertainties, 
including \ttbar\ modeling uncertainties.
}}
\label{fig:VRSR:t1_t2_pt}
\end{figure*}

\begin{figure*}[htbp]
\centering
\subfigure[]{ \includegraphics[width=0.45\textwidth]{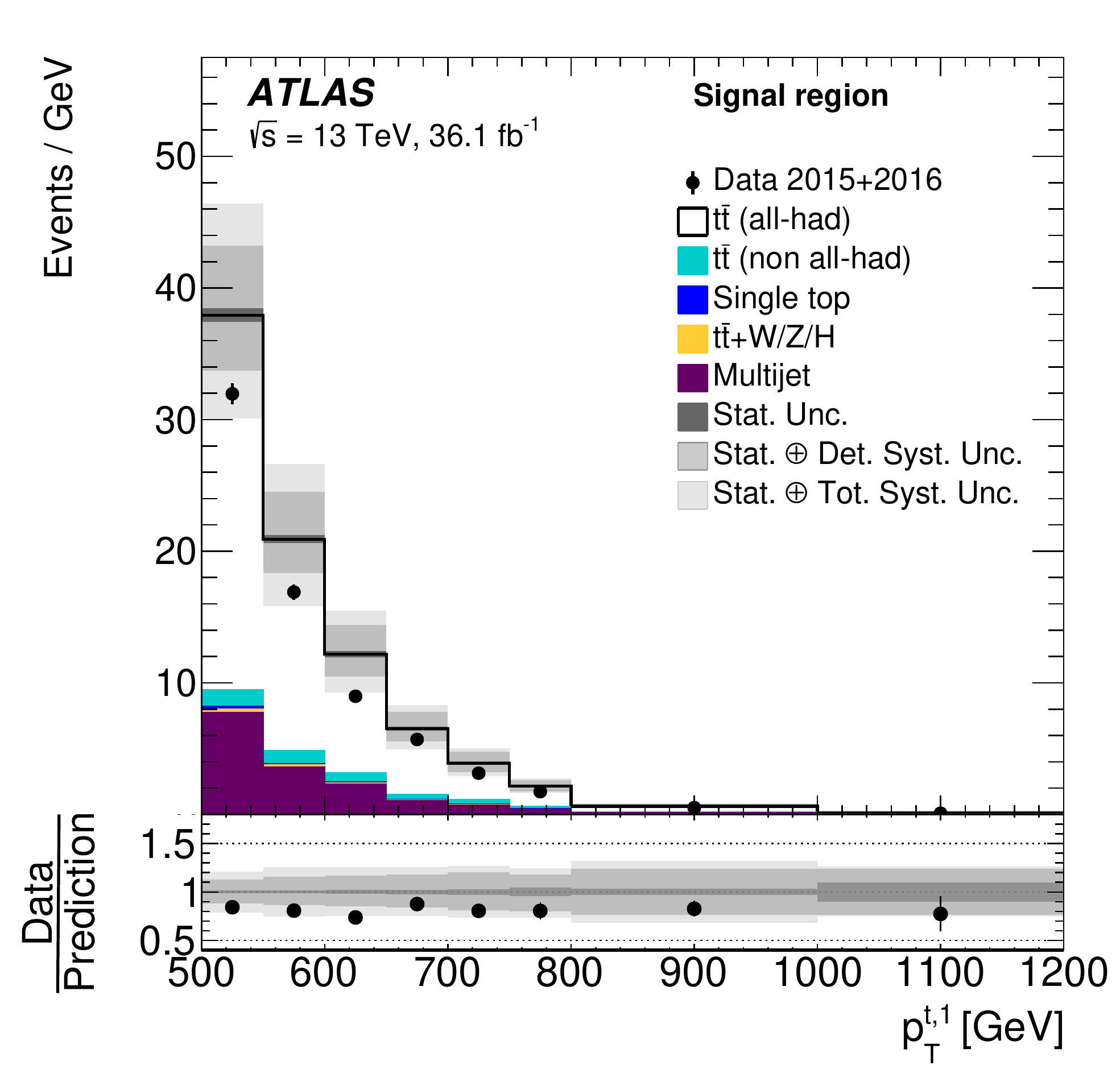}\label{fig:CP:SR:t1_pt}}
\subfigure[]{ \includegraphics[width=0.45\textwidth]{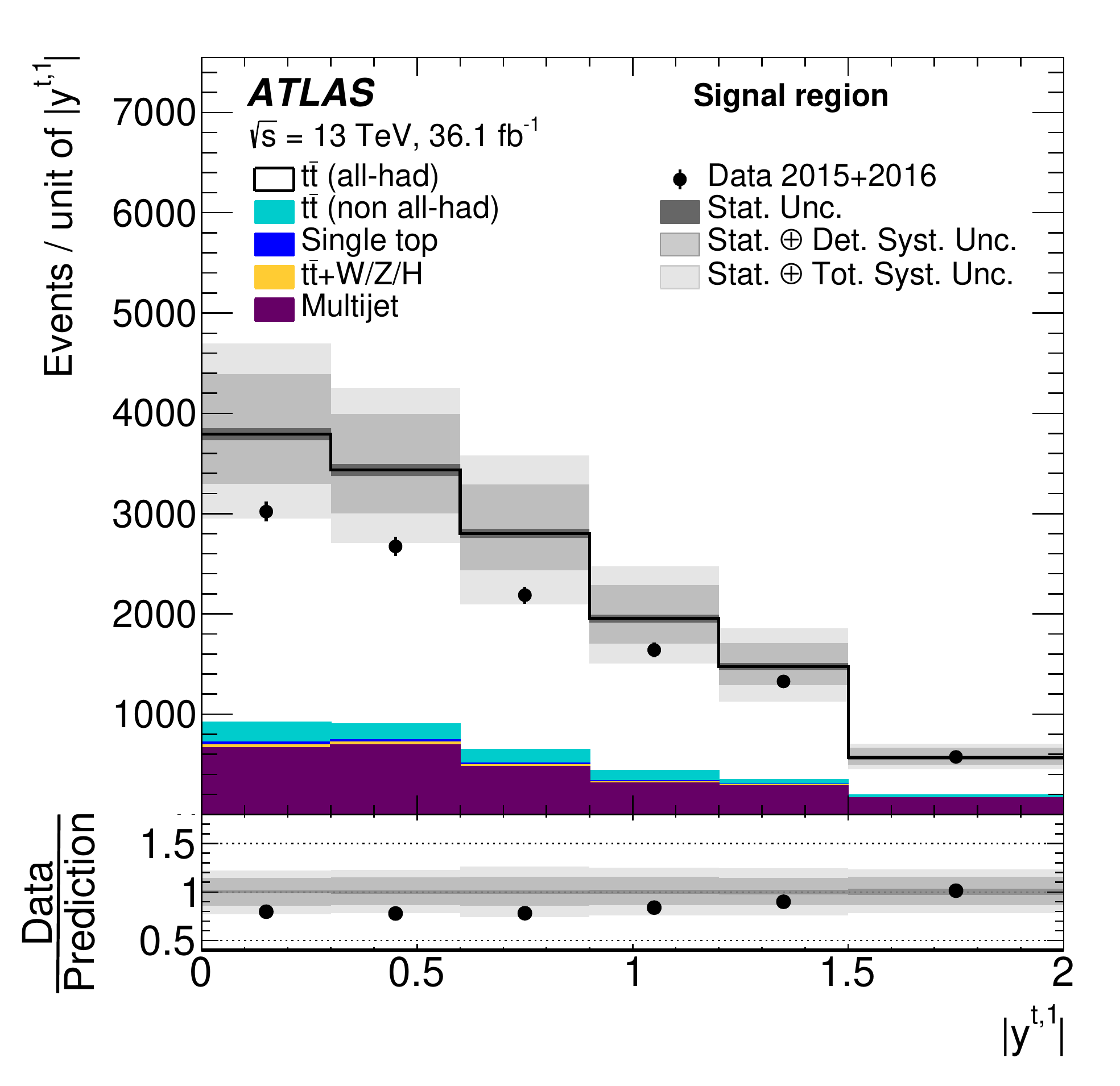}\label{fig:CP:SR:t1_y}}
\subfigure[]{ \includegraphics[width=0.45\textwidth]{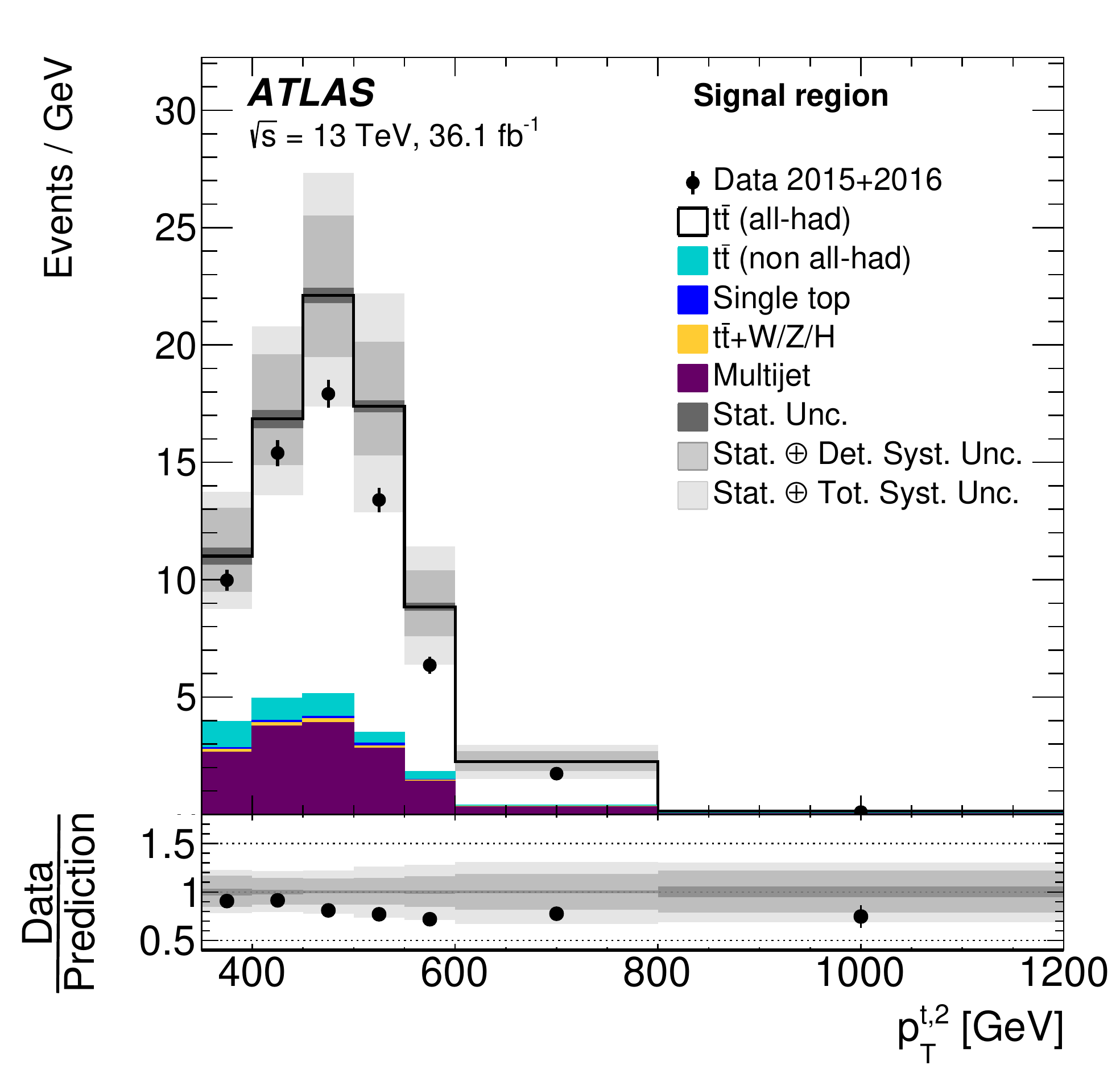}\label{fig:CP:SR:t2_pt}}
\subfigure[]{ \includegraphics[width=0.45\textwidth]{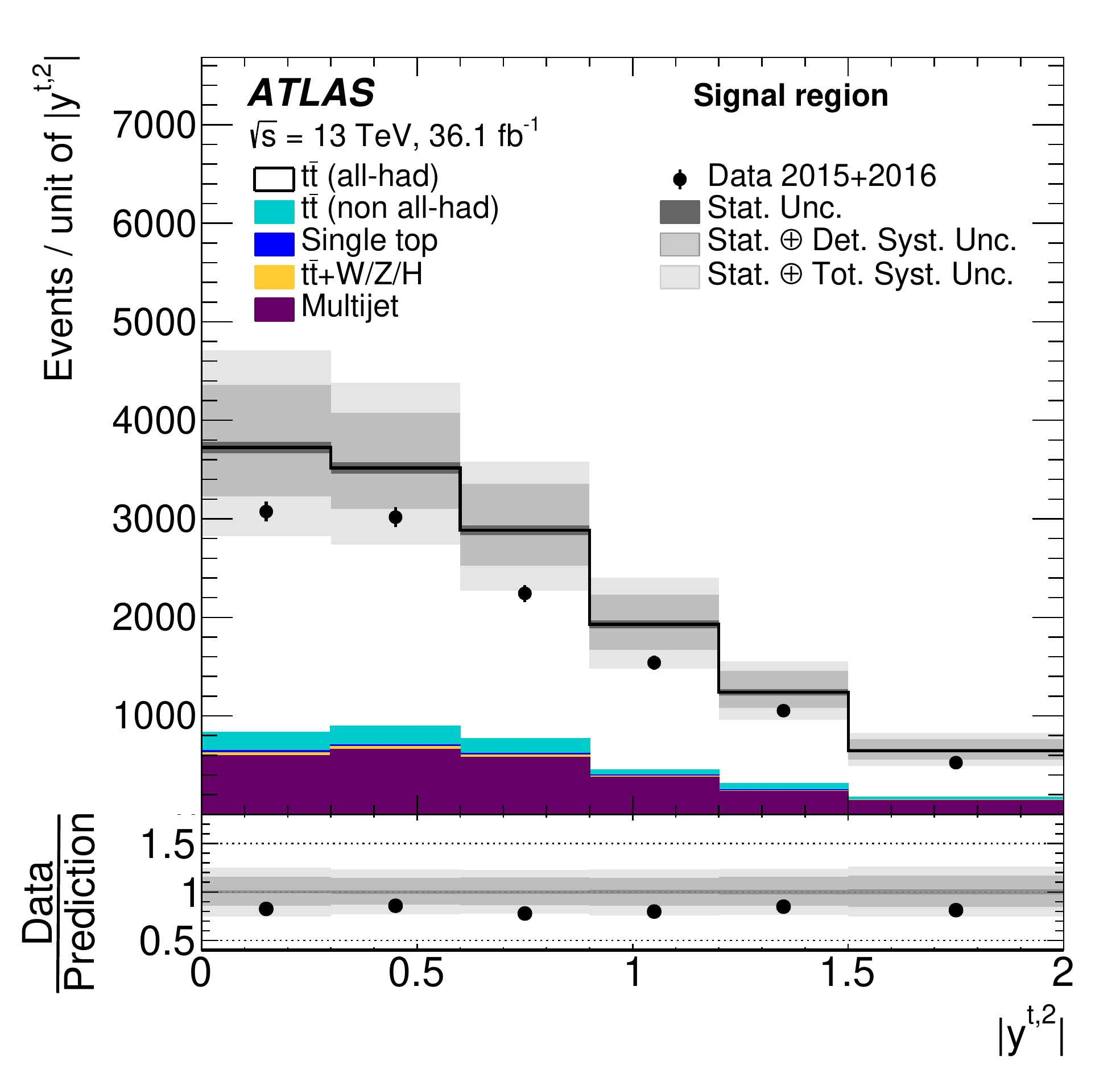}\label{fig:CP:SR:t2_y}}
\caption{\small{Kinematic distributions of top-quark candidate jets in the signal region S: 
\subref{fig:CP:SR:t1_pt}~transverse momentum and \subref{fig:CP:SR:t1_y} absolute value of the rapidity of the leading top-quark jet, 
\subref{fig:CP:SR:t2_pt}~transverse momentum and \subref{fig:CP:SR:t2_y} absolute value of the rapidity of the second-leading top-quark jet. 
The signal prediction (open histogram) is based on the \POWHEG{}+\PythiaEight{} simulation normalized to the NNLO+NLL cross-section.
The background is the sum of the data-driven multijet estimate (dark histogram) and the MC-based expectation for the contributions of 
non-all-hadronic \ttbar\ and single-top-quark processes. 
Events beyond the $x$-axis range are included in the last bin. The gray area indicates the combined statistical and systematic uncertainties in the total prediction, 
including \ttbar\ modeling uncertainties.
}}
\label{fig:CP:SR:top_quarks}
\end{figure*}

There is good agreement in the validation regions between the predicted and observed event yields, as 
well as in the shape of 
distributions that are sensitive to the proportion of \ttbar\ signal and multijet background. 
This is illustrated in Fig.~\ref{fig:VRSR:t1_M}, which compares the \largeR\ jet mass distributions and
the highest-\pT\ subjet mass distribution of the leading jets.
A shift between the measured and predicted jet mass distributions, 
shown in Figs.~\ref{fig:VR:t1_M} and ~\ref{fig:SR:t1_M}, is consistent
with the uncertainties arising from the calibration for \largeR\ jets~\cite{ATLAS-CONF-2016-035}.
The distributions for the leading and second-leading jet \pt\ and rapidity in regions N and L are shown in Fig.~\ref{fig:VRSR:t1_t2_pt}, and can be compared with the signal region
distributions in Fig.~\ref{fig:CP:SR:top_quarks}.

The level of agreement between the observed and predicted distributions in the signal region
can be seen in Fig.~\ref{fig:CP:SR:top_quarks}, which shows the
 distributions of the leading top-quark \pT\ and absolute value of rapidity, as well as the same distributions for the second-leading jet.  

The event yields are summarized in Table~\ref{tab:yields} for the simulated signal, the background
sources and the data sample.  

\begin{table}[t!]
\caption{Event yields in the signal region for the expected \ttbar\ signal process and the background processes.
The sum of these are compared to the observed yield.
The uncertainties represent the sum in quadrature of the statistical and systematic uncertainties in each subsample. 
Neither modeling uncertainties nor uncertainties in the inclusive \ttbar\ cross-section are included in the systematic 
uncertainties.
The single-top-quark background does not include the $t$-channel process.
}
\begin{center}
 \sisetup{
group-minimum-digits = 5,
table-format = 4.0,
separate-uncertainty,
table-align-uncertainty = true
}
 \begin{tabular}{l S}
      \toprule
    \vspace{.10cm}
\ttbar{} (all-hadronic) 	& \num{3250(470)} \\
 \vspace{.10cm}
\ttbar{} (non-all-hadronic) 	& \num{200(40)} \\
 \vspace{.10cm}
Single-top-quark 		& \num{24(12)}  \\ 
  \vspace{.10cm}
\ttbar{}$+W/Z/H$		& \num{33(10)} \\
  \vspace{.10cm}
Multijet events			& \num{810(50)} \\
\hline
\vspace{.10cm}
Prediction${}^{\phantom{1}^{\phantom{1}}}$	& \num{4320(530)} \\
  \vspace{.10cm}
Data (\lumitot) 	 					& \num{3541} \\
 \bottomrule
  \end{tabular}
\end{center}

\label{tab:yields}
\end{table}

\FloatBarrier

\section{Unfolding procedure}
\label{sec:unfolding}

The differential cross-sections are obtained from the data using an unfolding technique that corrects 
for detector effects
such as efficiency, acceptance and resolution. 
This correction is made to the particle level using a fiducial phase space that is defined to match the 
experimental acceptance and hence avoid large MC extrapolations.
The parton-level differential cross-sections are obtained using a similar procedure, but in this case the correction is made
to the top-quark parton after final-state radiation effects have been included in the generation process 
using a limited phase-space region matched to the kinematic acceptance of the analysis. 

In the following subsections, the particle-level fiducial phase space and the parton-level phase space are
defined and the algorithm used for the unfolding is described.

\subsection{Particle-level fiducial phase-space and parton-level phase-space regions}

The particle-level fiducial phase-space definition models the kinematic requirements used to select the \ttbar\ process.

In the MC signal sample, electrons and muons that do not originate from hadron decays are combined or ``dressed'' 
with any photons found in a cone of size $\Delta R = 0.1$ around the lepton direction. 
The four-momentum of each photon in the cone is added to the four-momentum of the lepton to produce the 
dressed lepton. 

Jets are clustered using all stable particles except those used in the definition of dressed electrons and muons 
and neutrinos not from hadron decays, using the anti-$k_t$ algorithm with a radius parameter $R=0.4$ and $R=1.0$ for 
\smallR\  and \largeR\ jets, respectively. 
The decay products of hadronically decaying $\tau$ leptons are included. 
These jets do not include particles from pileup events but do include those from the underlying event. 
Large-$R$ jets are required to have $p_{T}>350$~\GeV\ and a mass within $50$~\GeV\ of the top-quark mass.

The following requirements on particle-level electrons, muons and jets in the all-hadronic \ttbar\ MC events define the particle-level 
fiducial phase space:
\begin{itemize}
\item no dressed electrons or muons with $\pt>25$~\GeV\ and $|\eta|<2.5$\ be in the event,
\item at least two \antikt\ $R=1.0$\ jets with $\pt>350$~\GeV\ and $|\eta|<2.0$,
\item at least one \antikt\ $R=1.0$\ jet with $\pt>500$~\GeV\ and $|\eta|<2.0$,
\item the masses of the two \largeR\ jets be within 50~\GeV\ of the top-quark mass of 172.5~\GeV,
\item at least two \antikt\ $R=0.4$\ jets with $\pt>25$~\GeV\ and $|\eta|<2.5$\ and
\item the two leading $R=1.0$\ jets be matched to a $b$-hadron in the final state using a 
ghost-matching technique as described in Ref.~\cite{Cacciari:2008gn}\ (called top-quark particle jets).
\end{itemize}
 
The parton-level phase space is defined by requiring that the 
leading top quark have $\pT > 500$~\GeV\ and the second-leading
top quark have $\pT > 350$~\GeV. 
No rapidity or other kinematic requirements are made.
This definition avoids a large extrapolation in the unfolding procedure that results in
large systematic uncertainties. 

\subsection{Unfolding algorithm}
The iterative Bayesian method~\cite{unfold:bayes} as implemented in \textsc{RooUnfold}~\cite{Adye:2011gm} is used to correct the detector-level event distributions to their corresponding particle- and parton-level differential cross-sections.
The unfolding starts from the detector-level event distributions after subtraction of the estimated backgrounds.
An acceptance correction $f_{\rm acc}$ is applied that accounts for 
events that are generated outside the fiducial or parton phase space but pass the detector-level selection.

\begin{figure*}[t]
\centering
\subfigure[]{ \includegraphics[width=0.45\textwidth]{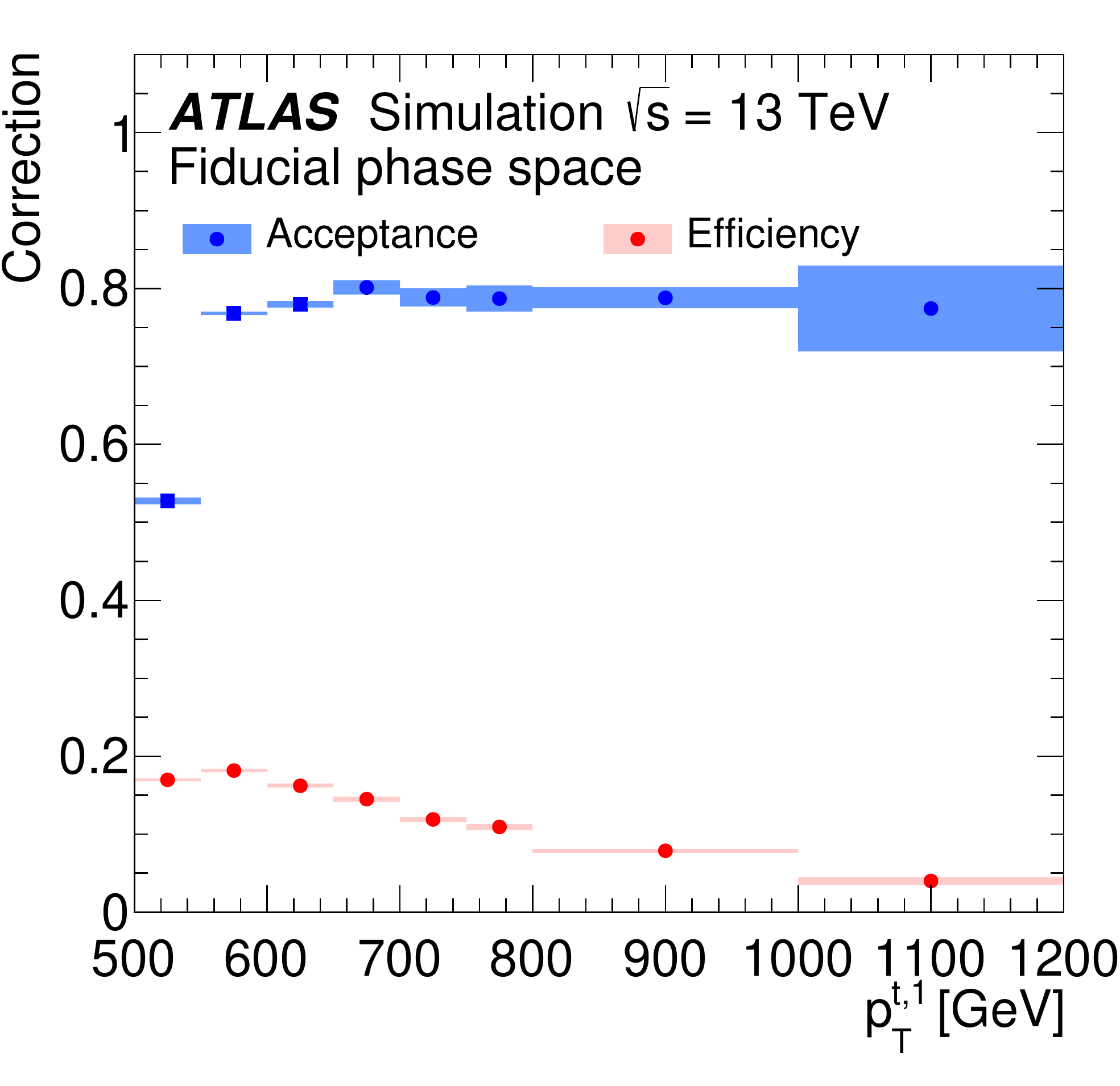}\label{fig:corrections:particle:t1_pt}}
\subfigure[]{ \includegraphics[width=0.45\textwidth]{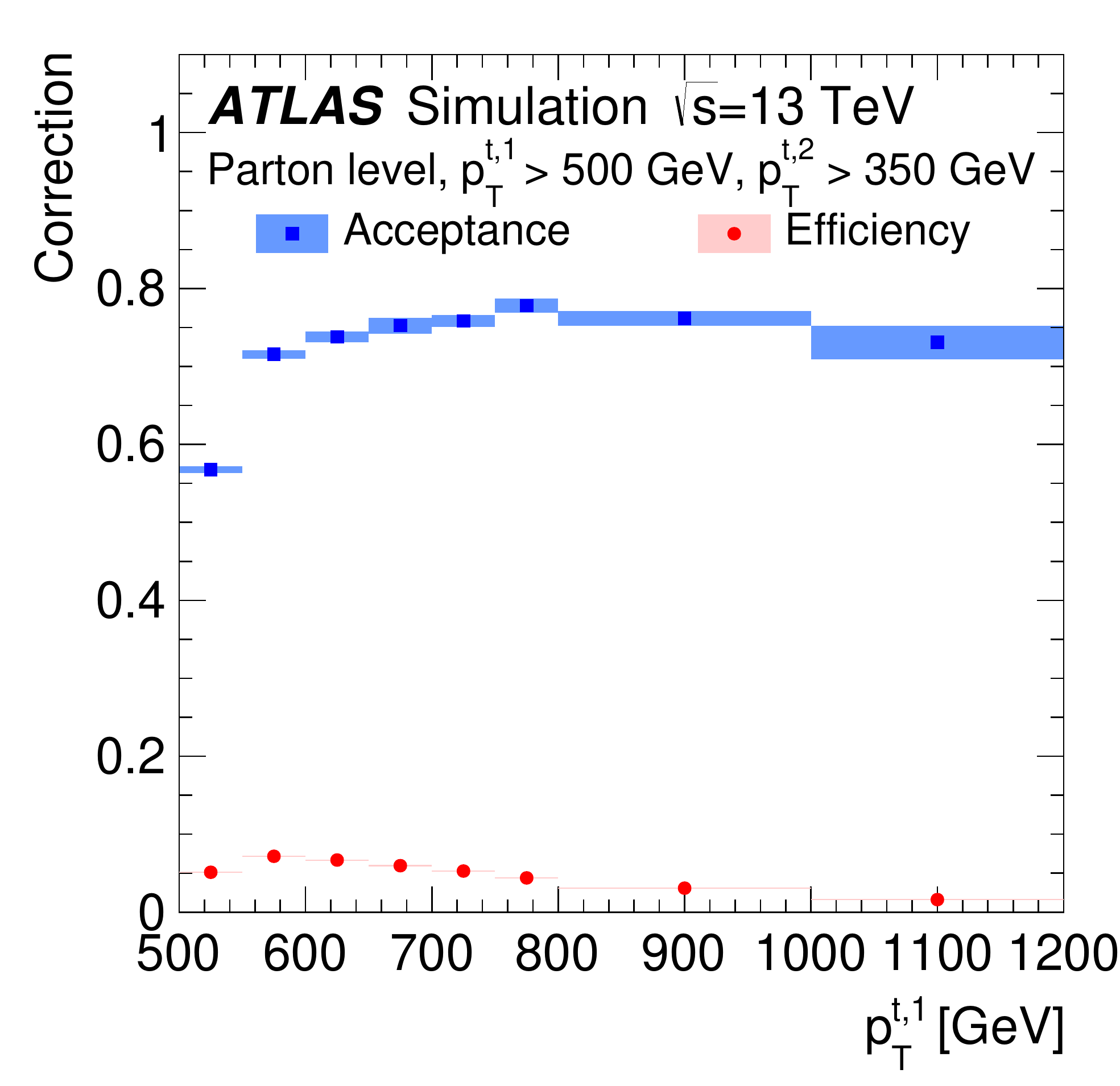}\label{fig:corrections:parton:t1_pt}}
\subfigure[]{ \includegraphics[width=0.45\textwidth]{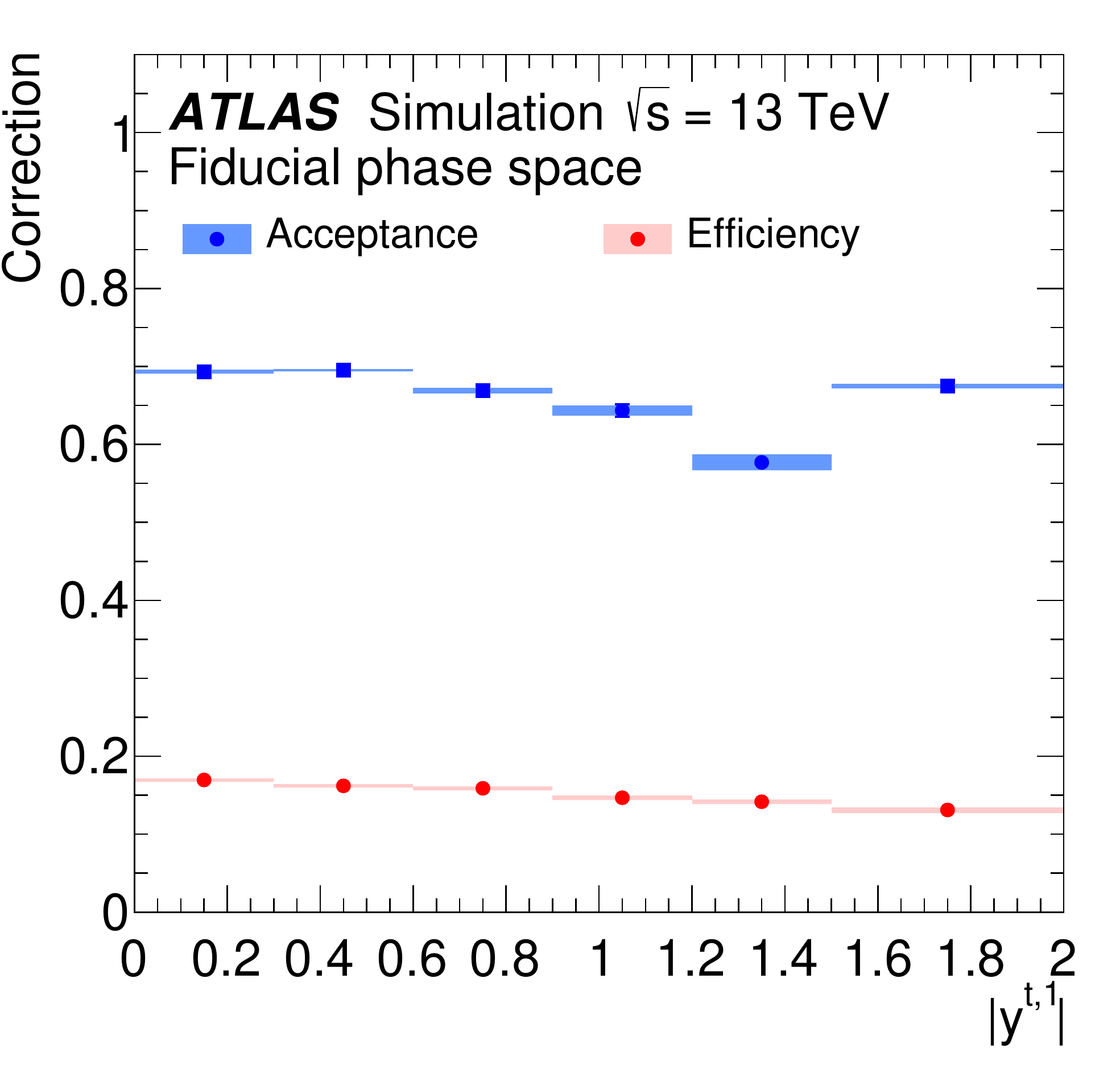}\label{fig:corrections:particle:t1_y}}
\subfigure[]{ \includegraphics[width=0.45\textwidth]{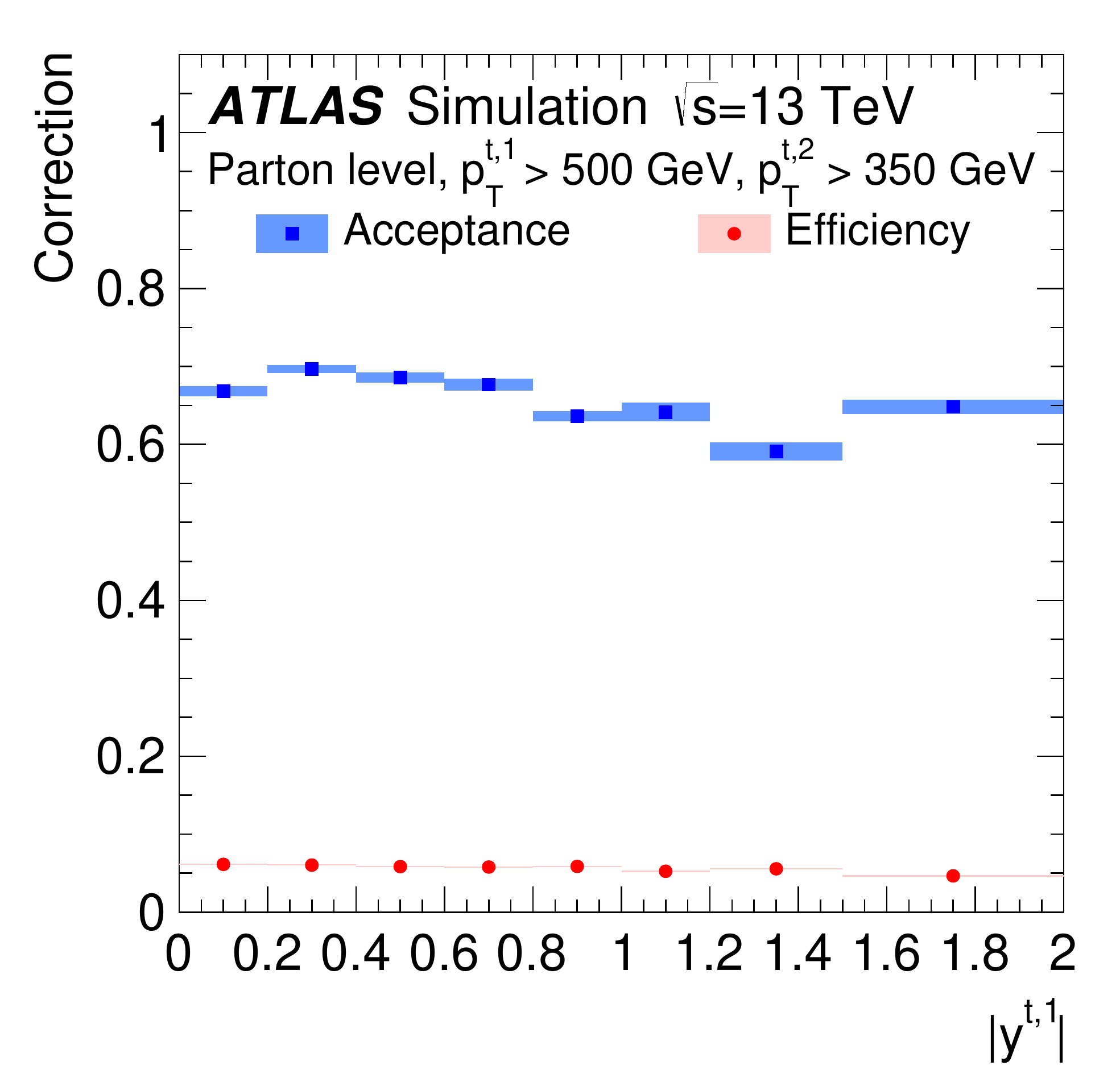}\label{fig:corrections:parton:t1_y}}
\caption{\small{Acceptance and efficiency corrections as a function of \pT\ and $|y|$\ of the leading top-quark jet for
the particle-level phase space in
\subref{fig:corrections:particle:t1_pt}\ and \subref{fig:corrections:particle:t1_y}\ and for the parton-level phase space in
\subref{fig:corrections:parton:t1_pt}\ and \subref{fig:corrections:parton:t1_y}.
The \POWHEG {}+\PythiaEight{} event generator is used as the nominal prediction to correct for detector effects. 
The blue and red areas represent statistical uncertainties.}}
\label{fig:corrections:particle:top_quarks}
\end{figure*}

In order to properly account for resolution and any combinatorial
effects, the detector-level and particle-level (parton-level) objects in MC events are
required to be  well-matched using the angular difference  $\Delta R$.
At particle (parton) level, each top-quark particle-level jet (top quark) is matched to the closest
detector-level jet within $\Delta R < 1.0$, a requirement that ensures
high matching efficiency. The resulting acceptance corrections $f_{\rm acc}^j$ are illustrated in Fig.~\ref{fig:corrections:particle:top_quarks}.

\begin{figure*}[htbp]
\centering
\subfigure[]{ \includegraphics[width=0.45\textwidth]{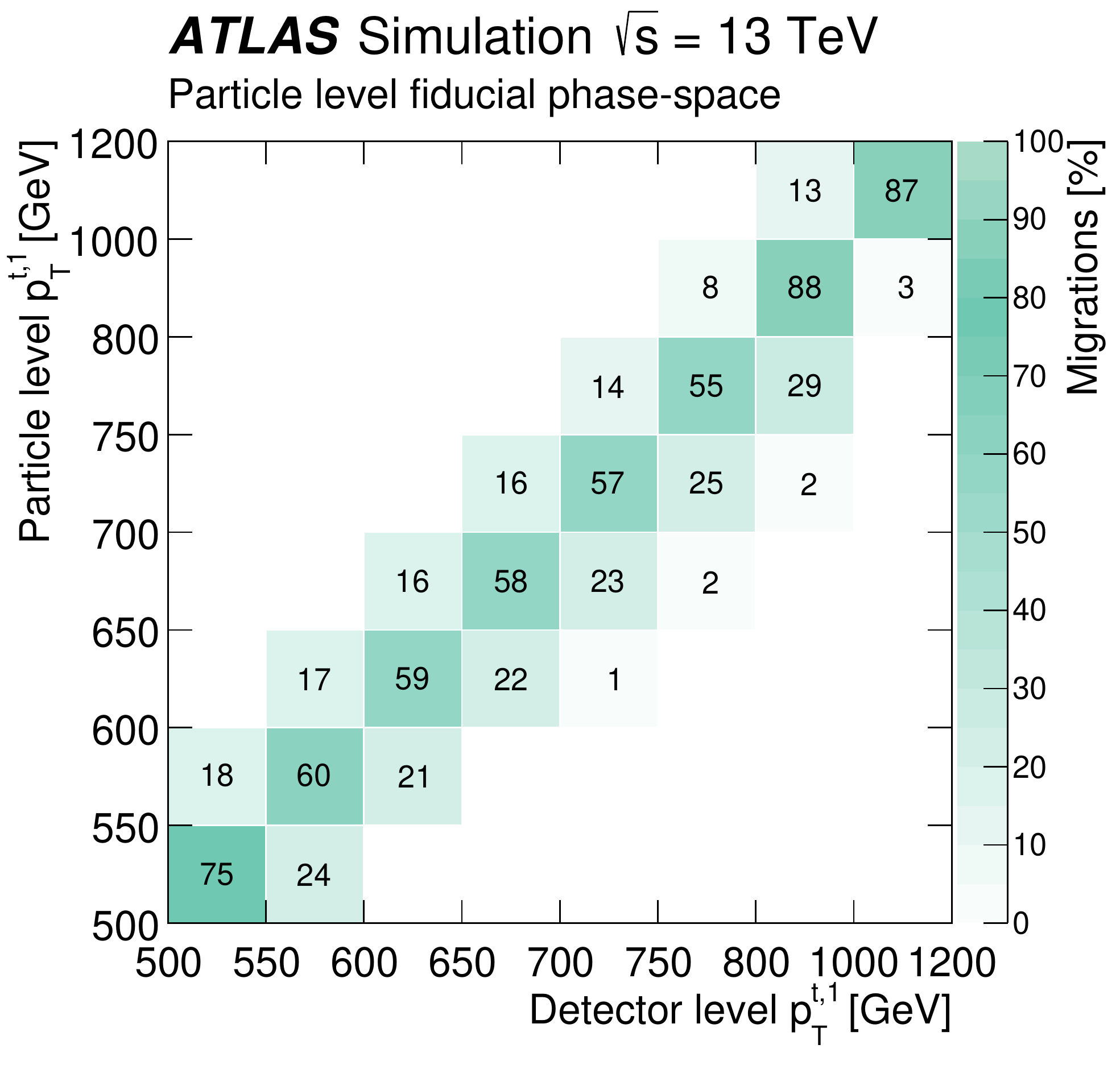}\label{fig:migrations:particle:t1_pt}}
\subfigure[]{ \includegraphics[width=0.45\textwidth]{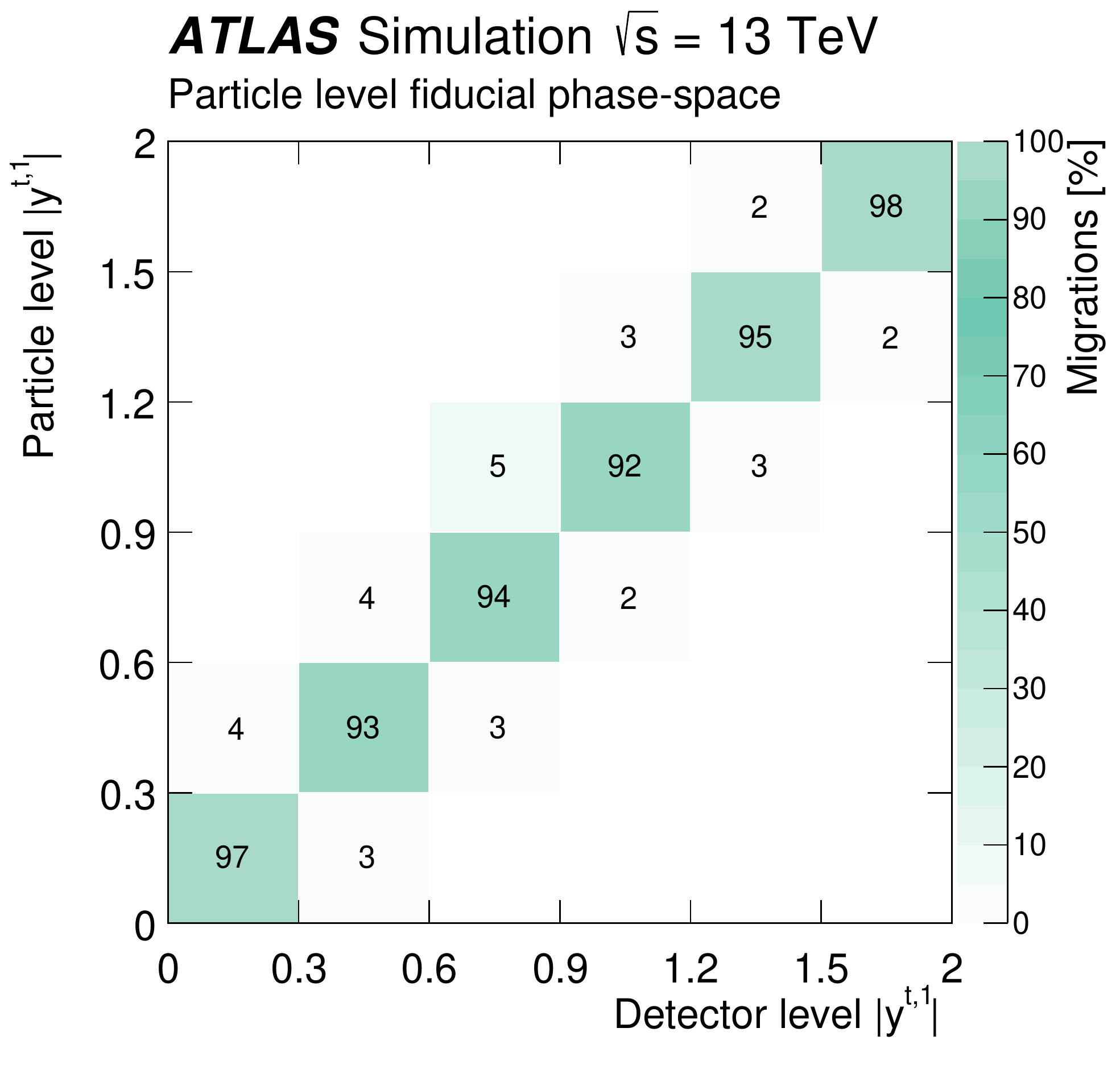}\label{fig:migrations:particle:t1_y}}
\subfigure[]{ \includegraphics[width=0.45\textwidth]{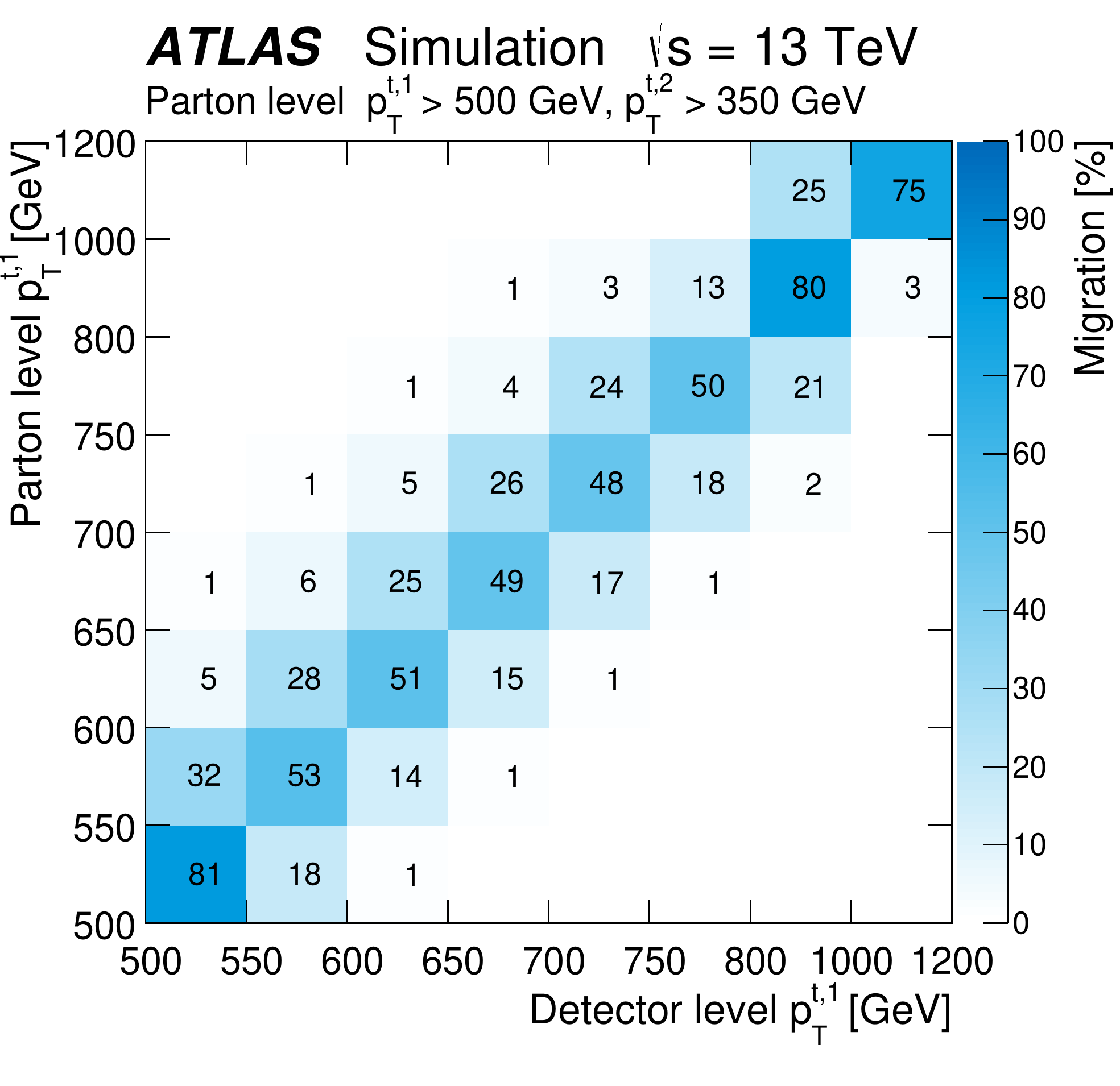}\label{fig:migrations:parton:t1_pt}}
\subfigure[]{ \includegraphics[width=0.45\textwidth]{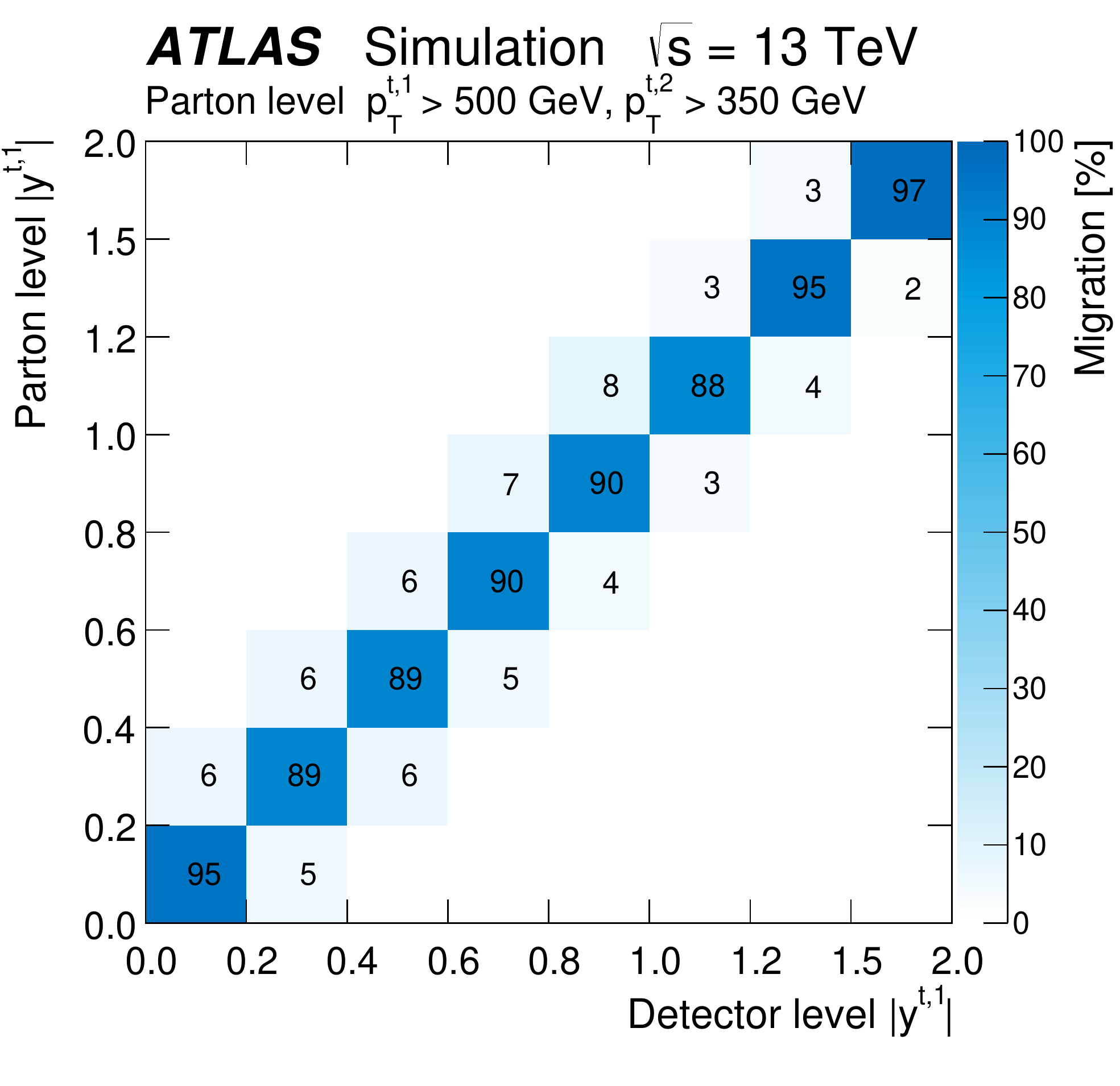}\label{fig:migrations:parton:t1_y}}
\caption{\small{Migration matrices for \pT\ and $|y|$\ of the leading top-quark jet 
in the particle-level fiducial phase space in \subref{fig:migrations:particle:t1_pt}\ and \subref{fig:migrations:particle:t1_y}
and parton-level phase space in
\subref{fig:migrations:parton:t1_pt}\ and \subref{fig:migrations:parton:t1_y}.
Each row is normalized to 100. The \POWHEG{}+\PythiaEight{} event generator is used as the nominal prediction.}}
\label{fig:migrations:particle:top_quarks}
\end{figure*}

The unfolding step uses a migration matrix ($\mathcal{M}$) derived from simulated \ttbar\ events with
matching detector-level jets by binning these events in the particle-level and 
parton-level phase spaces. 
The probability for particle-level (parton-level) events to remain in the same bin is therefore represented by the elements on the 
diagonal, and the off-diagonal elements describe the fraction of particle-level (parton-level) events that migrate into 
other bins. 
Therefore, the elements of each row add up to unity (within rounding) as shown in Fig. \ref{fig:migrations:particle:top_quarks}. 
The efficiency corrections $\epsilon_{\rm eff}$ correct for events
that are in the fiducial particle-level (parton-level) phase space but are not reconstructed 
at the detector level, and are illustrated in Fig.~\ref{fig:corrections:particle:top_quarks}.
The overall efficiency is largely determined by the working points of the $b$-tagging (70\%) and top-tagging (50\%) algorithms. 
The reduction in efficiency at higher top-quark candidate \pt\ arises primarily from the $b$-tagging 
requirements.
Examples of the migration matrices for several variables are shown in Fig.~\ref{fig:migrations:particle:top_quarks}.

The unfolding procedure for an observable $X$ at both particle and parton level is summarized by the expression
\begin{equation}
\frac{{d}\sigma^{\rm fid}}{{d}X^i} \equiv \frac{1}{\int\mathcal{L}\, dt \cdot \Delta X^i} \cdot  \frac{1}{\epsilon_{\rm eff}^i} \cdot \sum_j \mathcal{M}_{ij}^{-1} \cdot  f_{\rm acc}^j \cdot \left(N_{\rm reco}^j - N_{\rm bg}^j\right)\hbox{,}\nonumber
\end{equation}
where $N_{\rm reco}$ and $N_{\rm bg}$ refer to the number of reconstructed signal and background events, respectively; the index $j$ runs over bins of $X$ at detector level while the index $i$\ labels bins at particle and parton level; $\Delta X^i$ is the bin width while $\int\mathcal{L}\,dt$ is the integrated luminosity.
 The Bayesian unfolding is symbolized by $\mathcal{M}_{ij}^{-1}$.

The inclusive cross-section for \ttbar\ pairs in the fiducial (parton)
phase space, obtained by integrating the absolute differential cross-section, is used to determine the normalized differential 
cross-section $1/\sigma^{\rm fid}\cdot{d}\sigma^{\rm fid} / {d}X^i$.
This cross-section is not corrected for the all-hadronic \ttbar\ branching fraction of 0.457 \cite{PDG}.

Tests are performed at both particle and parton level to verify that the unfolding procedure is able to recover the 
generator-level distributions for input distributions that vary from the observed distributions or nominal 
predictions.
These closure tests show that the unfolding procedure results are unbiased so long as the features of the 
input distributions are consistent with the measurement resolution of the variable.

\FloatBarrier

\section{Systematic uncertainties}
\label{sec:uncertainties}

Systematic uncertainties resulting from electron, muon and jet reconstruction and
calibration, MC event generator modeling and background estimation, are
described below. 
The propagation of systematic uncertainties through the unfolding procedure is described in Sec. ~\ref {sec:prop_uncertainties}.

\subsection{Estimation of systematic uncertainties}

The systematic uncertainties in the measured distributions are estimated using MC data sets and 
the data satisfying the final selection requirements.

Estimates of \largeR\ jet uncertainties \cite{ATLAS-CONF-2016-035}\ 
are derived by studying tracking and calorimeter-based measurements and comparing these in data and MC simulations. 
These uncertainties also include the energy, mass and substructure response. 
The uncertainty in the \largeR\ jet mass resolution is incorporated by measuring the effect that an additional
 resolution degradation of 20\% has on the observables \cite{atlasJetSubstructure,ATLAS-CONF-2016-008}. 
The total uncertainty affecting the cross-section arising from jet calibration and reconstruction ranges 
from $11$\%\ to $30$\%\ for jet \pt\ over the range 350 to 900 \GeV.

The small-$R$ jet energy scale uncertainty is derived using a combination of simulations,
test-beam data and {\it in situ} measurements~\cite{jes:2013,jer_2,Aad:2012vm,ATL-PHYS-PUB-2015-015}. 
Additional uncertainty contributions from the jet flavor composition, calorimeter response to different jet flavors 
and pileup are taken into account. 
Uncertainties in the jet energy resolution are obtained with an {\it in situ}\ measurement of the jet response asymmetry 
in dijet events~\cite{jer:2013}. These small-$R$ jet uncertainties are typically below 1\% for all distributions. 

Uncertainties associated with pileup, the effect of additional interactions and the selection 
requirements used to mitigate them are
estimated using comparisons of data and MC samples and are approximately 1\%.
The efficiency to tag jets containing $b$-hadrons is corrected in
simulated events by applying $b$-tagging scale factors, extracted in  $\ttbar$ and dijet samples, in order to
account for the residual difference between data and simulation. 
The associated systematic uncertainties, computed by varying the $b$-tagging scale factors 
within their uncertainties~\cite{PERF-2012-04,ATL-PHYS-PUB-2016-012}, are
found to range from $\pm8$\%\ to $\pm17$\%\ for \largeR\ jet \pT increasing from 500 to 900~\GeV.
The uncertainties arising from lepton energy scale and resolution
 \cite{atlasElecPerf,ATLAS-CONF-2016-024,PERF-2015-10}\ are $<1$\%.

Systematic uncertainties affecting the multijet background estimates come from the subtraction of other 
background processes in the control regions
and from the uncertainties in the measured tagging correlations (which are statistical in nature).
The uncertainty in the subtraction of the all-hadronic \ttbar\ events in the control regions arises  
from the uncertainties in the \ttbar\ cross-section and 
$b$-matching algorithm.
Together, these result in background uncertainties ranging from $\pm2$\ to $\pm5$\%\ for \largeR\ jet \pT ranging from 
350 to 900~\GeV, respectively.
The uncertainty in the single-top-quark background rates comes from the uncertainties in the $Wt$\ production cross-section, the integrated luminosity, 
detection efficiency and the relative contribution of $t$-channel and $Wt$\ production, which is assigned an 
uncertainty of $\pm50$\%.

Other MC event generators are employed to assess modeling systematic uncertainties.
In these cases, the difference between the unfolded distribution of an alternative model and its own 
particle-level or parton-level distribution is used as the estimate of the corresponding systematic 
uncertainty in the unfolded differential cross-section.

To assess the uncertainty related to the matrix element calculation and matching to the parton shower, \MGAMCatNLO{}+\PythiaEight\ events are unfolded using the 
migration matrix and correction factors derived from the \POWHEG{}+\PythiaEight\ sample.
This uncertainty is found to be in the range $\pm10$--$15$\%, depending on the variable, increasing to 
$\pm20$--$30$\% at large \ptt, \mttbar, \ptttbar ~and \absyttbar\ where there are fewer data events.
To assess the uncertainty associated with the choice of parton shower
and hadronization model, 
a comparison is made of the unfolded and particle-level distributions of simulated events created 
with \POWHEG{} interfaced 
to the \HerwigSeven\ parton shower and hadronization model
using the nominal \POWHEG{}+\PythiaEight\ corrections and unfolding matrices.
The resulting systematic uncertainties, taken as the symmetrized difference, are found to be  
$\pm5$--$15$\%.
The uncertainty related to the modeling of initial- and final-state radiation is determined using two alternative 
\POWHEG{}+\PythiaEight\ \ttbar\ MC samples described in Sec.~\ref{sec:datasets}.
This uncertainty is found to be in the range $\pm10$--$15$\%, depending on the variable considered.
The uncertainty arising from the size of the nominal MC sample is approximately 1\%, scaling with the
statistical uncertainty of the data as a function of the measured variables.

The uncertainty arising from parton distribution functions is assessed using 
the \POWHEG{}+\PythiaEight\ \ttbar\ sample. 
An envelope of spectra is determined by reweighting the central prediction of the PDF4LHC 
PDF set \cite{PDF4LHC}\ and applying the relative variation to the nominal distributions. 
This uncertainty is found to be less than $1$\%.

The uncertainty in the integrated luminosity is $\pm2.1$\%.
It is derived, following a methodology similar to that detailed in Ref.
\cite{DAPR-2013-01}, from a calibration of the luminosity scale using $x$--$y$\ beam-separation scans performed in August 2015 and May 2016. 

Other sources of systematic uncertainty (e.g., the top-quark mass) are less than 1\%.

\subsection{Propagation of systematic uncertainties and treatment of correlations}
\label{sec:prop_uncertainties}

The statistical and systematic uncertainties are
propagated and combined in the same way for both the particle-level and parton-level
results using pseudoexperiments created from the nominal and alternative MC samples.

The effect of the data statistical uncertainty is incorporated by
creating pseudoexperiments in which independent Poisson fluctuations in each data bin are made.
The statistical uncertainty due to the size of the signal MC
samples used to correct the data is incorporated into the
pseudoexperiments by adding independent Poisson
fluctuations for a bin corresponding to the MC population in the bin.   

To evaluate the impact of each uncertainty after the unfolding, the simulated distribution is varied, then unfolded 
using corrections obtained with the nominal \POWHEG{}+\PythiaEight{}\ sample. 
The unfolded varied distribution is compared to the corresponding particle- or parton-level distribution.
For each systematic uncertainty, the correlation between the signal and background distributions is taken into account.
All detector- and background-related systematic uncertainties are estimated using the 
nominal \POWHEG{}+\PythiaEight{}\ sample. 
Alternative hard-scattering, parton shower and hadronization, ISR/FSR and PDF uncertainties are estimated by a comparison 
between the unfolded cross-section and the corresponding particle- or parton-level distribution produced using the 
corresponding alternative Monte Carlo event generator.

The systematic uncertainties for the particle-level fiducial phase-space total cross-section measurement described below 
are listed in Table~\ref{tab:systematics}.

\begin{table}[t]
\begin{center}
\caption{
Summary of the largest systematic and statistical relative uncertainties for the absolute particle-level fiducial phase-space 
cross-section measurement in percent.
Most of the uncertainties that are less than 1\%\ are not listed.
}
\sisetup{
  table-figures-integer = 2,
  table-figures-decimal = 1
}
\begin{tabular}{p{6.0cm} S[table-number-alignment = center]}
\hline
{Source} &  {Percentage}  \\
\hline
\hline
{Large-$R$ jet energy scale} 
&	5.9 \\
{Large-$R$ jet mass calibration} 
&	1.4 \\
{Large-$R$ jet top-tagging}
&	12 \\
{Small-R jets} 
&	0.3 \\
{Pileup} 
&	0.6 \\
{Flavor tagging}
&	8.3 \\
{Background}
&	0.9 \\
{Luminosity} 
&	2.0 \\
{Monte Carlo statistical uncertainty}
&	0.9 \\
{Alternative hard-scattering model}
&     11 \\
{Alternative parton-shower model}
&	14   \\
{ISR/FSR + scale} 
&	1.1 \\
\hline
{Total systematic uncertainty}
&	24 \\
{Data statistical uncertainty}
&	2.3 \\
{Total uncertainty}
&	24 \\
\hline
\end{tabular}
\label{tab:systematics}
\end{center}
\end{table}

\begin{figure*}[htbp]
\centering
\subfigure[]{ \includegraphics[width=0.45\textwidth]{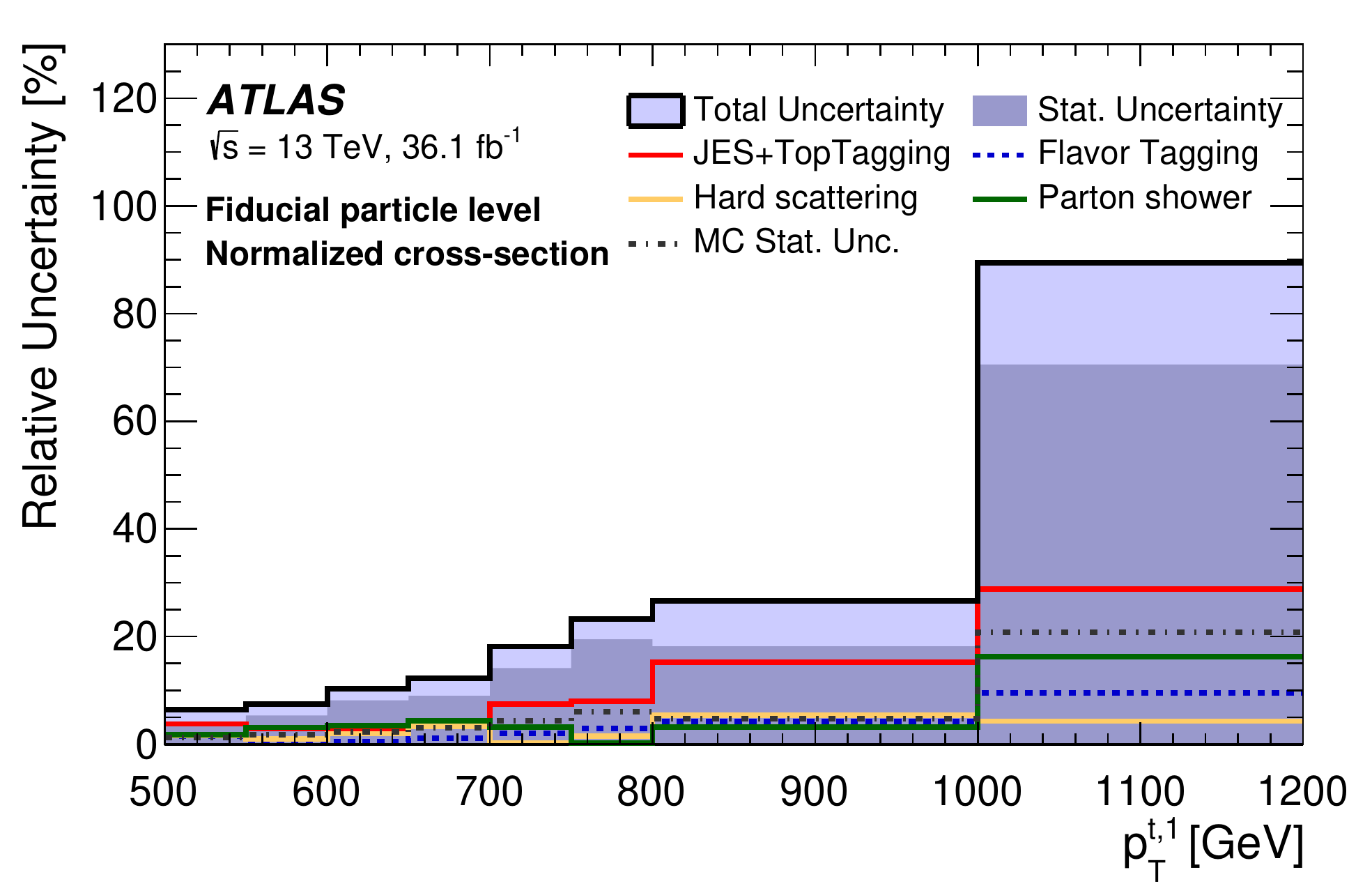}\label{fig:particle:t1_pt:rel:unc}}
\subfigure[]{ \includegraphics[width=0.45\textwidth]{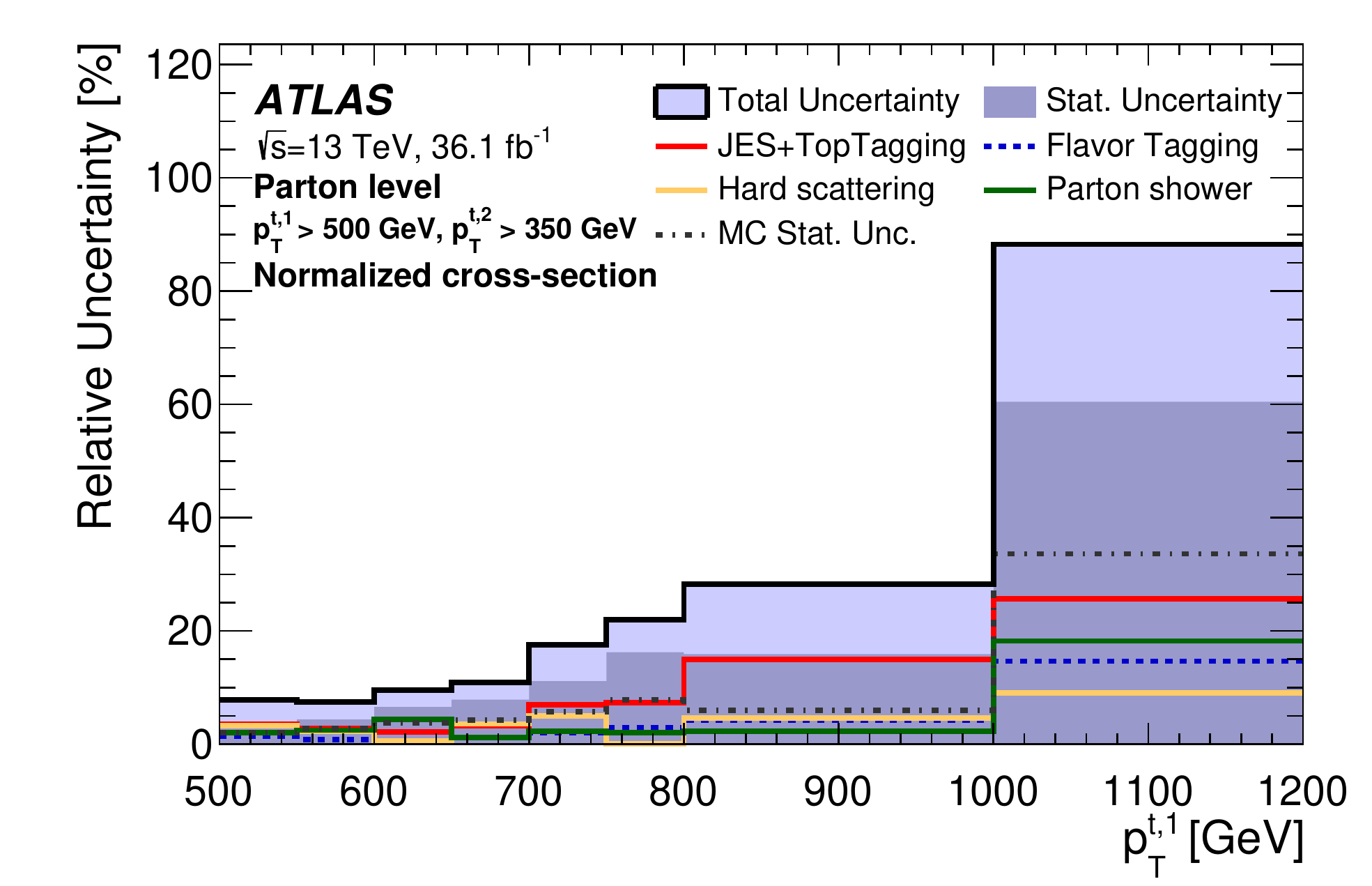}\label{fig:parton:t1_pt:rel:unc}}
\subfigure[]{ \includegraphics[width=0.45\textwidth]{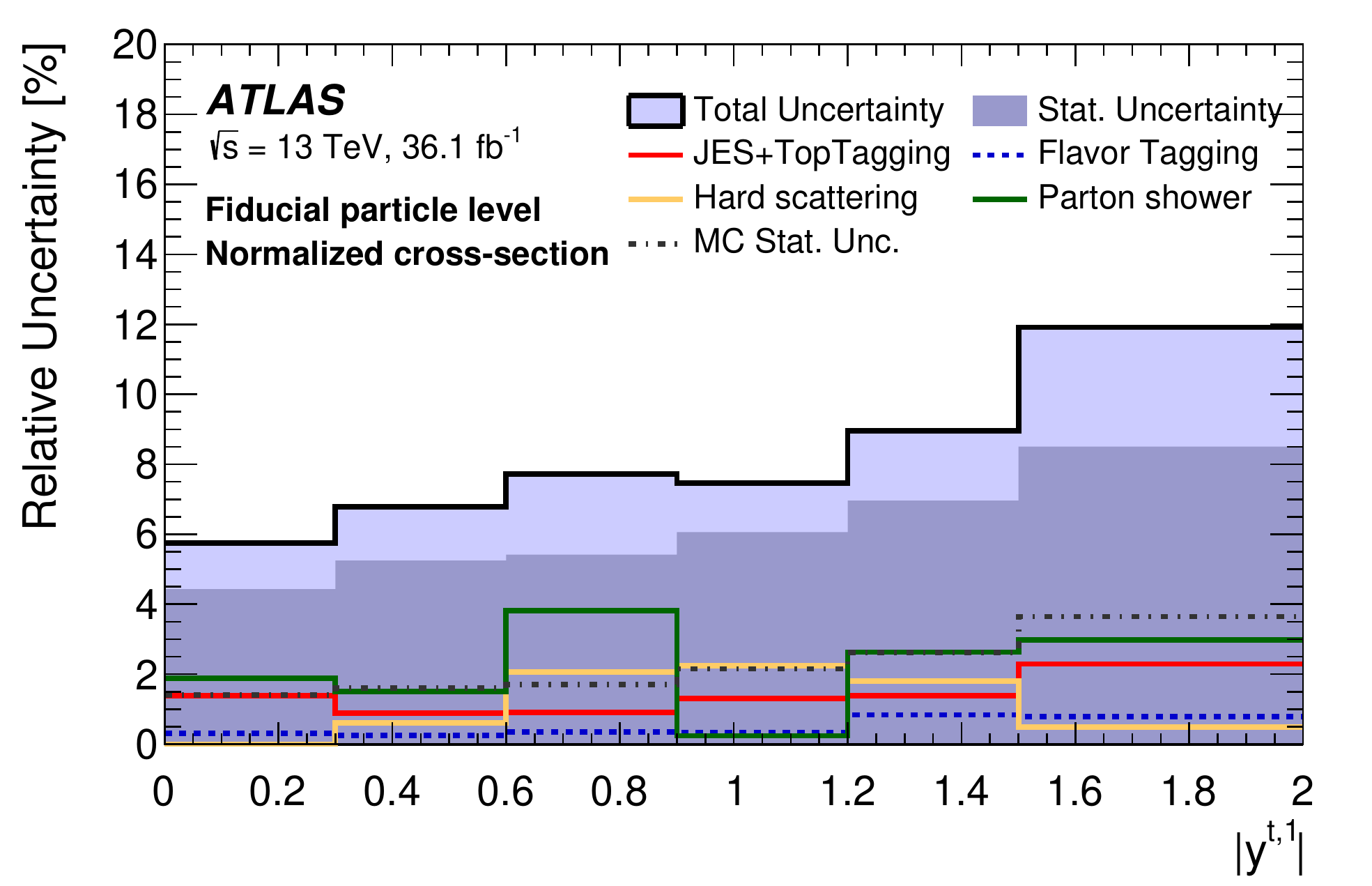}\label{fig:particle:t1_y:rel:unc}}
\subfigure[]{ \includegraphics[width=0.45\textwidth]{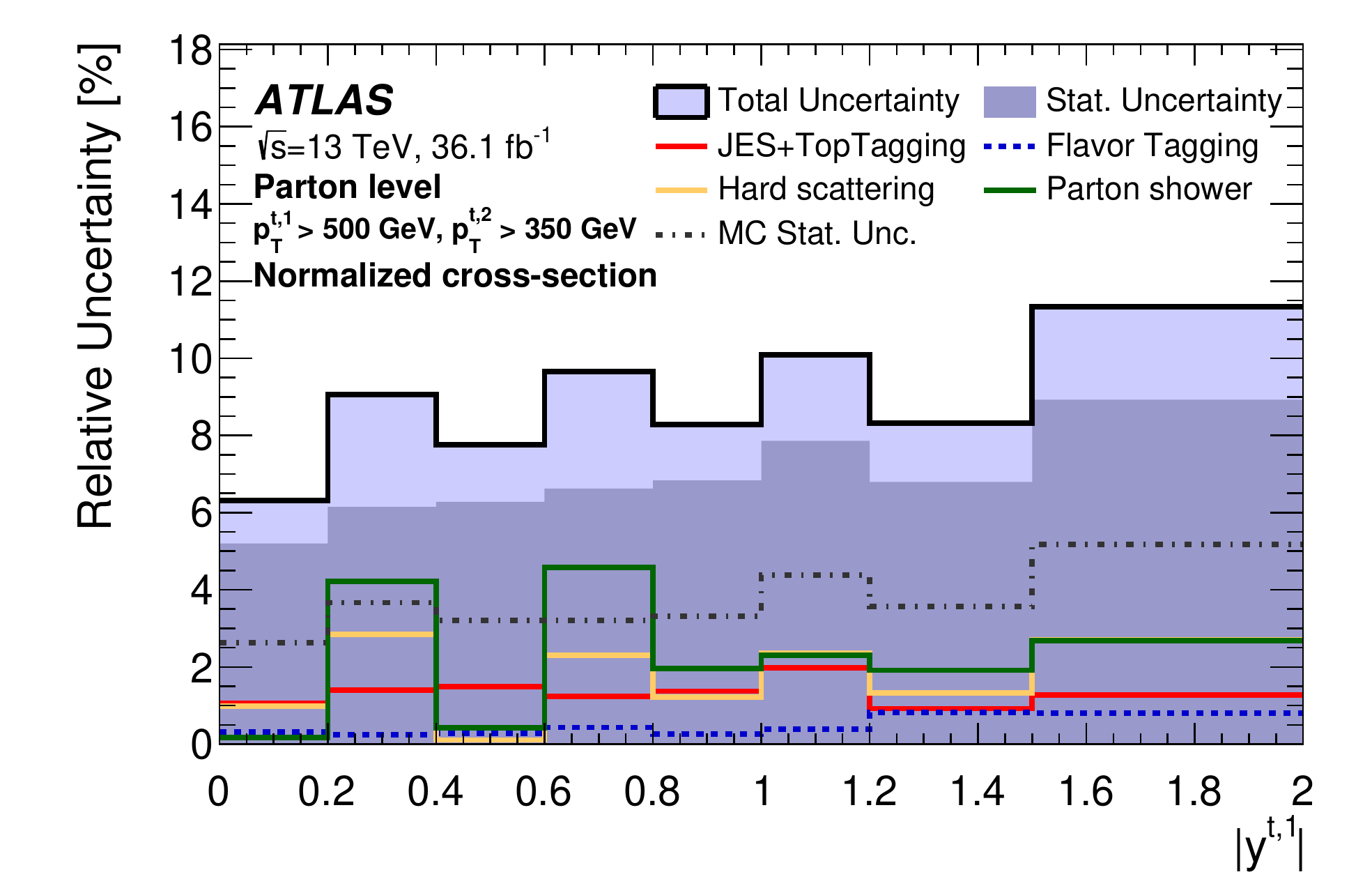}\label{fig:parton:t1_y:rel:unc}}
\caption{\small{Relative uncertainties in the normalized differential cross-sections as a function of the leading top-quark 
jet transverse momentum and rapidity at particle level and parton level. 
The light and dark blue areas represent the total and statistical uncertainty, respectively. 
The \POWHEG{}+\PythiaEight\ event generator is used as the nominal prediction to correct for detector effects. 
}}

\label{fig:fractional_unc:t1:rel}
\end{figure*}

Figure~\ref{fig:fractional_unc:t1:rel} shows a summary of the relative size of the systematic uncertainties for the 
leading top-quark jet transverse momentum and rapidity at particle level and parton level. 

\FloatBarrier

A covariance matrix is constructed for each differential cross-section to include the effect of all uncertainties to 
allow quantitative comparisons with theoretical predictions. 
This covariance matrix is derived by summing
two covariance matrices following the same approach used
in Refs.~\cite{atlasDiff4,atlasDiff8}.

The first covariance matrix incorporates statistical uncertainties and 
systematic uncertainties from detector effects and background estimation by using
pseudoexperiments to convolve the sources. 
In each pseudoexperiment, the detector-level data distribution is varied following a Poisson distribution. 
For each systematic uncertainty effect, Gaussian-distributed shifts are 
coherently included by scaling each Poisson-fluctuated bin content with its
expected relative variation from the associated systematic uncertainty.
Differential cross-sections are obtained by unfolding the
varied distribution with the nominal corrections, and the
distribution of the resulting changes in the unfolded distributions are used to compute 
this first covariance matrix.

The second covariance matrix is obtained by summing four separate
covariance matrices corresponding to the effects of the \ttbar\ event generator,
parton shower and hadronization, ISR/FSR and PDF uncertainties. 
The bin-to-bin
correlation values are set to unity for all these matrices. 

The comparison between the measured differential cross-sections and a
variety of MC predictions is quantified by calculating $\chi^{2}$ values
employing the covariance matrix and by calculating the
corresponding $p$-values (probabilities that the $\chi^2$ is larger
than or equal to the observed value assuming that the measured and
predicted distributions are statistically equivalent) from the $\chi^2$ and the number of degrees of freedom (NDF).
The $\chi^2$ values are obtained using
\begin{linenomath*}
\begin{equation*}
\chi^2 = V_{N_{\textrm b}}^{\textrm T} \cdot {\textrm{Cov}}_{N_{\textrm b}}^{-1} \cdot V_{N_{\textrm b}} \, ,
\end{equation*} 
\end{linenomath*}
where $V_{N_{\textrm b}}$ is the vector of differences between measured differential cross-section values and 
predictions, and $ {\textrm{Cov}}_{N_{\textrm b}}^{-1}$ is the inverse of the covariance matrix.

The normalization constraint used to derive the normalized
differential cross-sections lowers the NDF to one less that the rank of the
$N_{\textrm b} \times N_{\textrm b}$ covariance matrix, 
where $N_{\textrm b}$ is the number of bins in the unfolded distribution. 
The $\chi^2$ for the normalized
differential cross-sections is
\begin{linenomath*}
\begin{equation*}
\chi^2 = V_{N_{\textrm b}-1}^{\textrm T} \cdot {\textrm{Cov}}_{N_{\textrm b}-1}^{-1} \cdot V_{N_{\textrm b}-1} \, ,
\end{equation*} 
\end{linenomath*}
where $V_{N_{\textrm b}-1}$ is the vector of differences between measurement
and prediction obtained by discarding one of the $N_{\textrm b}$
elements and ${\textrm{Cov}}_{N_{\textrm b}-1}$ is the $(N_{\textrm
  b}-1) \times (N_{\textrm b}-1)$ sub-matrix derived from the
covariance matrix by discarding the corresponding row and column.

\section{Measurement of the differential cross-sections}
\label{sec:measurements}
The differential cross-sections are obtained from the data using the unfolding technique described above.
In the following subsections, 
the resulting particle-level and parton-level differential cross-sections are presented.

\subsection{Particle-level fiducial phase-space differential cross-section}
\label{sec:unfoldedresults}

\begin{figure*}[t]
\centering
\subfigure[]{ \includegraphics[width=0.45\textwidth]{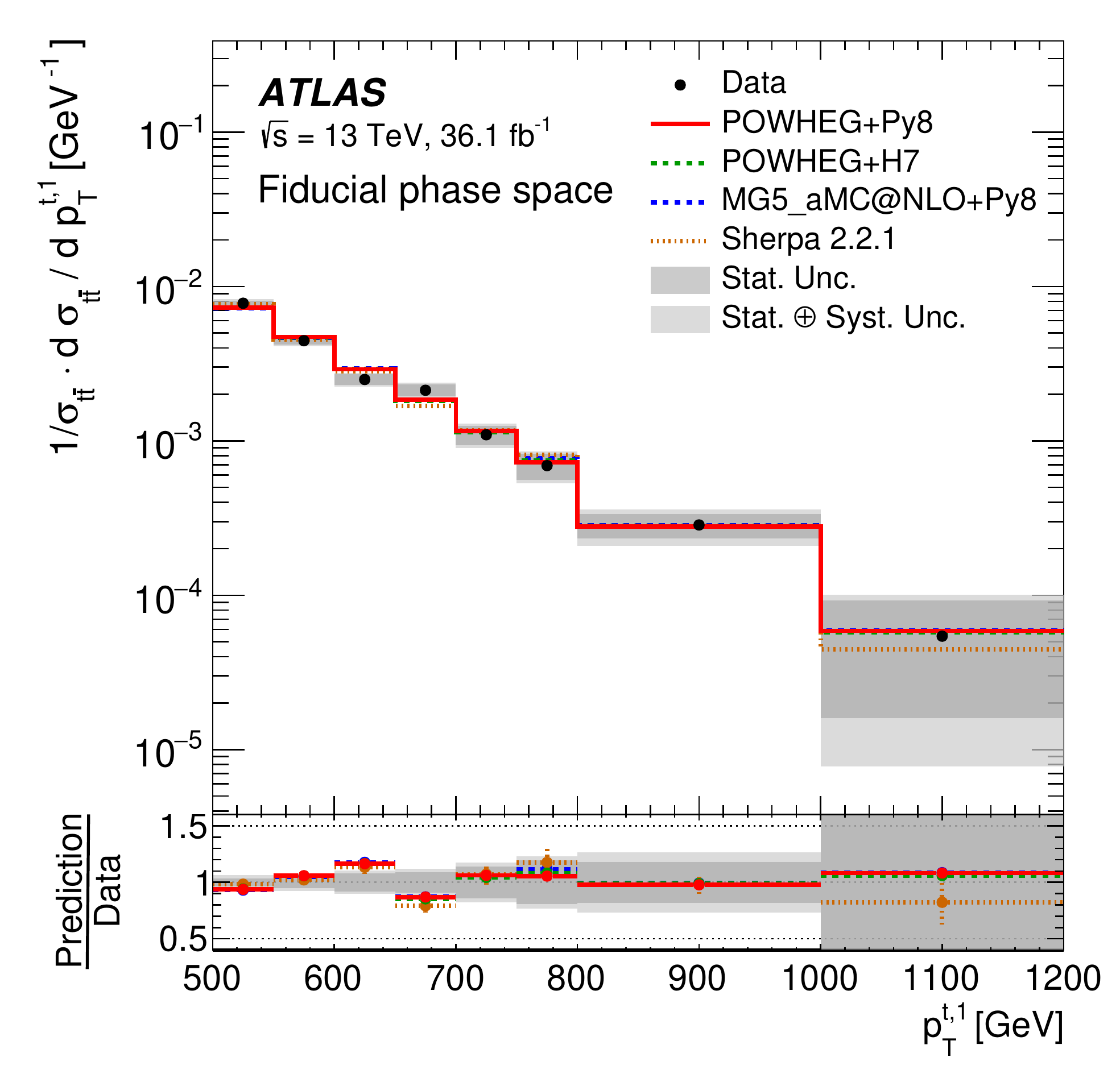}\label{fig:particle:t1_pt:rel}}
\subfigure[]{ \includegraphics[width=0.45\textwidth]{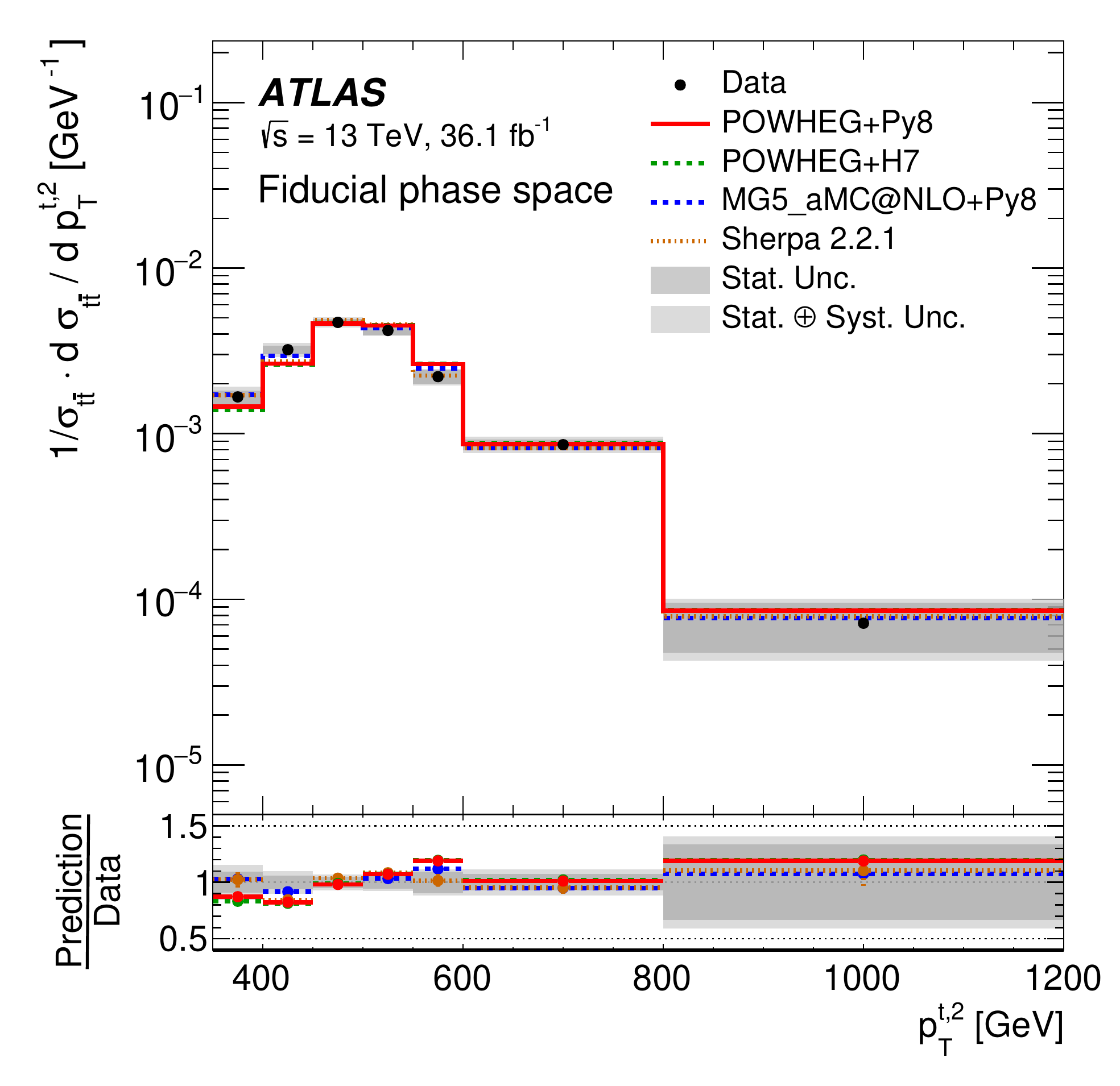}\label{fig:particle:t2_pt:rel}}
\subfigure[]{ \includegraphics[width=0.45\textwidth]{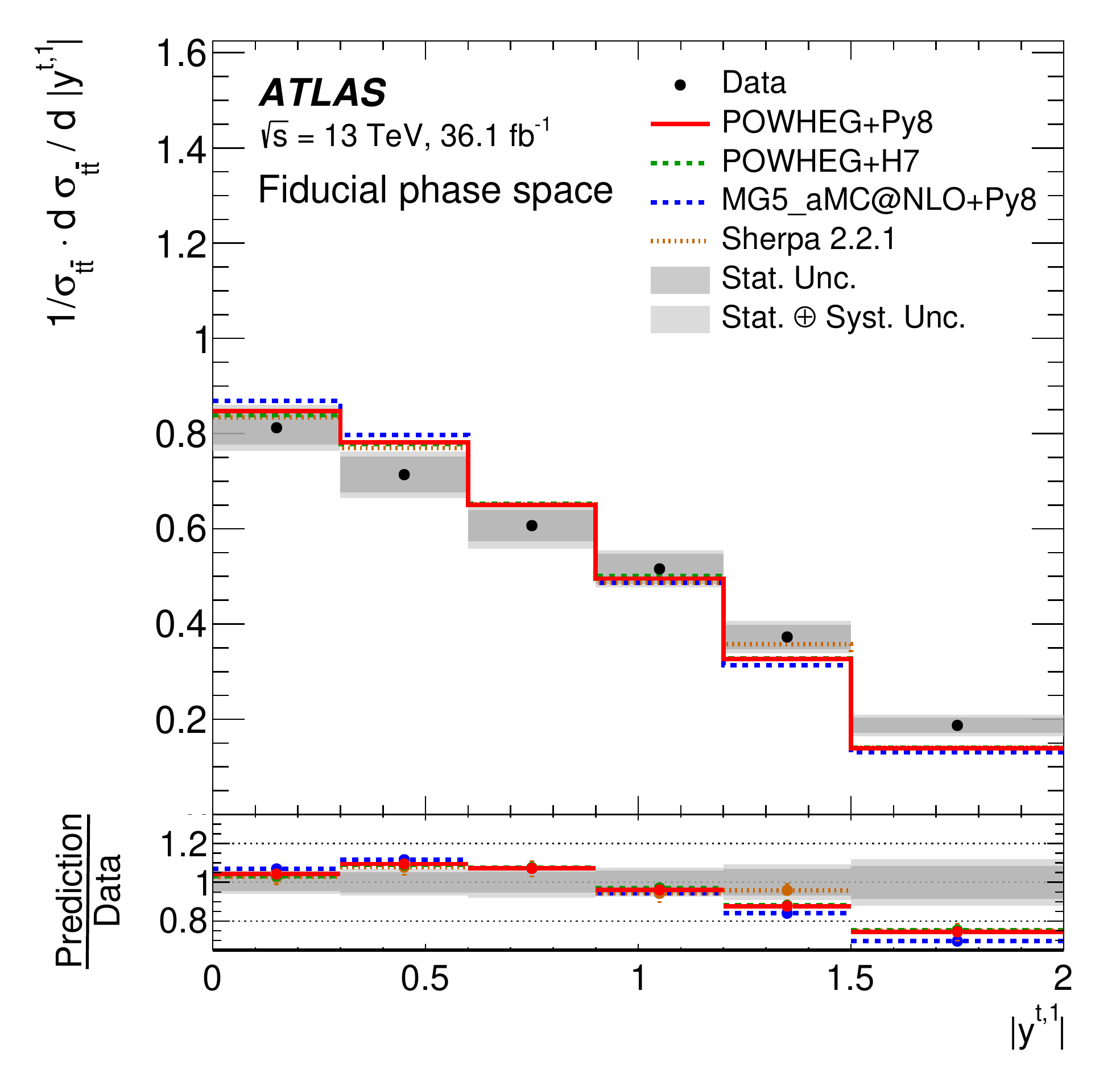}\label{fig:particle:t1_y:rel}}
\subfigure[]{ \includegraphics[width=0.45\textwidth]{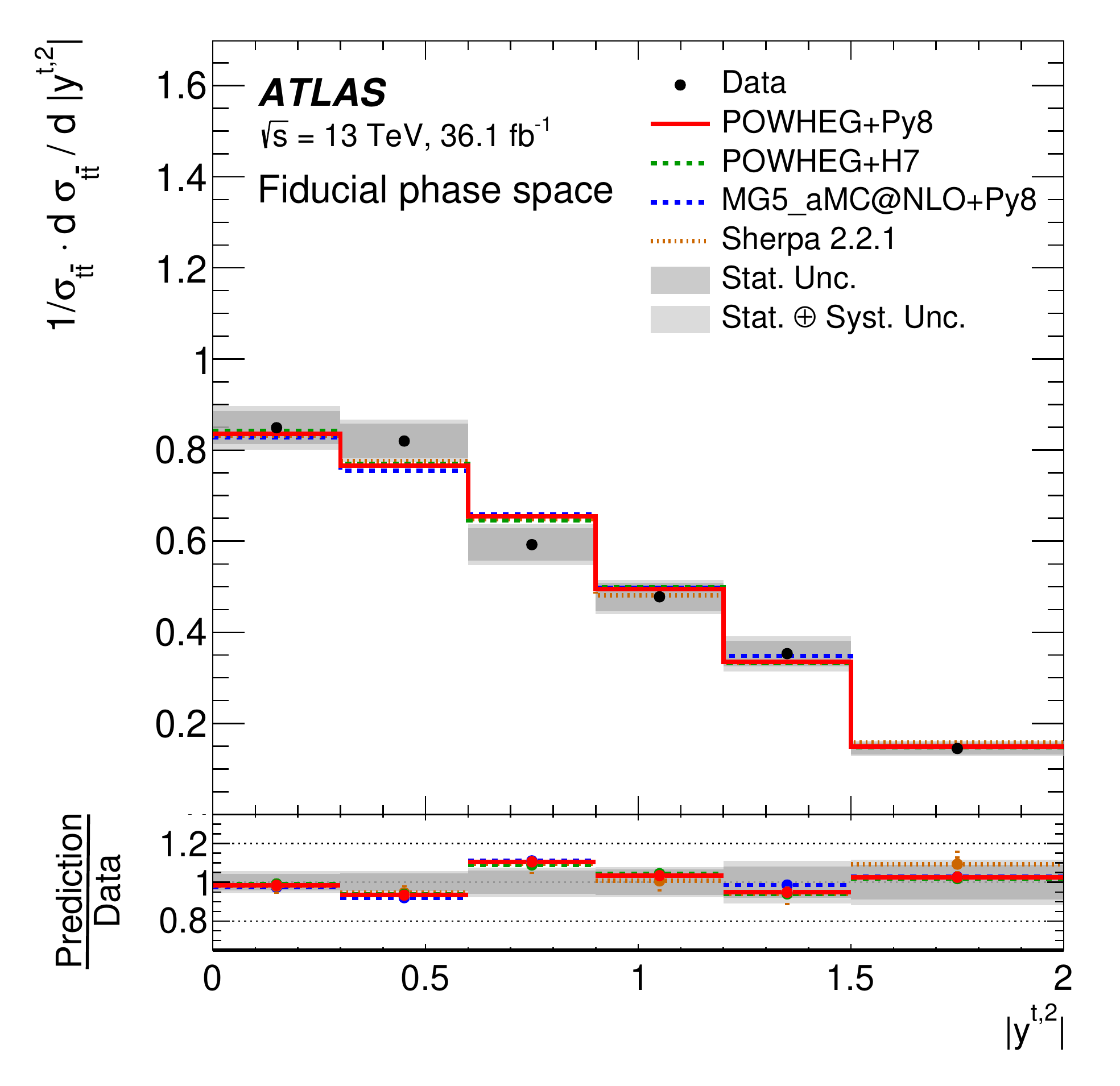}\label{fig:particle:t2_y:rel}}
\caption{\small{Normalized particle-level fiducial phase-space differential cross-sections as a function of 
\subref{fig:particle:t1_pt:rel}~transverse momentum of the leading top-quark jet, 
\subref{fig:particle:t2_pt:rel}~transverse momentum of the second-leading top-quark jet, 
\subref{fig:particle:t1_y:rel} absolute value of the rapidity of the leading top-quark jet and
\subref{fig:particle:t2_y:rel} absolute value of the rapidity of the second-leading top-quark jet. 
The gray bands indicate the total uncertainty in the data in each bin. The vertical bars indicate the statistical uncertainties in the theoretical models.
The \POWHEG{}+\PythiaEight\ event generator is used as the nominal prediction. Data points are placed at the center of each bin.}}
\label{fig:particle:top_quarks_kinematics:rel}
\end{figure*}

\begin{figure*}[htbp]
\centering
\subfigure[]{ \includegraphics[width=0.45\textwidth]{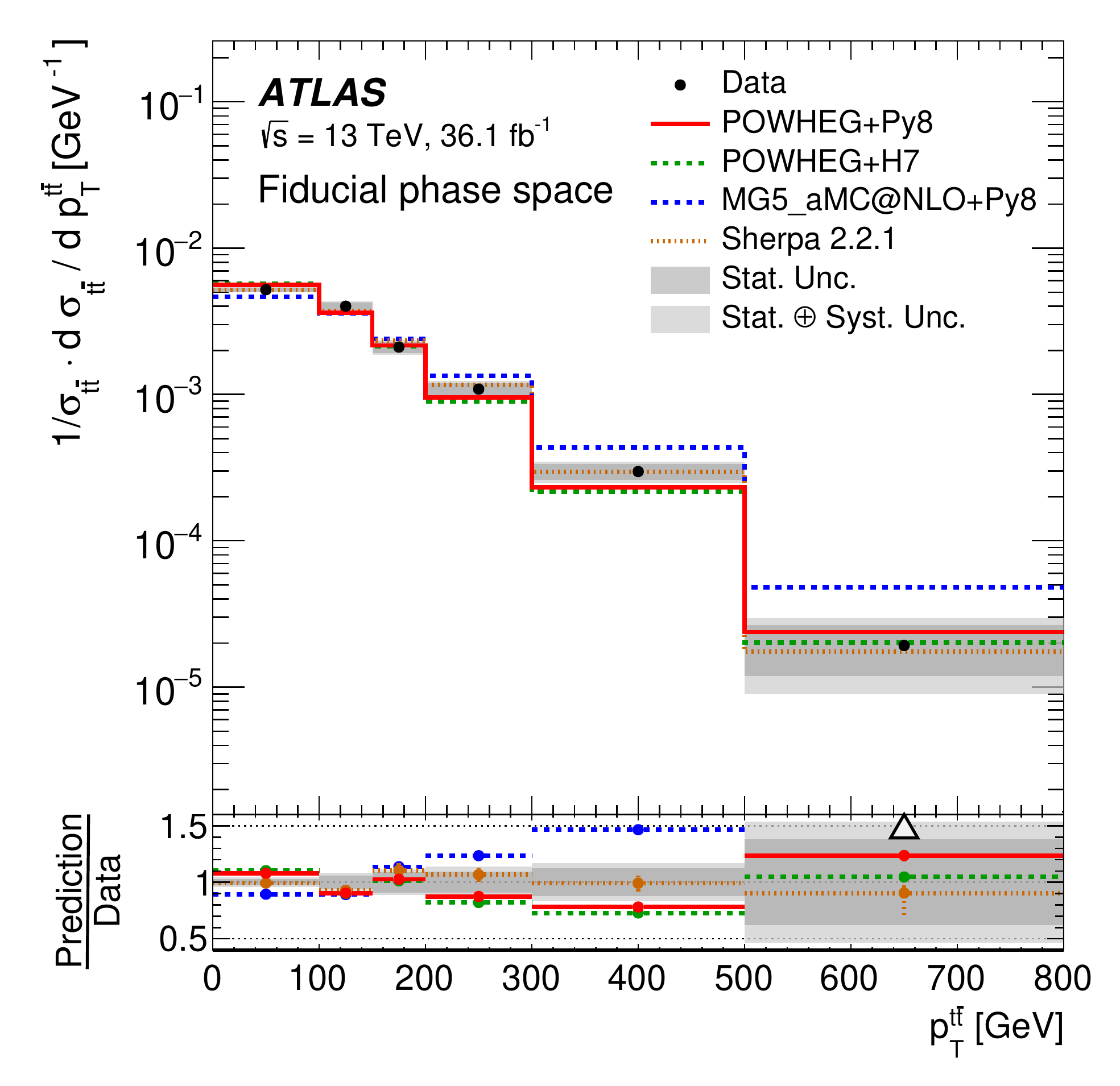}\label{fig:particle:tt_pt:rel}}
\subfigure[]{ \includegraphics[width=0.45\textwidth]{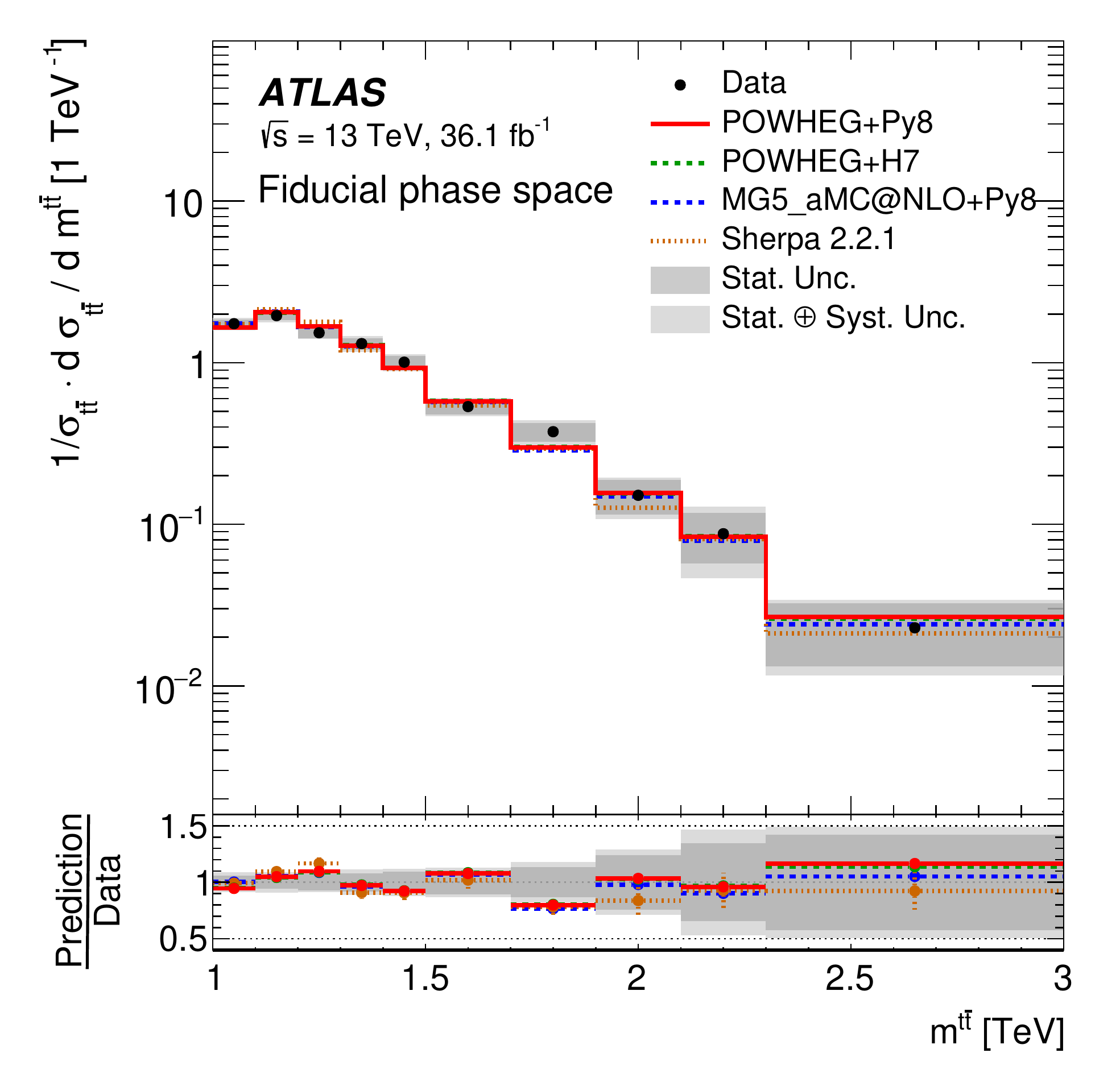}\label{fig:particle:tt_m:rel}}
\subfigure[]{ \includegraphics[width=0.45\textwidth]{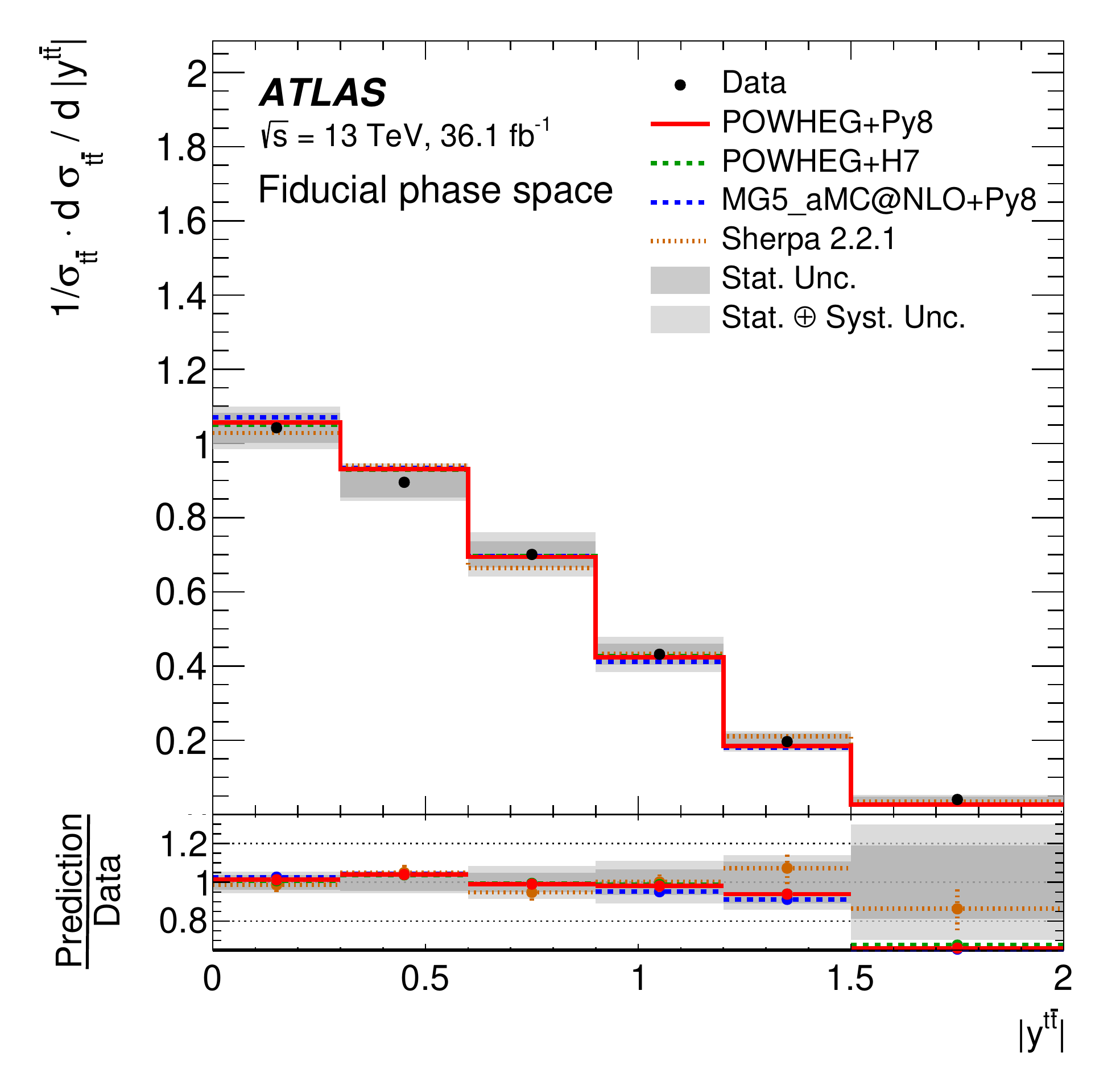}\label{fig:particle:tt_y:rel}}
\caption{\small{Normalized particle-level fiducial phase-space differential cross-sections as a function of 
\subref{fig:particle:tt_pt:rel}~transverse momentum, 
\subref{fig:particle:tt_m:rel} invariant mass and 
\subref{fig:particle:tt_y:rel} absolute value of the rapidity of the \ttbar{}~system. 
The gray bands indicate the total uncertainty in the data in each bin. The vertical bars indicate the statistical uncertainties in the theoretical models.
The \POWHEG{}+\PythiaEight\ event generator is used as the nominal prediction. Data points are placed at the center of each bin.}}
\label{fig:particle:ttbar_system_baseline:rel}
\end{figure*}

\begin{figure*}[htbp]
\centering
\subfigure[]{ \includegraphics[width=0.45\textwidth]{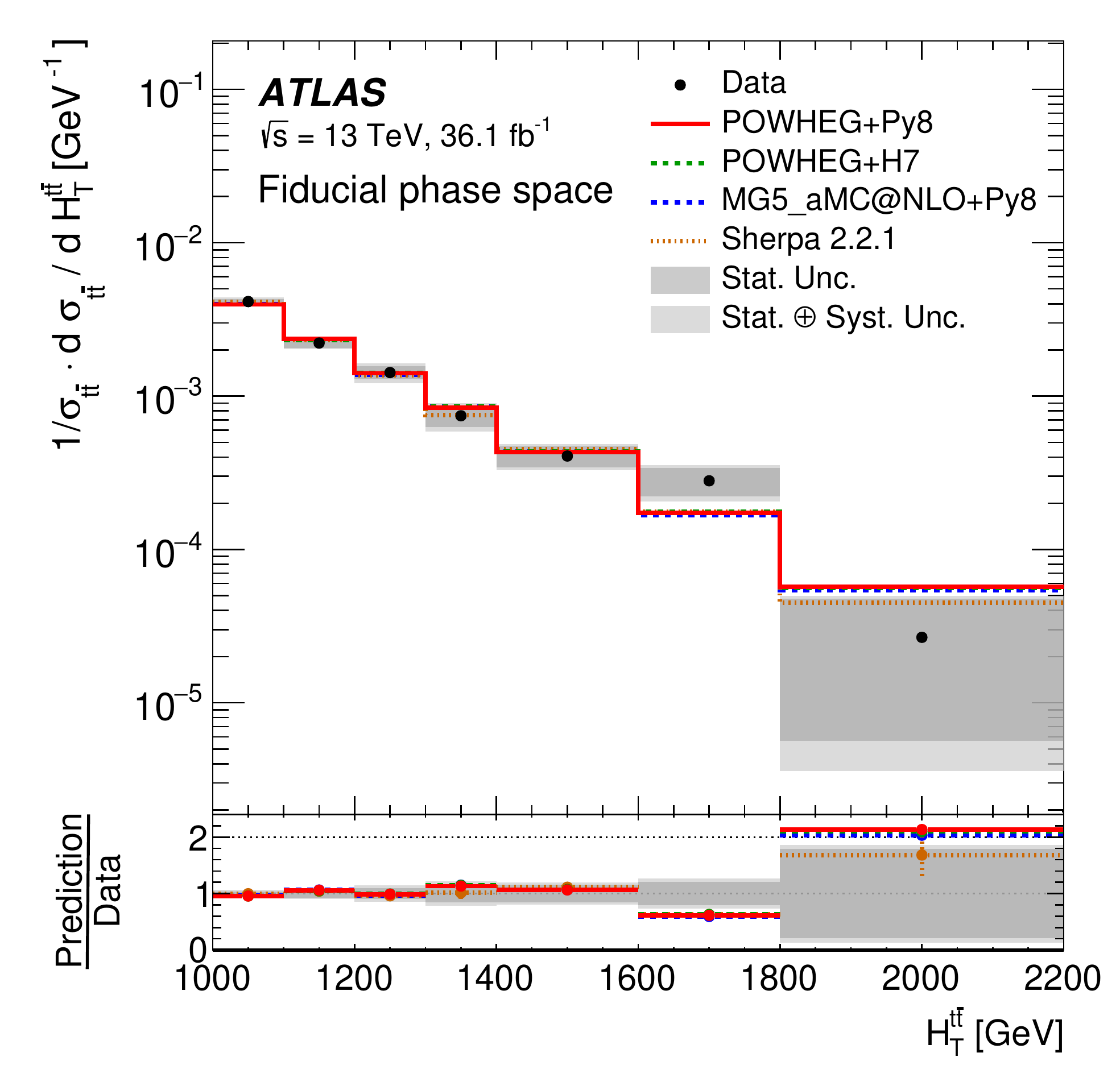}\label{fig:particle:tt_HTttbar:rel}}
\subfigure[]{ \includegraphics[width=0.45\textwidth]{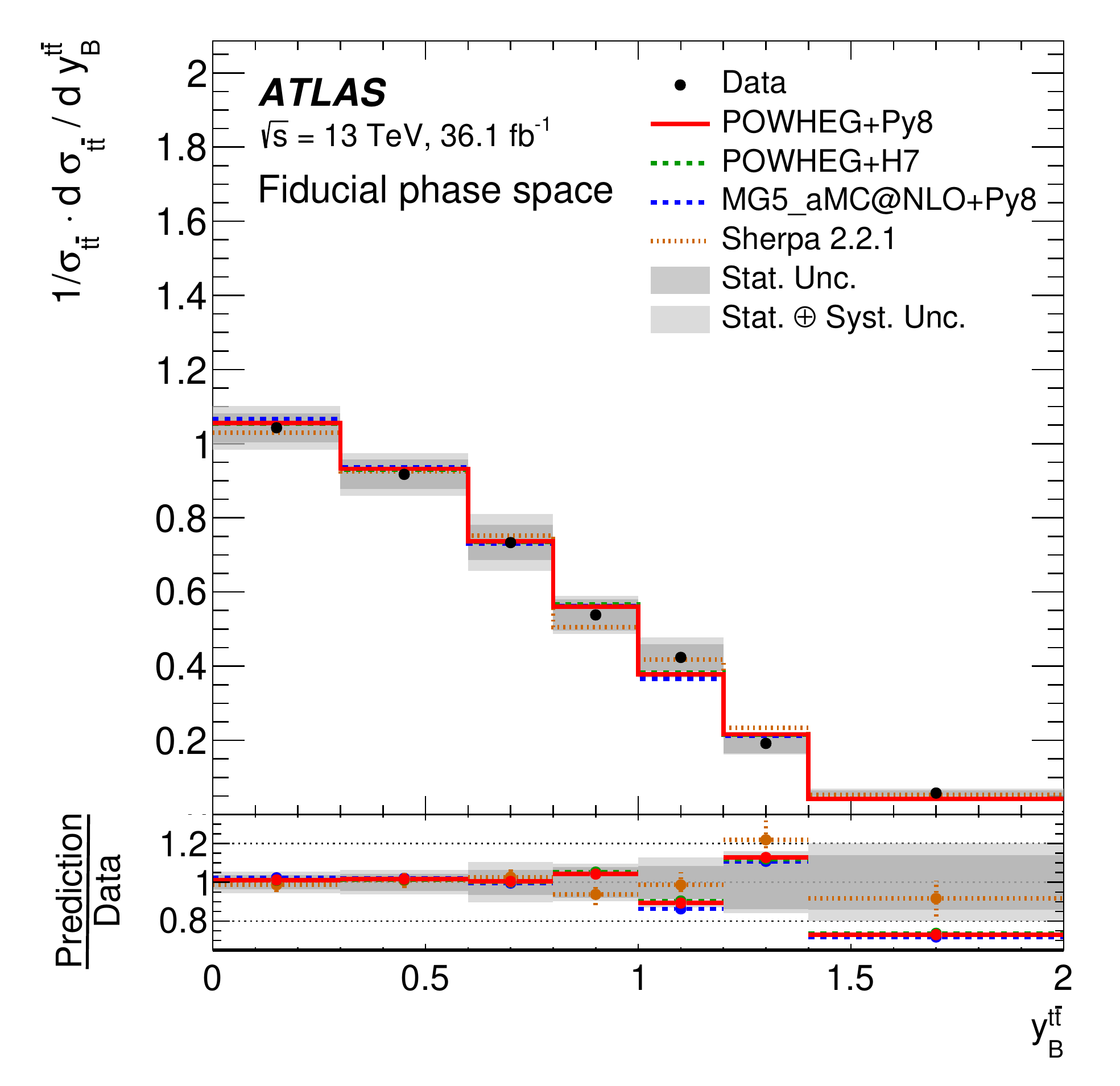}\label{fig:particle:tt_yboost:rel}}
\caption{\small{Normalized particle-level fiducial phase-space differential cross-sections as a function of
\subref{fig:particle:tt_HTttbar:rel}~the scalar sum of the transverse momenta of the two top-quark jets and
\subref{fig:particle:tt_yboost:rel}~the longitudinal boost \boostttbar.
The gray bands indicate the total uncertainty in the data in each bin. The vertical bars indicate the statistical uncertainties in the theoretical models. 
The \POWHEG{}+\PythiaEight\ event generator is used as the nominal prediction. Data points are placed at the center of each bin.}}
\label{fig:particle:ttbar_system_extra:rel1}
\end{figure*}

\begin{figure*}[htbp]
\centering
\subfigure[]{ \includegraphics[width=0.45\textwidth]{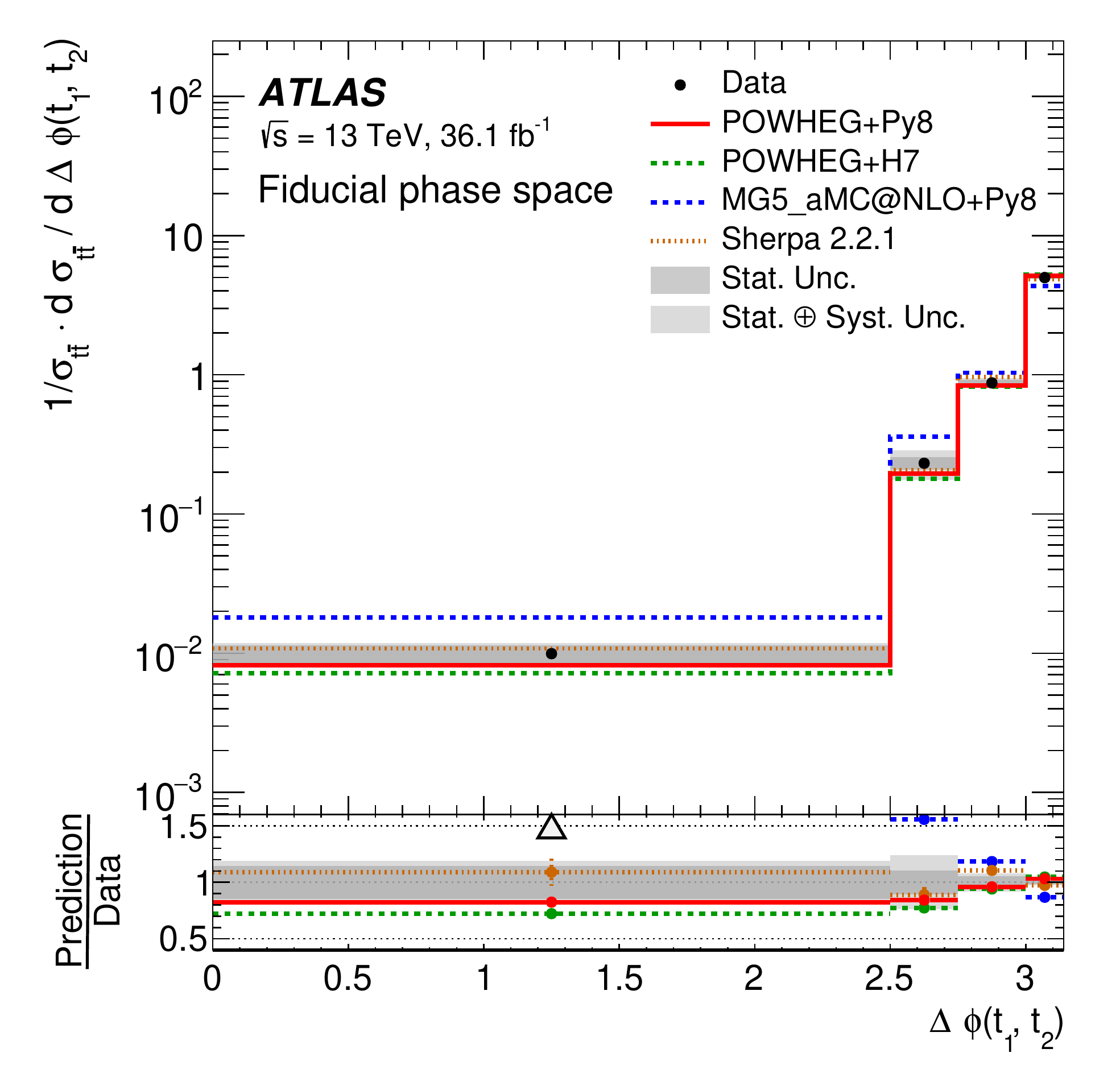}\label{fig:particle:tt_dPhittbar:rel}}
\subfigure[]{ \includegraphics[width=0.45\textwidth]{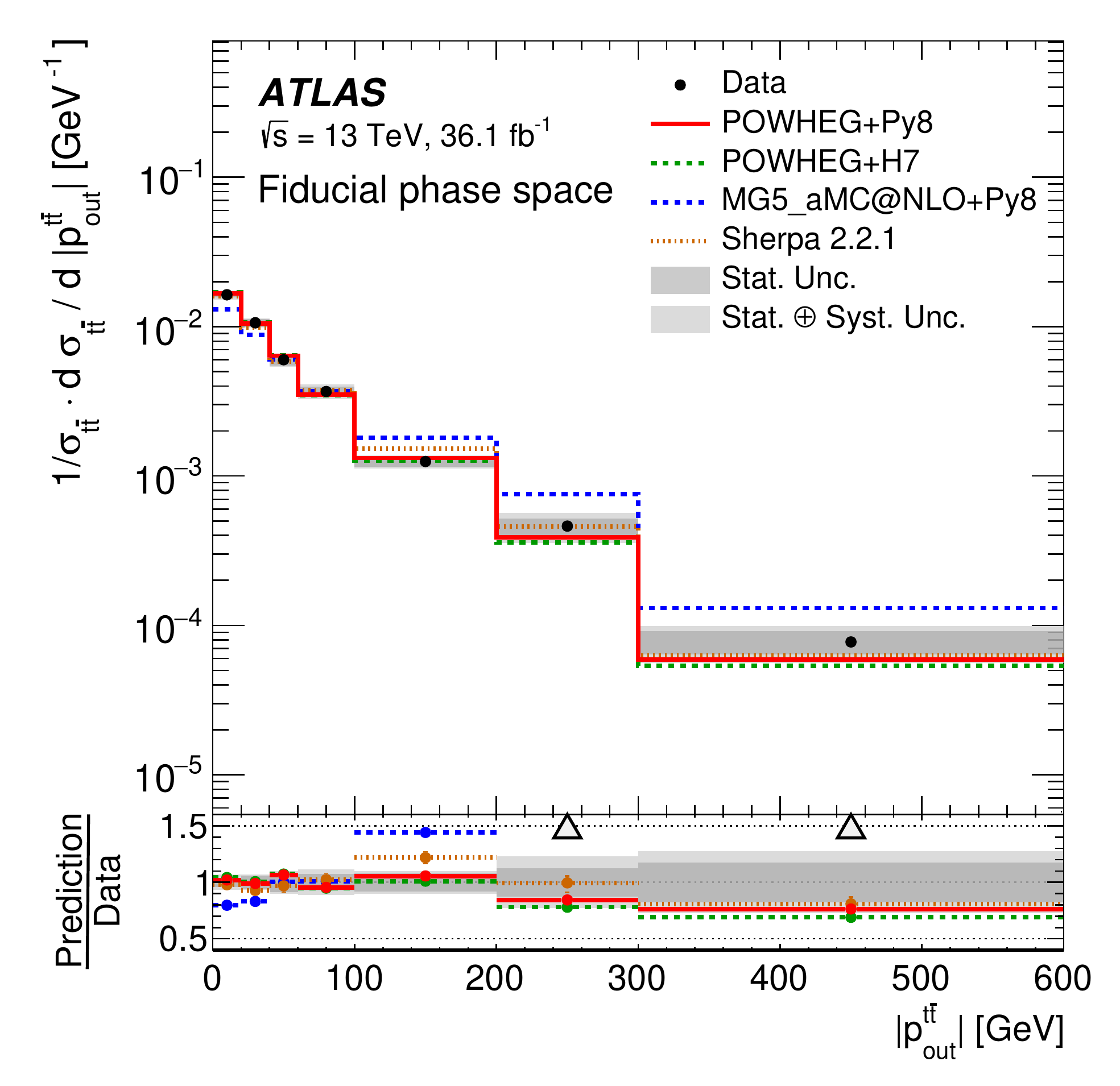}\label{fig:particle:tt_pout:rel}}
\caption{\small{Normalized particle-level fiducial phase-space differential cross-sections as a function of 
\subref{fig:particle:tt_dPhittbar:rel}~the azimuthal angle between the two top-quark jets \dPhittbar~ and
\subref{fig:particle:tt_pout:rel}~the absolute value of the out-of-plane momentum \Poutttbar.
The gray bands indicate the total uncertainty in the data in each bin. The vertical bars indicate the statistical uncertainties in the theoretical models. 
The \POWHEG{}+\PythiaEight\ event generator is used as the nominal prediction. Data points are placed at the center of each bin.}}
\label{fig:particle:ttbar_system_extra:rel2}
\end{figure*}

\begin{figure*}[htbp]
\centering
\subfigure[]{ \includegraphics[width=0.45\textwidth]{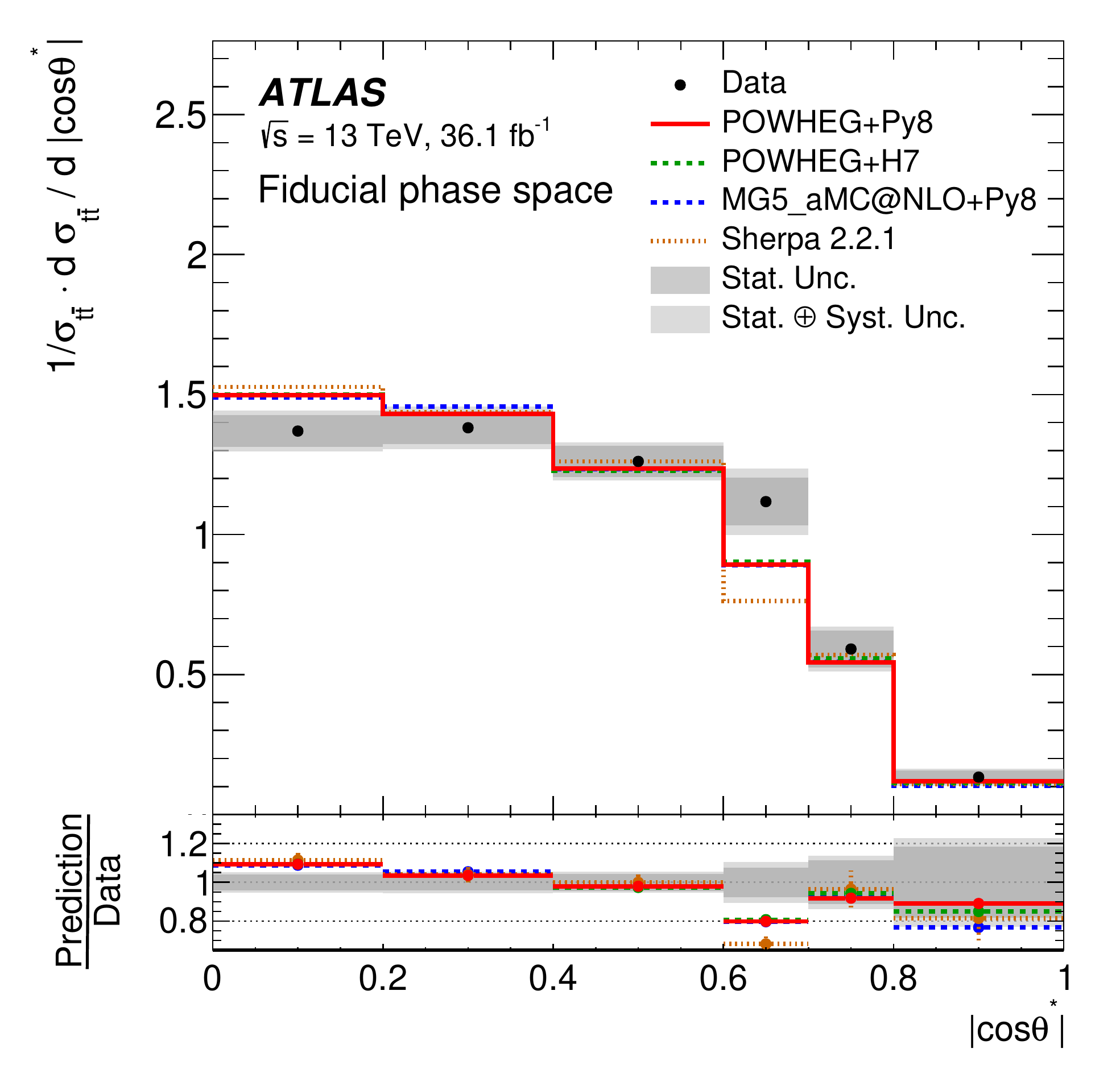}\label{fig:particle:tt_cosThetaStar:rel}}
\subfigure[]{ \includegraphics[width=0.45\textwidth]{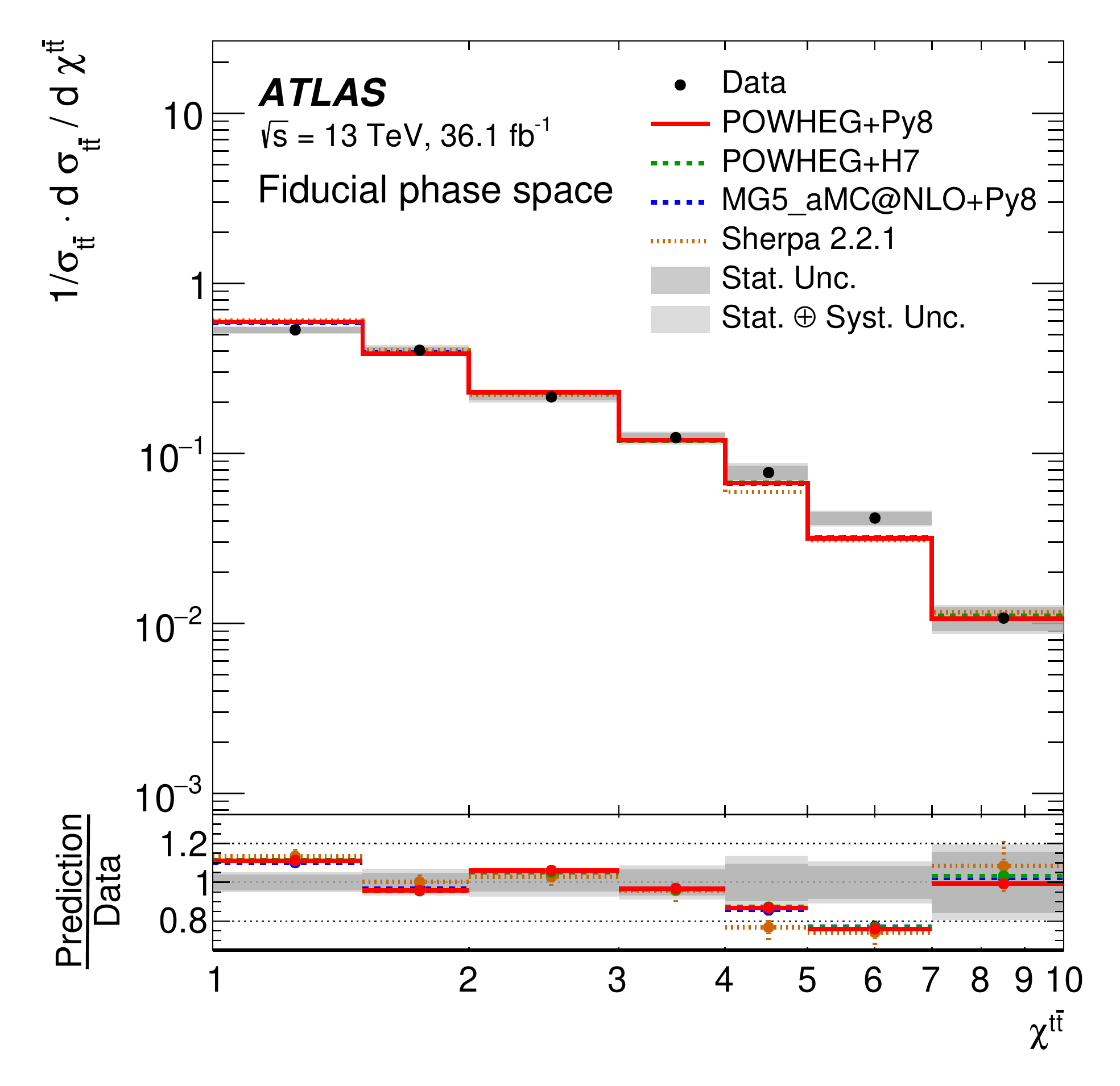}\label{fig:particle:tt_chittbar:rel}}
\caption{\small{Normalized particle-level fiducial phase-space differential cross-sections as a function of 
\subref{fig:particle:tt_cosThetaStar:rel}~the production angle in the Collins--Soper reference frame \ and
\subref{fig:particle:tt_chittbar:rel}~the variable \chittbar.
The gray bands indicate the total uncertainty in the data in each bin. The vertical bars indicate the statistical uncertainties in the theoretical models. 
The \POWHEG{}+\PythiaEight\ event generator is used as the nominal prediction. Data points are placed at the center of each bin.}}
\label{fig:particle:ttbar_system_extra:rel3}
\end{figure*}

The unfolded differential cross-sections, normalized to the total cross-section for the fiducial 
phase space,  are shown in Fig.~\ref{fig:particle:top_quarks_kinematics:rel}\ for the 
\pt\ and rapidity of the  leading and second-leading top-quark jets, 
and in Fig.~\ref{fig:particle:ttbar_system_baseline:rel}\ for 
the \pt, mass and rapidity of the \ttbar\ system.
The unfolded differential cross-sections are shown in Figs.~\ref{fig:particle:ttbar_system_extra:rel1}--\ref{fig:particle:ttbar_system_extra:rel3}\ for 
the \ttbar\ production angle in the Collins--Soper reference frame,
the scalar sum of the transverse momenta of the two top quarks, \HTttbar,
the longitudinal boost, \boostttbar,
the azimuthal angle between the two top-quark jets, \dPhittbar,
the variable related to the rapidity difference between the two top-quark jets, \chittbar, and
the absolute value of the out-of-plane momentum, \Poutttbar.
These are compared with 
SM predictions obtained using the NLO MC event generators described in Sec.~\ref{sec:datasets}. 

This analysis is sensitive to top-quark jets produced with \pt\ up to approximately 1~\TeV\ and  to a rapidity $|\yt| < 2.0$.
The differential cross-section falls by two orders of 
magnitude as a function of top-quark jet transverse momentum over a \pT\ range from 500 \GeV\ to 1 \TeV. 
The production cross-section decreases as a function of top-quark jet rapidity by approximately 30\%\ from $\yt=0$ to $\yt=\pm1$.
The differential cross-section as a function of \pT\ for the second-leading top-quark jet reflects the effect
of the \pT\ requirement on the leading top-quark jet and the strong correlation in \pT\ of the two top-quark jets arising
from the pair-production process.

The \ttbar\ system is centrally produced with 
a transverse momentum typically below 200~\GeV, an invariant mass below 1.5 \TeV\ and a rapidity $|\yt| < 1.5$. In particular, the \mttbar\ distribution falls smoothly, with a sensitivity that extends up to $\sim 2$~\TeV.

\FloatBarrier

\subsection{Parton-level phase-space differential cross-sections}

\begin{figure*}[htbp]
\centering
\subfigure[]{\includegraphics[width=0.45\textwidth]{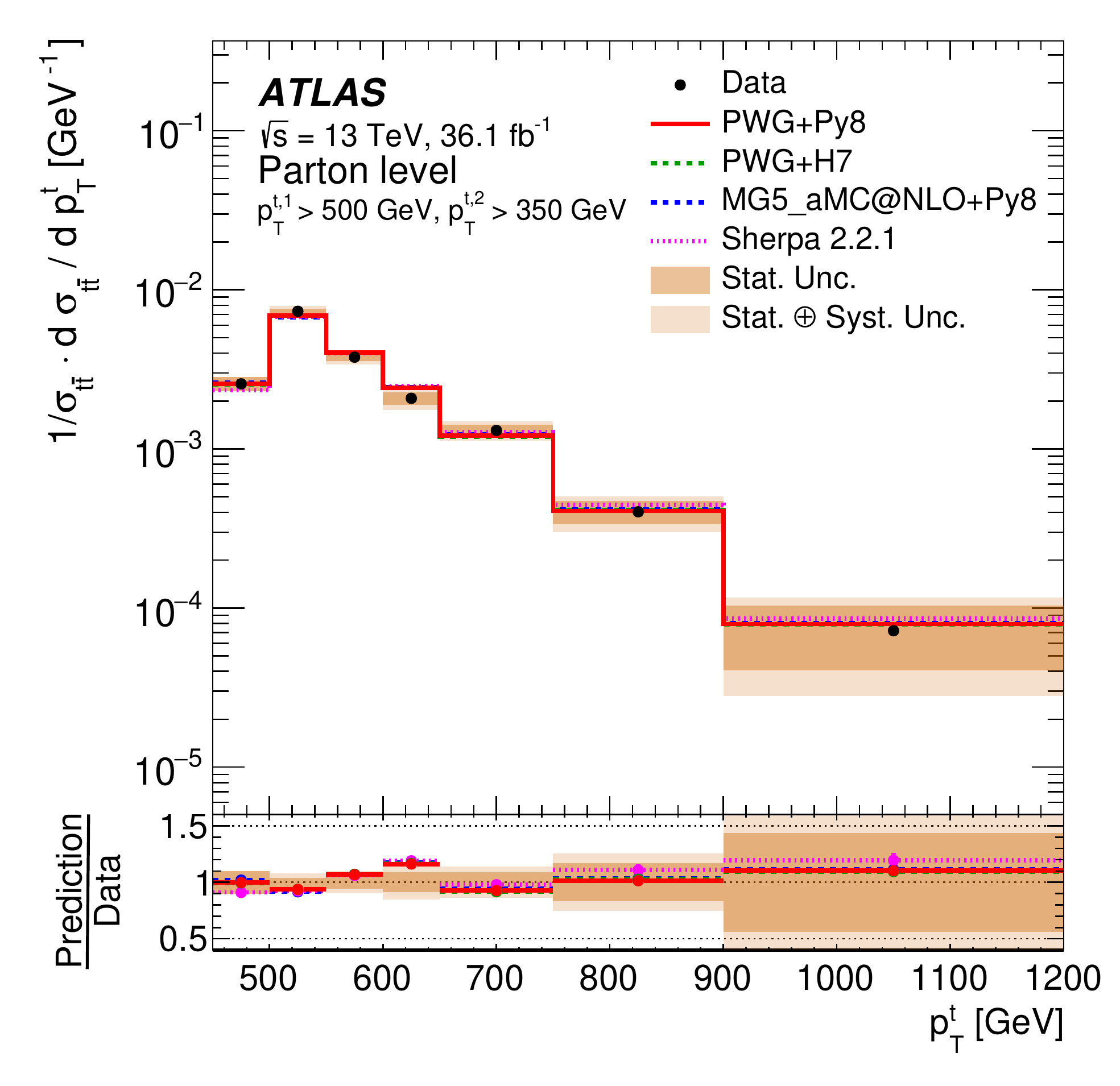}\label{fig:results:parton:top_pt_rel}}%
\subfigure[]{\includegraphics[width=0.45\textwidth]{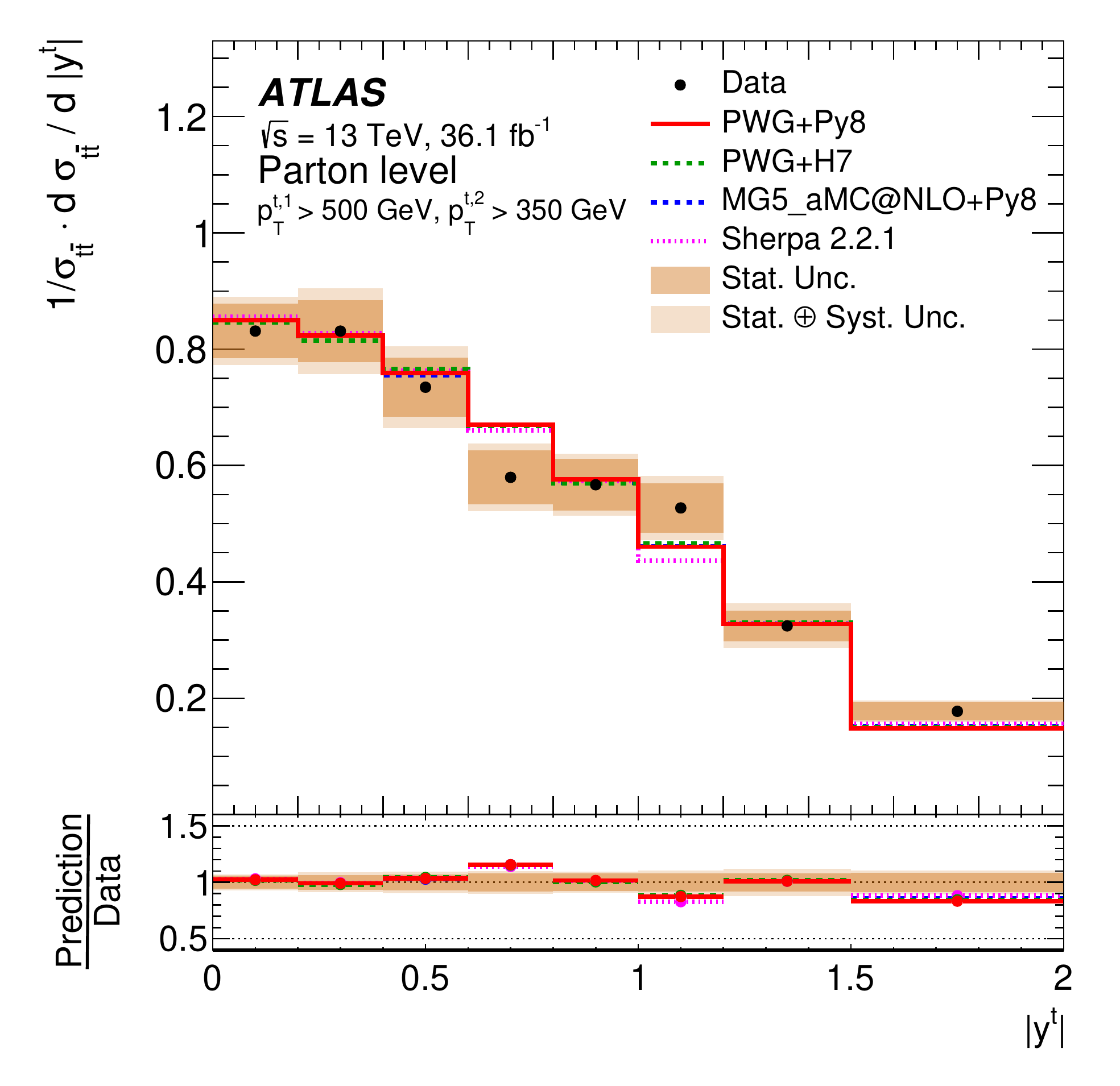}\label{fig:results:parton:randomTop_y_rel}}%
\caption{\small{The normalized parton-level differential cross-sections 
as a function of 
\subref{fig:results:parton:top_pt_rel}~the transverse momentum and
\subref{fig:results:parton:randomTop_y_rel}~the rapidity of the top quark.
The orange bands indicate the total uncertainty in the data in each bin. The vertical bars indicate the statistical uncertainties in the theoretical models. 
The \POWHEG{}+\PythiaEight\ event generator is used as the nominal prediction to correct for detector effects, parton showering and hadronization. Data points are placed at the center of each bin.
The unfolding has required the leading top-quark $\pT > 500$~\GeV~ and the second-leading top-quark $\pT > 350$~\GeV.
}}
\label{fig:parton:random_top_quark_kinematics:rel}
\end{figure*}

\begin{figure*}[htbp]
\centering
\subfigure[]{\includegraphics[width=0.45\textwidth]{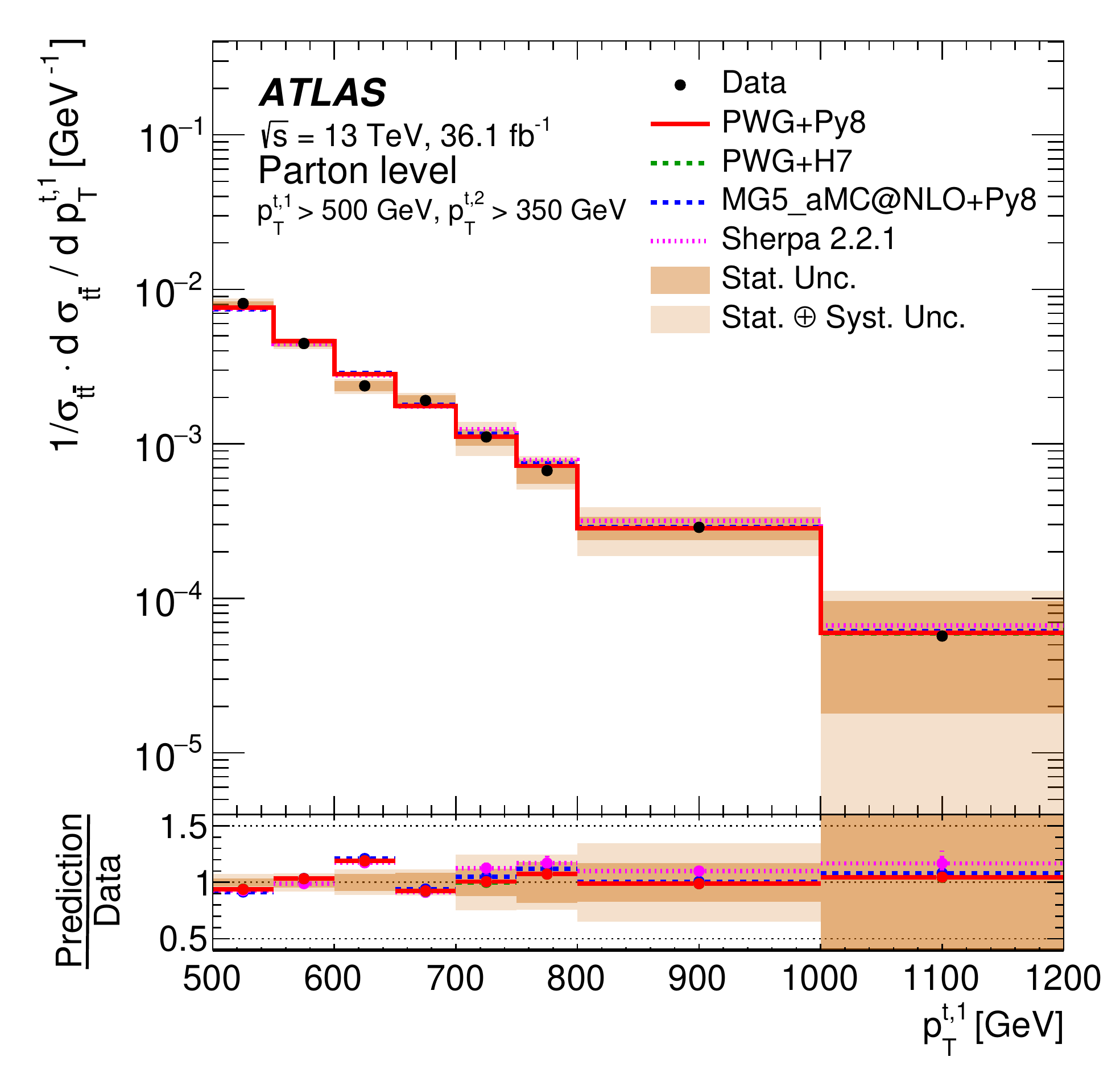}\label{fig:results:parton:leadingTop_pt_rel}}
\subfigure[]{\includegraphics[width=0.45\textwidth]{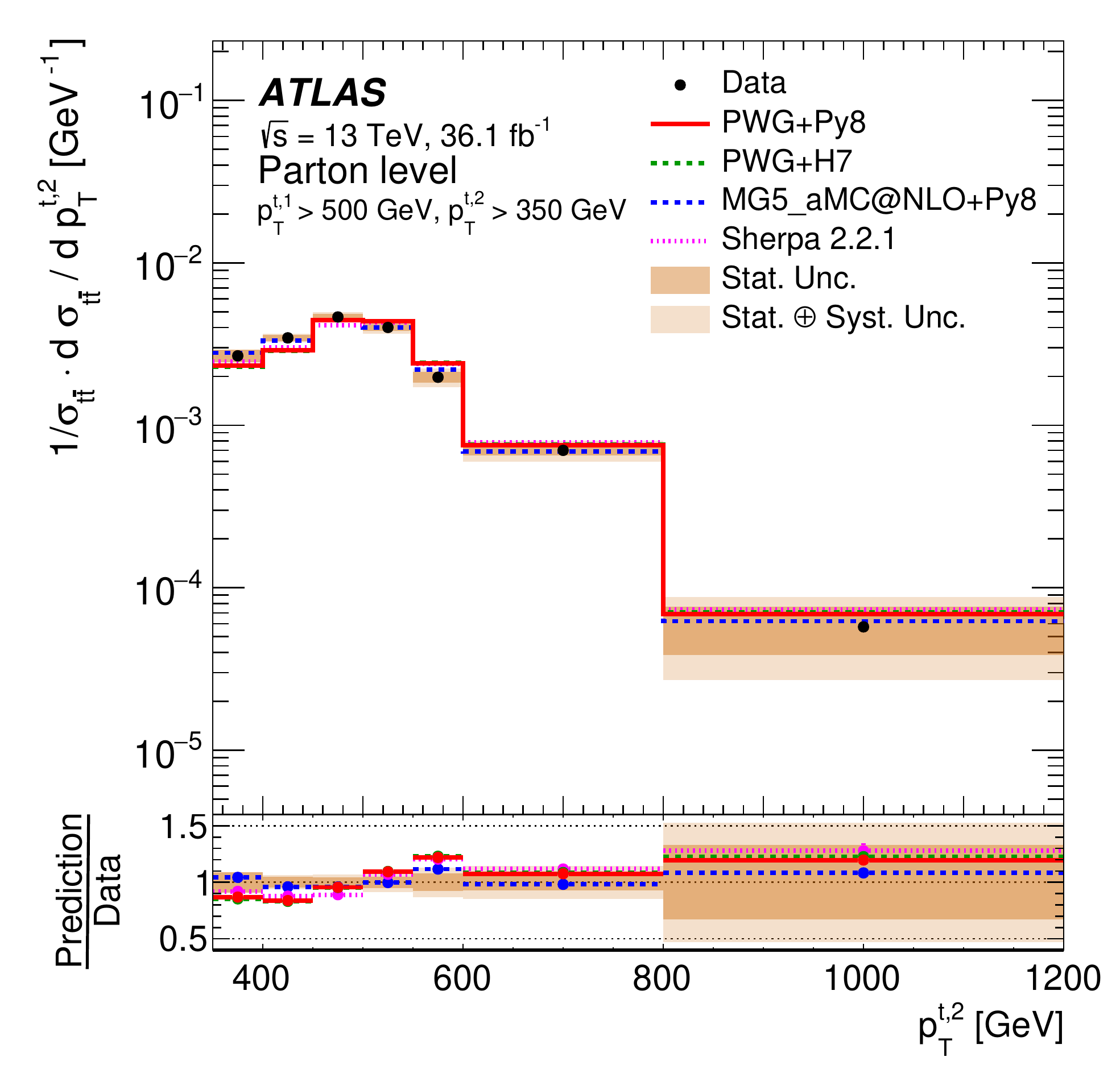}\label{fig:results:parton:subleadingTop_pt_rel}}
\subfigure[]{\includegraphics[width=0.45\textwidth]{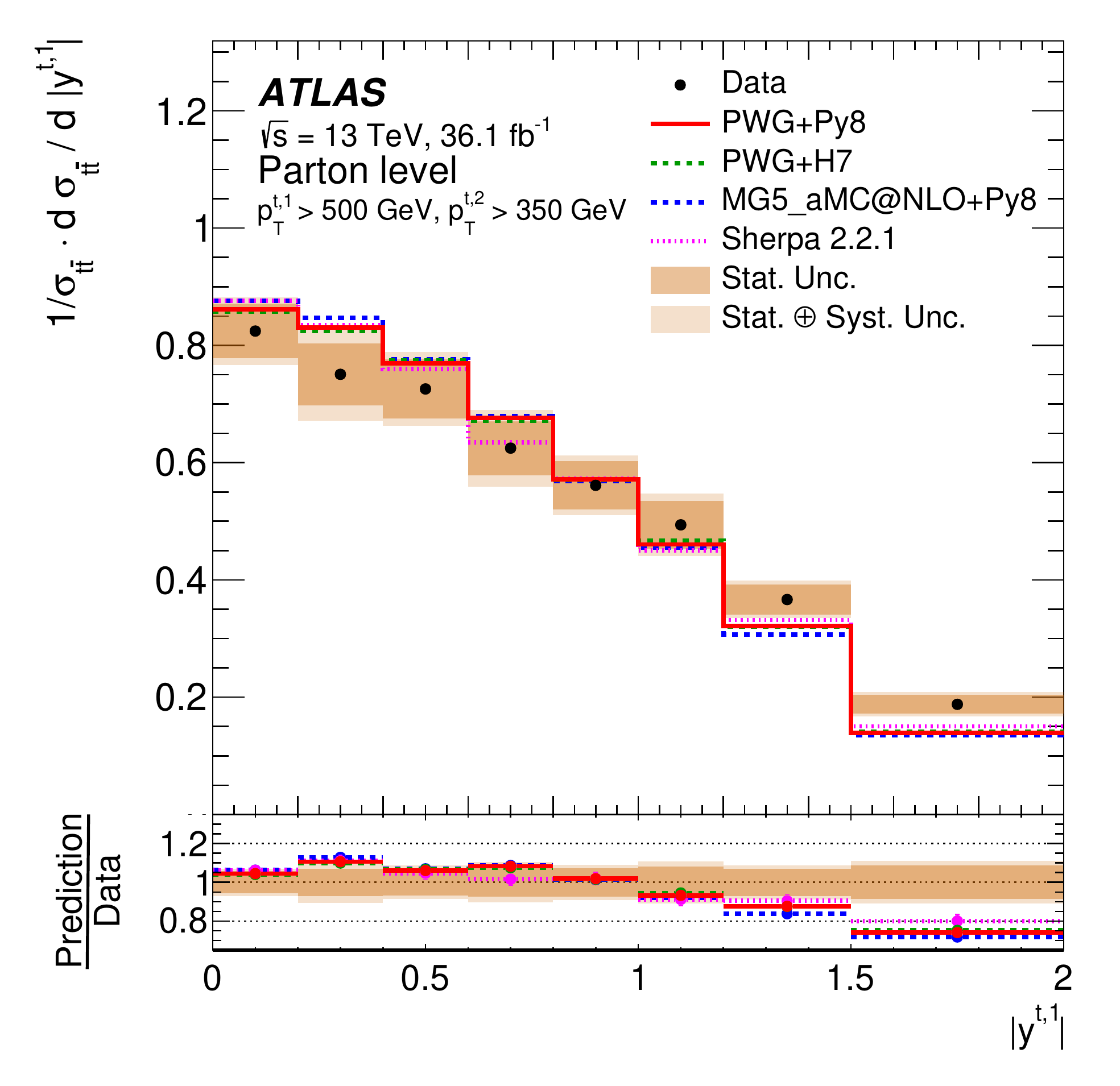}\label{fig:results:parton:leadingTop_y_rel}}
\subfigure[]{\includegraphics[width=0.45\textwidth]{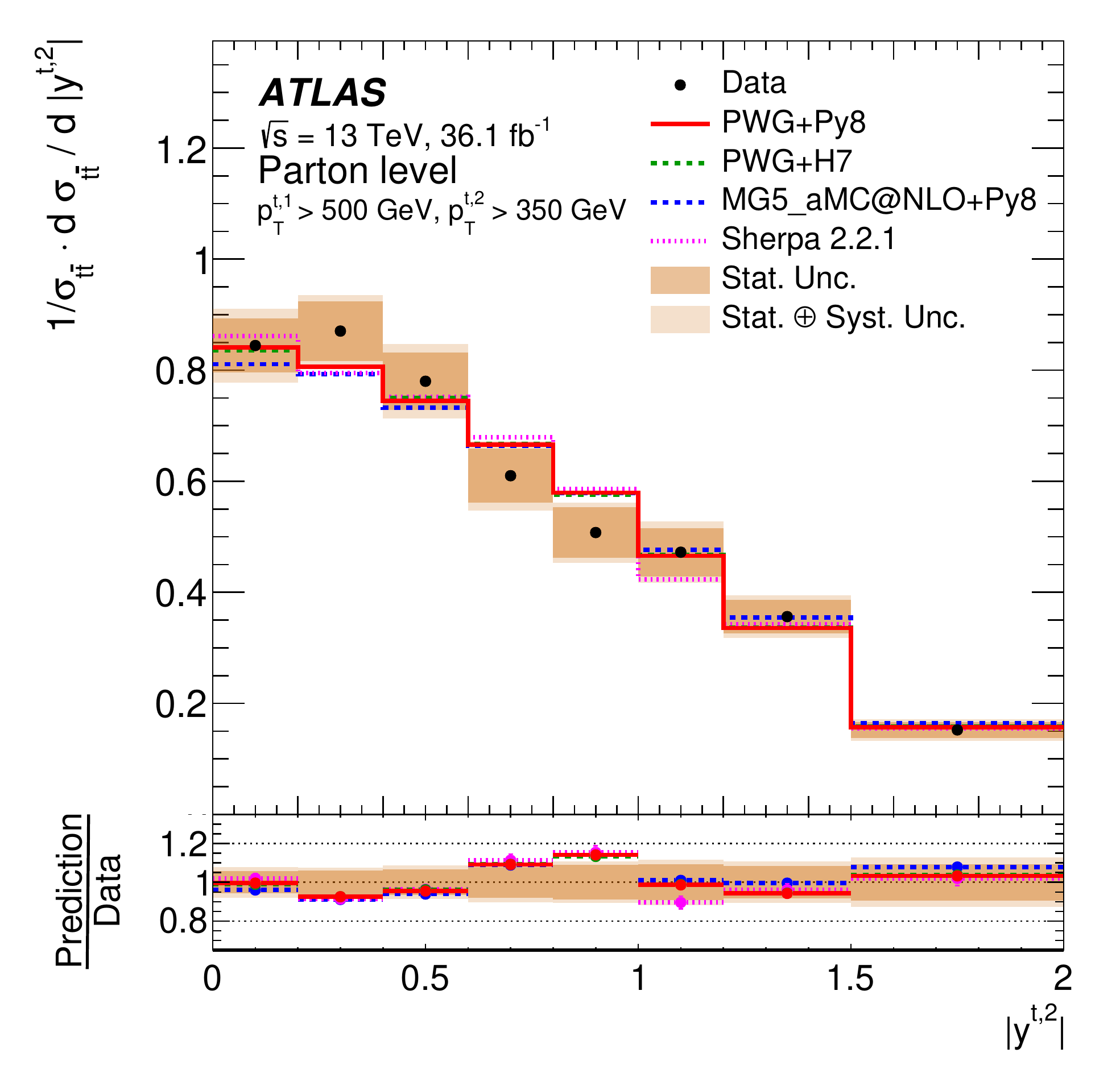}\label{fig:results:parton:subleadingTop_y_rel}}
\caption{\small{The normalized parton-level differential cross-sections 
as a function of 
\subref{fig:results:parton:leadingTop_pt_rel}~the transverse momentum of the leading top quark, 
\subref{fig:results:parton:subleadingTop_pt_rel}~the transverse momentum of the second-leading top quark, 
\subref{fig:results:parton:leadingTop_y_rel} the absolute value of the rapidity of the leading top quark and
\subref{fig:results:parton:subleadingTop_y_rel} the absolute value of the rapidity of the second-leading top quark. 
The orange bands indicate the total uncertainty in the data in each bin. The vertical bars indicate the statistical uncertainties in the theoretical models. 
The \POWHEG{}+\PythiaEight\ event generator is used as the nominal prediction to correct for detector effects, parton showering and hadronization. Data points are placed at the center of each bin.
The unfolding has required the leading top-quark $\pT > 500$~\GeV~ and the second-leading top-quark $\pT > 350$~\GeV.
}}
\label{fig:parton:top_quarks_kinematics:rel}
\end{figure*}

\begin{figure*}[htbp]
\centering
\subfigure[]{\includegraphics[width=0.45\textwidth]{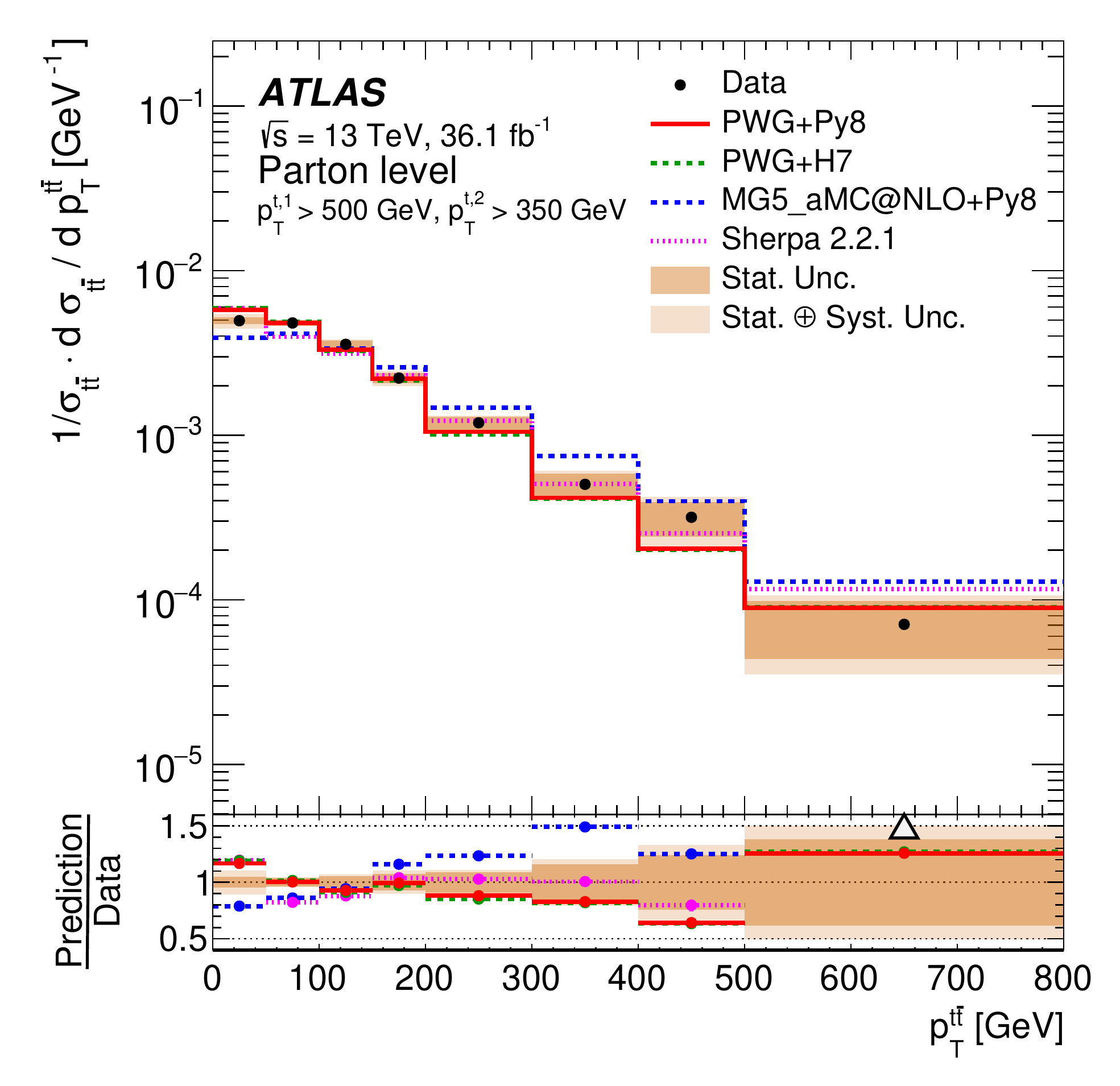}\label{fig:results:parton:ttbar_pt_rel}}
\subfigure[]{\includegraphics[width=0.45\textwidth]{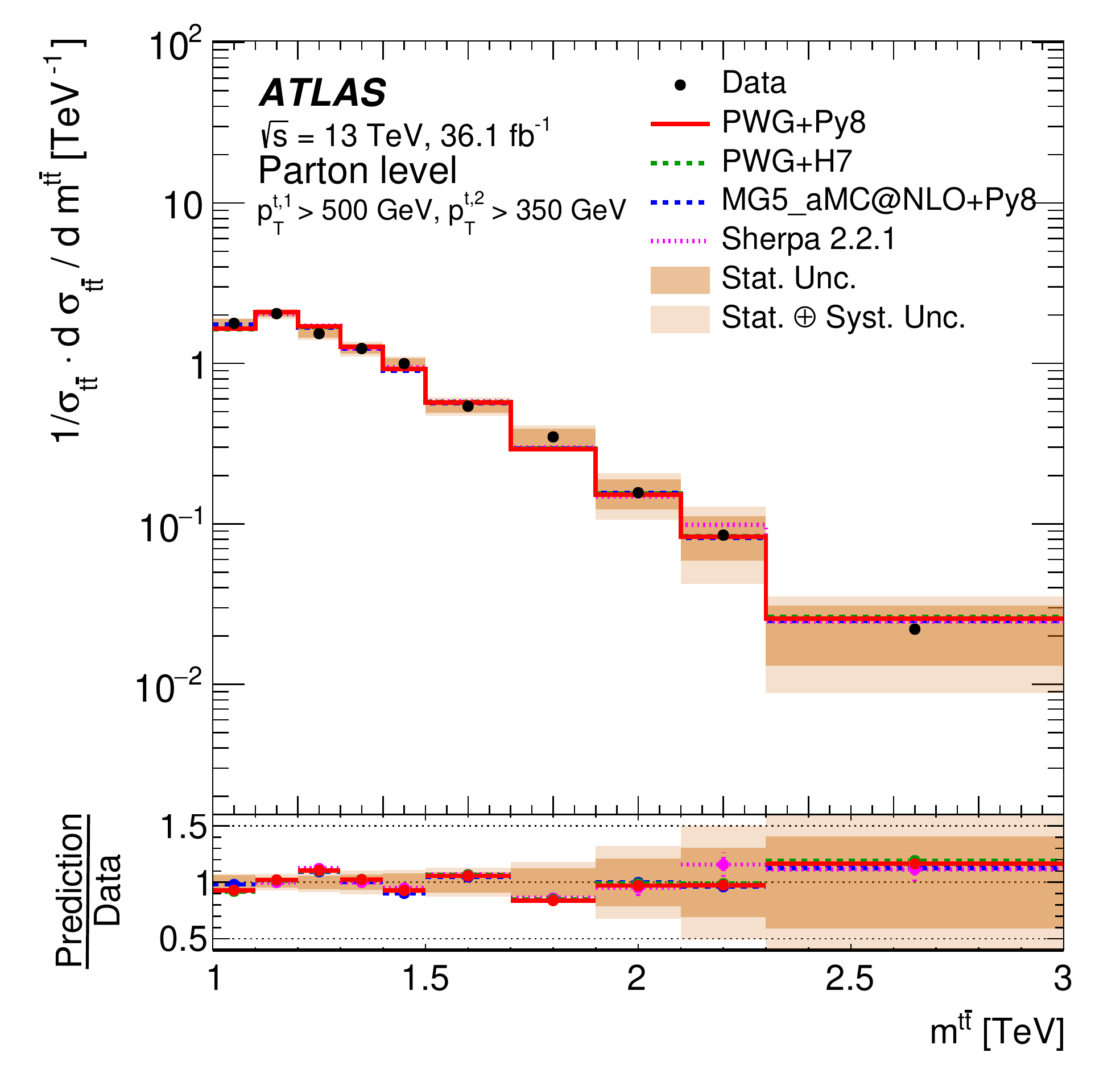}\label{fig:results:parton:ttbar_mass_rel}}
\subfigure[]{\includegraphics[width=0.45\textwidth]{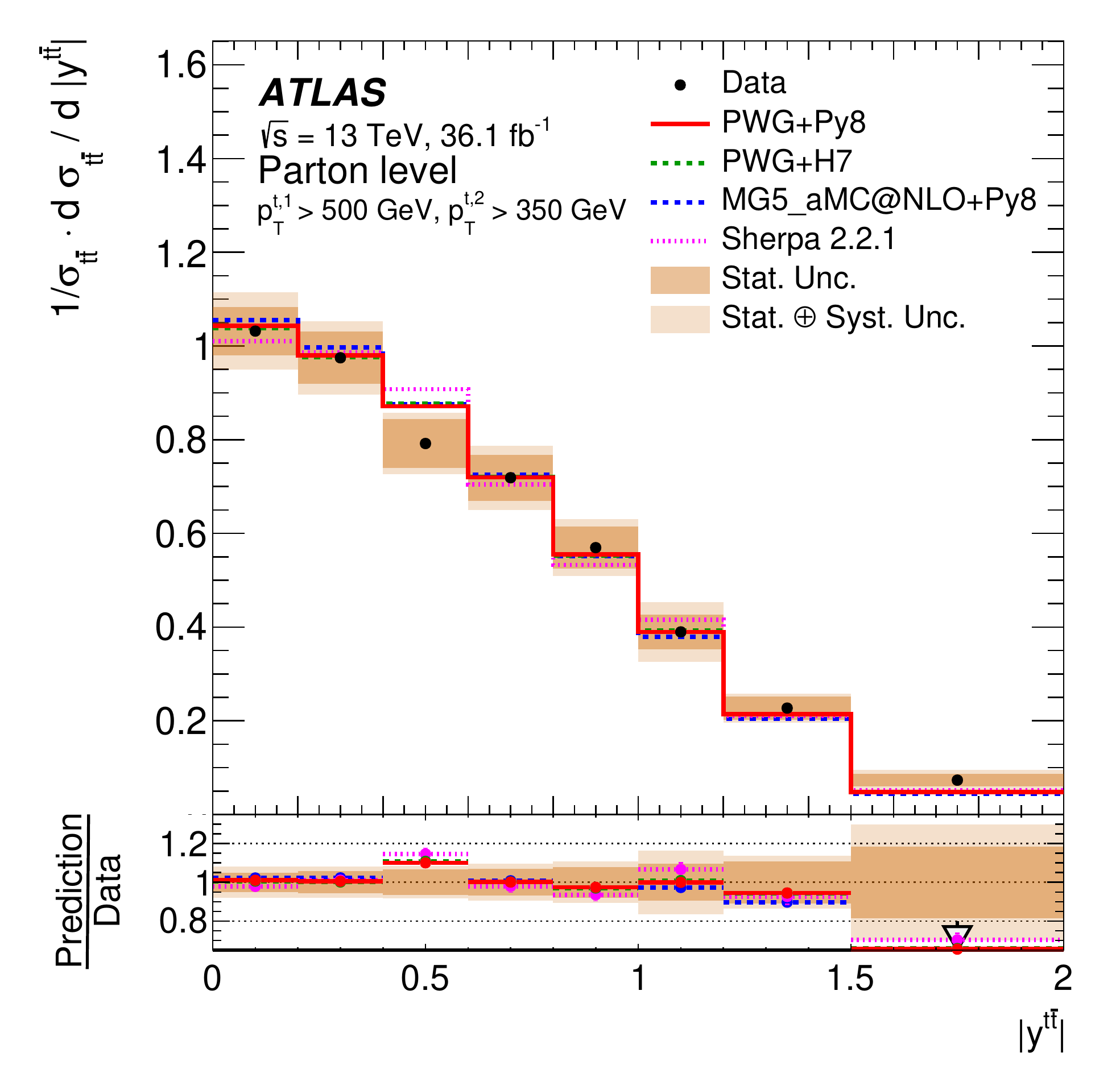}\label{fig:results:parton:ttbar_y_rel}}
\caption{\small{
The normalized parton-level differential cross-sections as a function of
\subref{fig:results:parton:ttbar_pt_rel}~the~\ttbar~\pt,
\subref{fig:results:parton:ttbar_mass_rel}\ the \ttbar\ invariant mass and
\subref{fig:results:parton:ttbar_y_rel}\ the absolute value of \ttbar\ rapidity.
The orange bands indicate the total uncertainty in the data in each bin. The vertical bars indicate the statistical uncertainties in the theoretical models. 
The \POWHEG{}+\PythiaEight\ event generator is used as the nominal prediction to correct for detector effects, parton showering and hadronization. Data points are placed at the center of each bin.
The unfolding has required the leading top-quark $\pT > 500$~\GeV~ and the second-leading top-quark $\pT > 350$~\GeV.
}}
\label{fig:parton:ttbar_system_baseline:rel}
\end{figure*}

\begin{figure*}[htbp]
\centering
\subfigure[]{\includegraphics[width=0.45\textwidth]{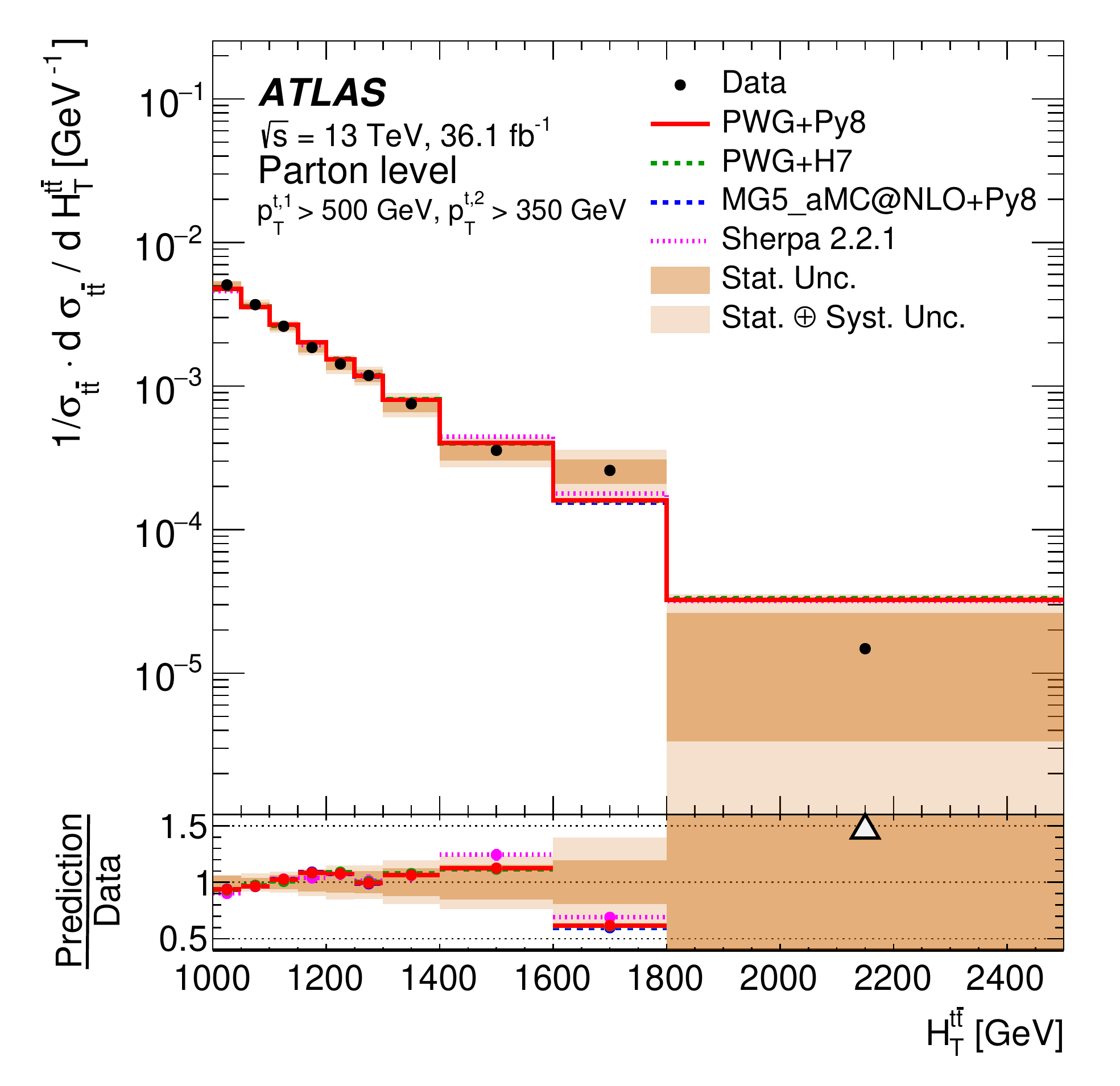}\label{fig:results:parton:H_tt_rel}}%
\subfigure[]{\includegraphics[width=0.45\textwidth]{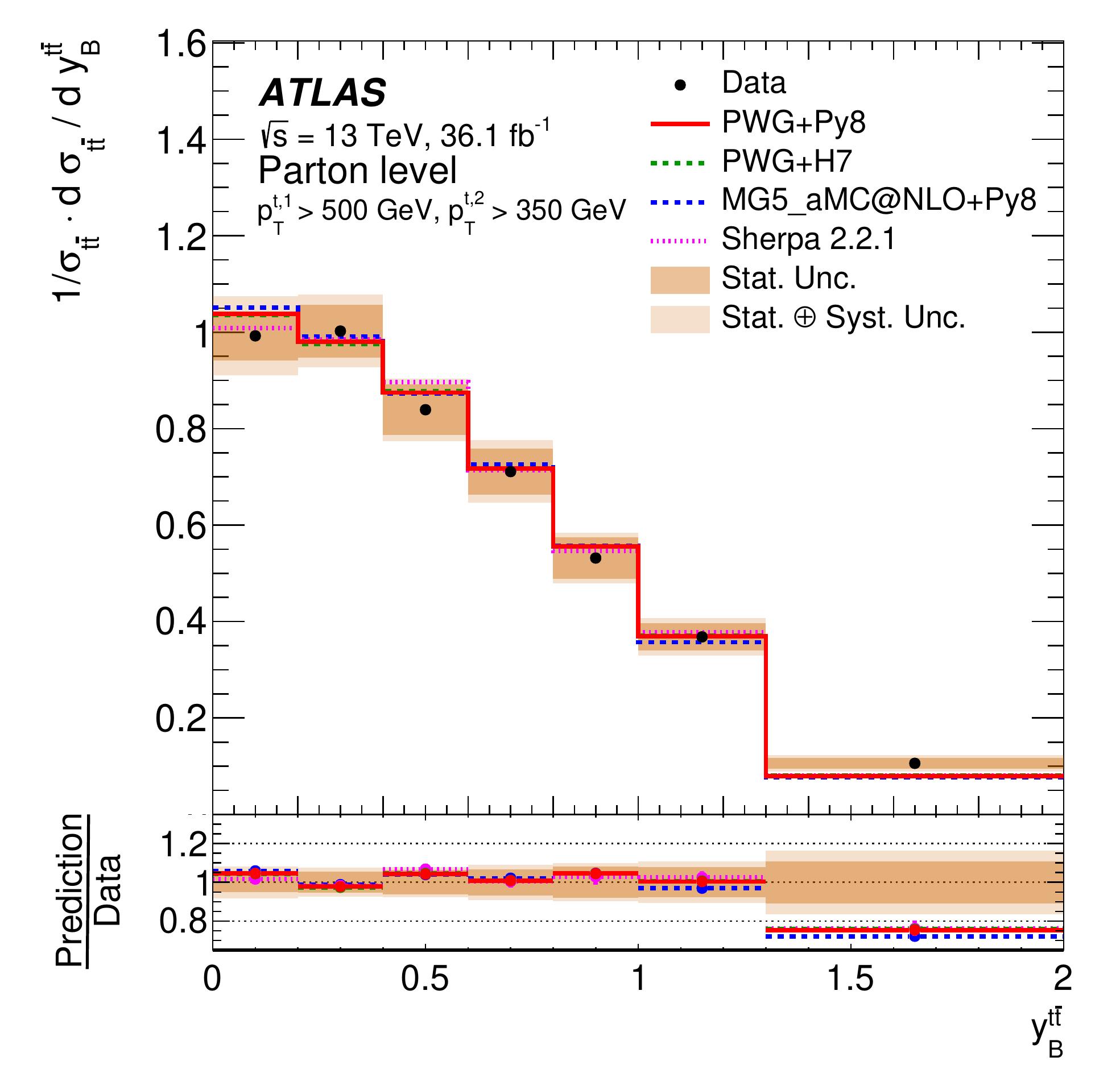}\label{fig:results:parton:y_boost_rel}}%
\caption{\small{
The normalized parton-level differential cross-sections as a function of
\subref{fig:results:parton:H_tt_rel}\  $H^{\ttbar}$ and 
\subref{fig:results:parton:y_boost_rel} $y_{B}^{t\bar{t}}$.
The orange bands indicate the total uncertainty in the data in each bin. The vertical bars indicate the statistical uncertainties in the theoretical models. 
The \POWHEG{}+\PythiaEight\ event generator is used as the nominal prediction to correct for detector effects, parton showering and hadronization. Data points are placed at the center of each bin.
The unfolding has required the leading top-quark $\pT > 500$~\GeV~ and the second-leading top-quark $\pT > 350$~\GeV.
}}
\label{fig:parton:ttbar_system_extra:rel1}
\end{figure*}

\begin{figure*}[htbp]
\centering
\subfigure[]{\includegraphics[width=0.45\textwidth]{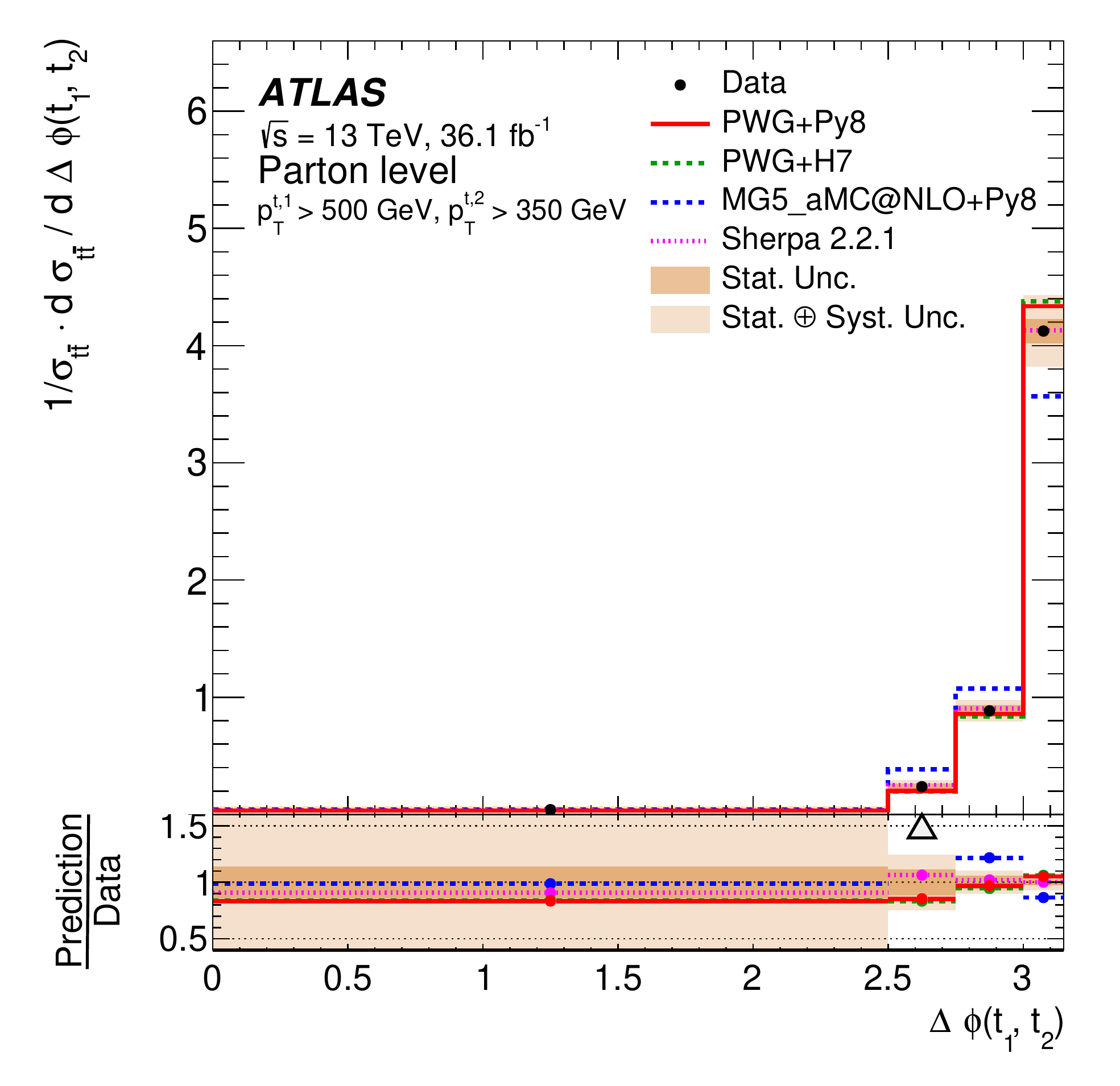}\label{fig:results:parton:delta_phi_rel}}
\subfigure[]{\includegraphics[width=0.45\textwidth]{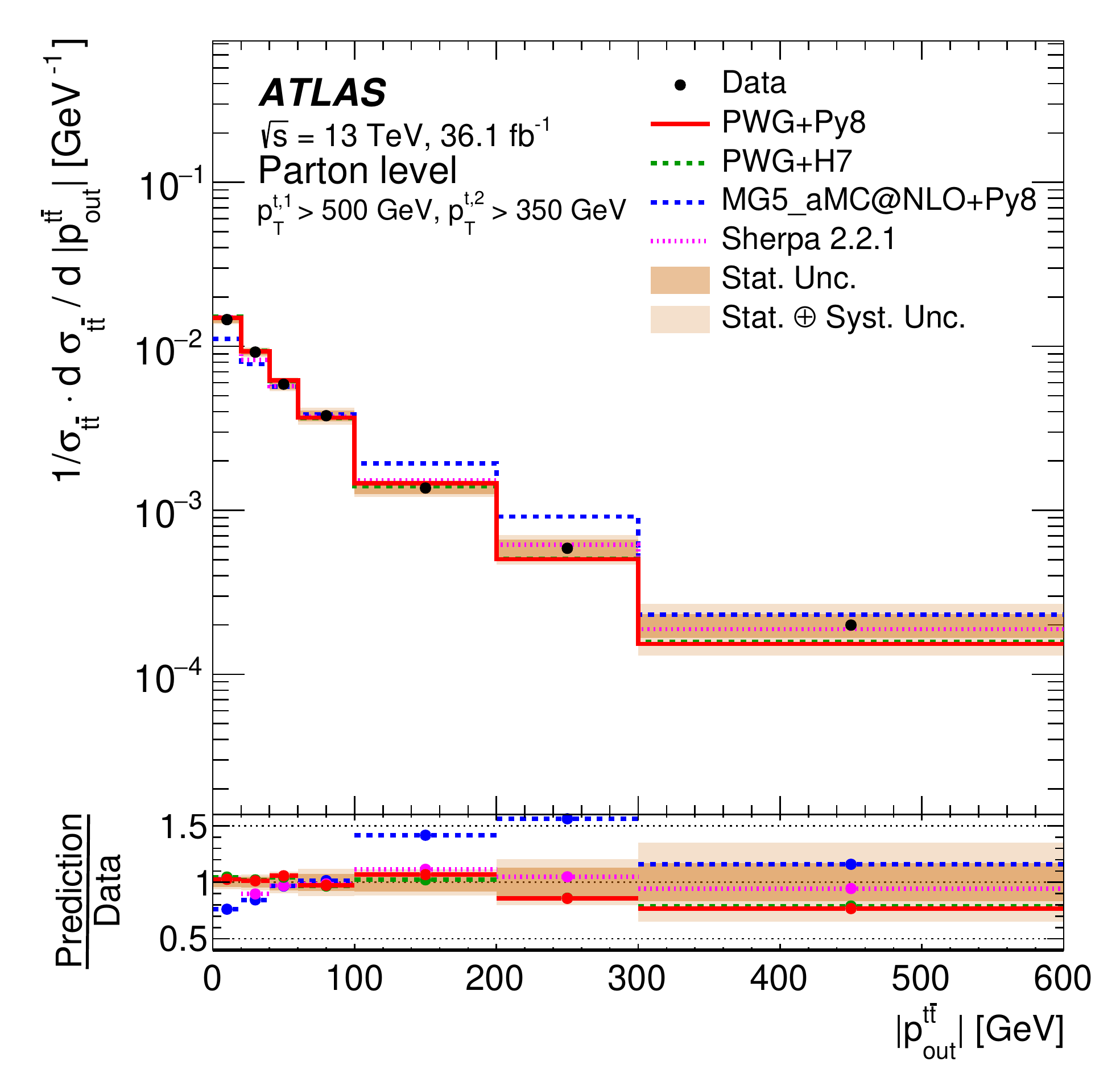}\label{fig:results:parton:pout_rel}}%
\caption{\small{
The normalized parton-level differential cross-sections as a function of
\subref{fig:results:parton:delta_phi_rel}\ $\Delta\phi (t_{1},t_{2})$\ and
\subref{fig:results:parton:pout_rel}\ \absPoutttbar.
The orange bands indicate the total uncertainty in the data in each bin. The vertical bars indicate the statistical uncertainties in the theoretical models. 
The \POWHEG{}+\PythiaEight\ event generator is used as the nominal prediction to correct for detector effects, parton showering and hadronization. Data points are placed at the center of each bin.
The unfolding has required the leading top-quark $\pT > 500$~\GeV~ and the second-leading top-quark $\pT > 350$~\GeV.
}}
\label{fig:parton:ttbar_system_extra:rel2}
\end{figure*}

The unfolded parton-level phase-space differential cross-sections are shown in 
Figs.\ref{fig:parton:random_top_quark_kinematics:rel}--\ref{fig:parton:ttbar_system_extra:rel2}\
for the kinematical variables describing the top quark, leading top quark, second-leading top quark and the \ttbar\ system.

To measure the average top-quark \pT\ distribution that can be compared with NNLO+NNLL calculations \cite{Czakon:2017wor,NNLO_calc,nnloMtt}, 
the data are unfolded by randomly selecting 
one of the two top-quark candidates at the detector level for each event.
The normalized average top-quark \pT\ and rapidity differential cross-sections are shown in Fig.~\ref{fig:results:parton:top_pt_rel} and 
Fig.~\ref{fig:results:parton:randomTop_y_rel}, respectively. 

\begin{figure*}[htbp]
\centering
\subfigure[]{\includegraphics[width=0.45\textwidth]{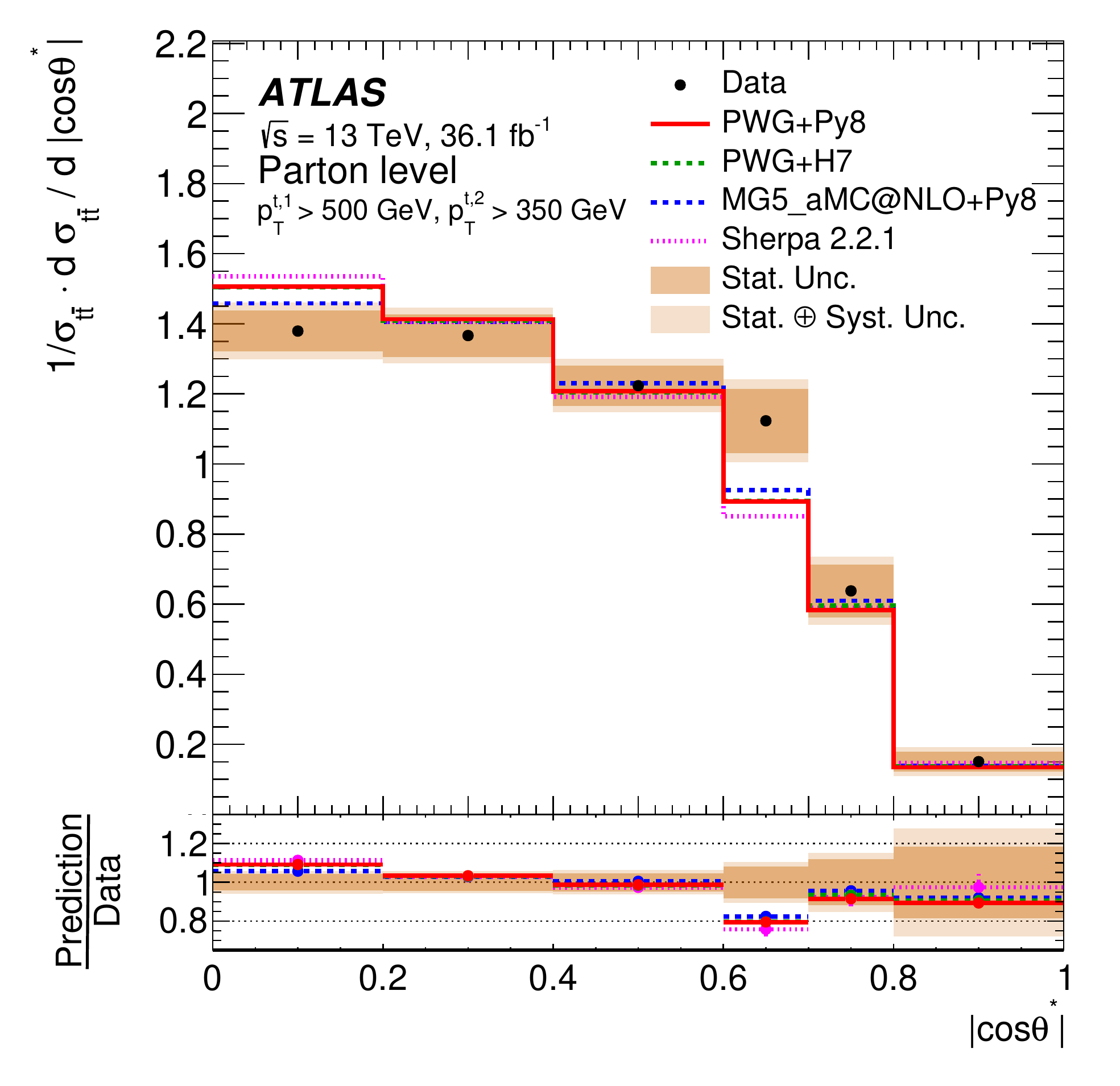}\label{fig:results:parton:cos_theta_star_rel}}%
\subfigure[]{\includegraphics[width=0.45\textwidth]{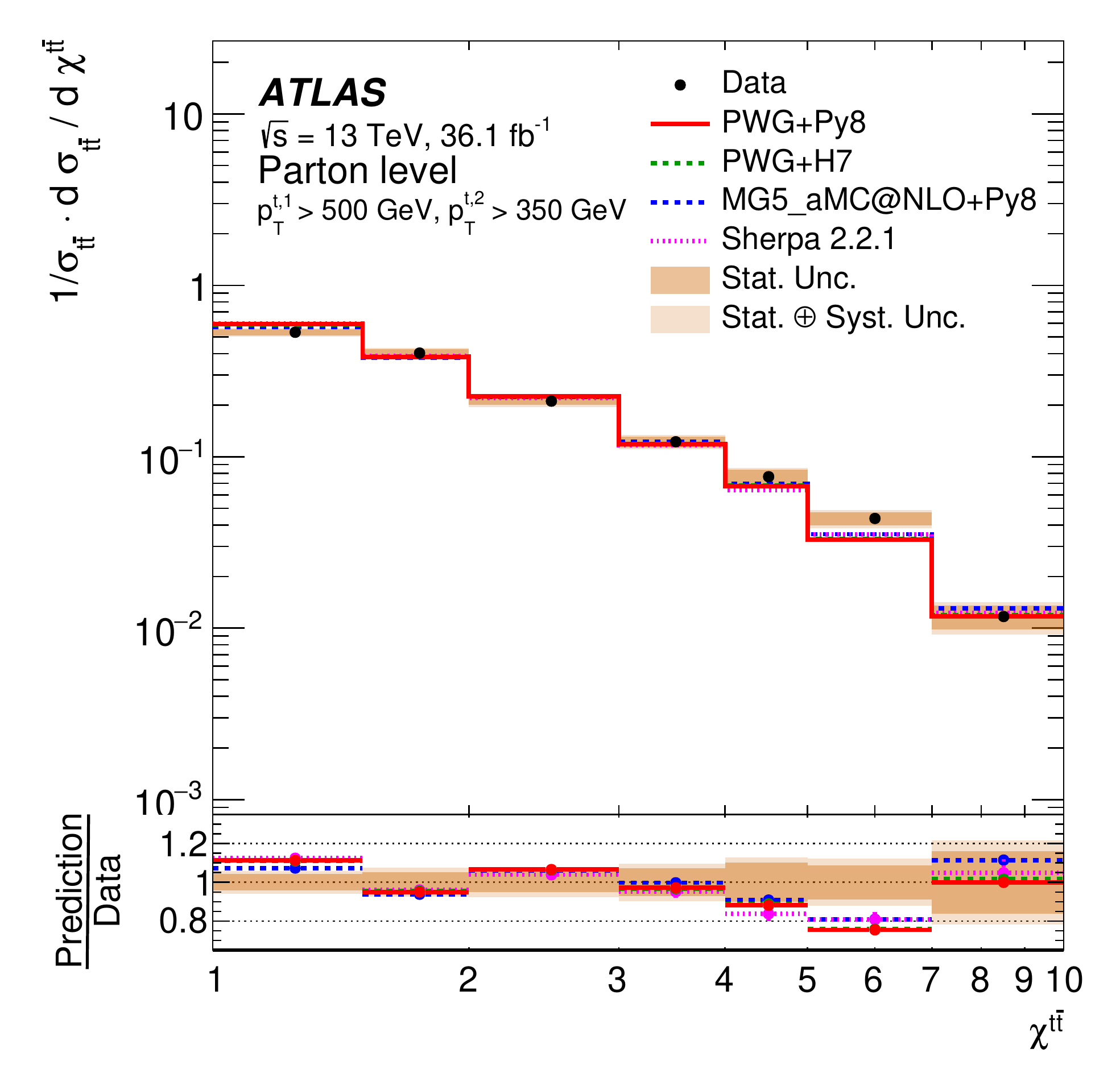}\label{fig:results:parton:chi_ttbar_rel}}%
\caption{\small{
The normalized parton-level differential cross-sections as a function of
\subref{fig:results:parton:cos_theta_star_rel} $\cos \theta^{*}$~ and
\subref{fig:results:parton:chi_ttbar_rel}\ $\chi^{\ttbar}$.
The orange bands indicate the total uncertainty in the data in each bin. The vertical bars indicate the statistical uncertainties in the theoretical models.
The \POWHEG{}+\PythiaEight\ event generator is used as the nominal prediction to correct for detector effects, parton showering and hadronization. Data points are placed at the center of each bin.
The unfolding has required the leading top-quark $\pT > 500$~\GeV~ and the second-leading top-quark $\pT > 350$~\GeV.
}}
\label{fig:parton:ttbar_system_extra:rel3}
\end{figure*}

\FloatBarrier

\subsection{Fiducial phase-space inclusive cross-section}

The cross-section of \ttbar\ production in the fiducial phase space defined in this analysis is determined 
using the same methodology employed to obtain the unfolded differential cross-sections at particle level, with the exception that all events are grouped into a single bin.
The inclusive fiducial cross-section is:
\begin{eqnarray}
\sigmafid = 292 \pm 7\ \rm{(stat)} \pm 76\ \rm{(syst)}~\rm{fb}.\nonumber
\end{eqnarray}
The systematic uncertainties in this measurement, which are dominated by tagging and modeling uncertainties, 
are summarized in Table~\ref{tab:systematics}.

\begin{figure*}[htbp]
\centering
\includegraphics[width=0.7\textwidth]{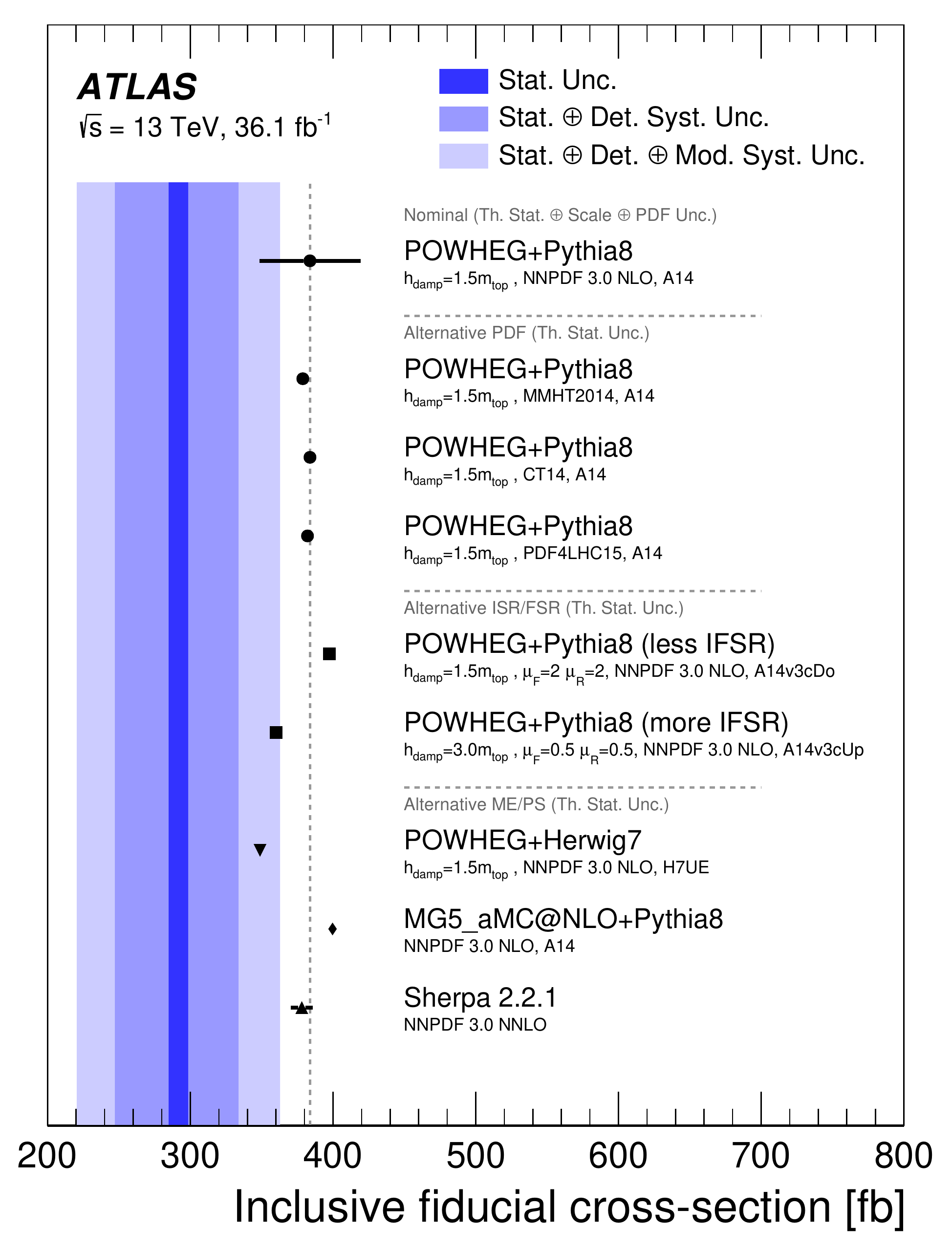}
\caption{\small{Particle-level fiducial phase-space cross-section. 
The shaded (blue) bands indicate the statistical, detector and modeling uncertainties in the measurement. 
The \POWHEG{}+\PythiaEight{} event generator is used as the nominal prediction to correct for detector effects. 
The uncertainty associated with the \POWHEG{}+\PythiaEight{} signal model is the sum in quadrature of statistical, 
scale and PDF uncertainties as well as the uncertainty in the inclusive cross-section prediction. 
Other predictions show only the statistical uncertainty of the MC sample.}}
\label{fig:particle:inclusive:abs}
\end{figure*}

The resulting inclusive fiducial cross-section measurement is 
shown in Fig.~\ref{fig:particle:inclusive:abs}\ and compared
with various MC predictions. 
The measured value is below all of the predictions, and
in particular is below the \POWHEG+\PythiaEight\ prediction of $384\pm36$~fb.
The uncertainty in this MC prediction is the sum in quadrature of statistical, 
scale and PDF uncertainties, including the uncertainty in the {NNLO+NNLL} total cross-section prediction. 
The scale uncertainty is estimated by determining the envelope of predictions when the 
factorization $\mu_{\rm F}$ and renormalization $\mu_{\rm R}$ scales are varied by factors of $0.5$\ and $2.0$. 
The PDF uncertainty is obtained using the PDF4LHC prescription with 30 eigenvectors. 
All of the predictions are normalized to the {NNLO+NNLL} total \ttbar\ cross-section.

\FloatBarrier

\section{Comparisons with Standard Model predictions}
\label{sec:comparisons}

The particle-level fiducial phase-space differential cross-sections and the parton-level differential cross-sections are
compared with several Standard Model calculations. 

The predicted total particle-level cross-section for top-quark pair production in the fiducial 
phase-space region is larger than the one observed. 
However, the effect is not statistically significant due to the large systematic
uncertainties. 
A better agreement is found for \POWHEG{}+\HerwigSeven{} and to a lesser extent for the predictions of
\POWHEG{}+\PythiaEight{} with more initial- and final-state radiation.  

The information provided by the shapes of the observed differential cross-section measurements is compared to the 
predictions using the $\chi^2$\ test described in Sec.~\ref{sec:prop_uncertainties}, which takes into account the 
correlations between the measured quantities.  
The largest correlations at the detector-level arise from sources of uncertainty that affect all bins equally, so that 
the most effective comparison is made using the normalized differential cross-sections where many of the common
detector-level uncertainties largely cancel.
The $\chi^{2}$ values and associated $p$-values that quantify the
level of agreement between the measurements and the predictions are shown in Table \ref{tab:chi2:particle:rel}\
for the normalized particle-level fiducial phase-space differential
cross-sections and in Table \ref{tab:chi2:parton:rel} for the
normalized parton-level differential cross-sections.

The particle-level differential cross-sections are
generally well-described by the \POWHEG+\PYTHIA{8},
\POWHEG+\HerwigSeven, \MGAMCatNLO{}+\PYTHIA{8} and \SHERPA\
event generator predictions. 
The \ttbar\ differential cross-section as a function of the absolute value of the
leading top-quark rapidity (Fig.~\ref{fig:particle:t1_y:rel}) is broader in the data than 
the predictions of all Monte Carlo event generators.
 A similar effect is observed in the \ttbar\ system rapidity differential cross-section
(Fig.~\ref{fig:particle:tt_y:rel}).
However, the $p$-values arising from the $\chi^2$\ comparisons are
mostly within 0.15 to 0.55, reflecting the overall reasonable agreement of the predictions with the measured 
differential cross-sections. 
There are modest differences in the
distributions of the production angle $\cos\theta^{*}$\ (Fig.~\ref{fig:particle:tt_cosThetaStar:rel}) and the
variable \chittbar\ (Fig.~\ref{fig:particle:tt_chittbar:rel}), both
showing $p$-values that are generally below 0.2.

The most significant deviations are in the \MGAMCatNLO\ particle-level fiducial phase-space differential cross-sections 
as a function of \ptttbar\ (Fig.~\ref{fig:particle:tt_pt:rel}), \deltaPhittbar\
(Fig.~\ref{fig:particle:tt_dPhittbar:rel}) and
\absPoutttbar\ (Fig.~\ref{fig:particle:tt_pout:rel}) for which
the \MGAMCatNLO{}+\PYTHIA{8}\ MC event generator predicts a harder \ptttbar\ spectrum, 
a broader azimuthal opening angle differential cross-section than what is measured and
a slower decline than observed as a function of \absPoutttbar.

There is similar good agreement between the parton-level differential 
cross-sections and the \POWHEG{}+\PYTHIA{8}, \POWHEG{}+\HerwigSeven,
\MGAMCatNLO{} +\PYTHIA{8}\ and \SHERPA\ predictions, 
confirming the results of the fiducial phase-space measurements, but with larger uncertainties. 
As shown in Fig.~\ref{fig:results:parton:ttbar_pt_rel}, the
\POWHEG{}+\PYTHIA{8}\ and \POWHEG{} +\HerwigSeven\ event generators predict a
softer \pT\ spectrum of the  \ttbar\ system, while the \MGAMCatNLO{} +
\PYTHIA{8}\ event generator predicts a harder spectrum.
The \SHERPA\ event generator offers a good description of the differential cross-section behavior for \ptttbar\ in the 
range $100$\ to $500$~\GeV\ but predicts a steeper distribution for lower momenta and a higher rate for $\ptttbar > 500$~\GeV\
than observed. 

The modeling uncertainties generally play a dominant role in
determining the significance of the difference between the measurements and the nominal \POWHEG{}+\PYTHIA{8}\ prediction.
It suggests that future work should seek the sources for this potential discrepancy,
considering variations in parton shower and hadronization models as well as the matching of 
higher-order matrix elements with the parton shower model.

\begin{table*} [t!]
\footnotesize
\caption{\small{
Comparison between the measured normalized particle-level fiducial phase-space differential cross-sections and the predictions 
from several SM event generators.
For each variable and prediction, a $\chi^2$ and a $p$-value are calculated using the covariance matrix described in the text,
which includes all sources of uncertainty.
The number of degrees of freedom (NDF) is equal to $N_{\rm b}-1$, where $N_{\rm b}$ is the number of bins in the distribution.
}}
\resizebox{\textwidth}{!}{
\centering
\footnotesize
\centering
\begin{tabular}{|c | r @{/} l r<{\hspace{4pt}}  | r @{/} l r<{\hspace{4pt}}  | r @{/} l r<{\hspace{4pt}}  | r @{/} l r<{\hspace{4pt}}  | r @{/} l r<{\hspace{4pt}}  | r @{/} l r<{\hspace{4pt}} | }
\hline
Observable  & \multicolumn{3}{|c|}{\textsc{PWG+PY8}} & \multicolumn{3}{|c|}{\textsc{\AMCatNLO+PY8}} & 
\multicolumn{3}{|c|}{\textsc{PWG+H7}} & 
\multicolumn{3}{|c|}{$\begin{array}{c} \textsc{PWG+PY8} \\ \textrm{(more IFSR)} \end{array}$} & 
\multicolumn{3}{|c|}{$\begin{array}{c} \textsc{PWG+PY8} \\ \textrm{(less IFSR)} \end{array}$} & 
\multicolumn{3}{|c|}{\textsc{Sherpa~2.2.1}} \\
& \multicolumn{2}{|c}{$\chi^{2}$/NDF} & $p$-value & \multicolumn{2}{|c}{$\chi^{2}$/NDF} &  $p$-value  & \multicolumn{2}{|c}{$\chi^{2}$/NDF} &  $p$-value  & \multicolumn{2}{|c}{$\chi^{2}$/NDF} &  $p$-value  & \multicolumn{2}{|c}{$\chi^{2}$/NDF} &  $p$-value  & \multicolumn{2}{|c}{$\chi^{2}$/NDF} & $p$-value  \\
\hline
\hline
$p_{\text{T}}^{t,1}$ &
{\ } 7.7 & 7 & 0.36  &  {\ } 8.2 & 7 & 0.32  &  {\ } 8.0 & 7 & 0.33  &  {\ } 9.1 & 7 & 0.24  &  {\ } 8.7 & 7 & 0.27  &  {\ } 9.3 & 7 & 0.23 \\
$           |y^{t,1}|$ &
{\ } 7.5 & 5 & 0.18  &  {\ } 12.2 & 5 & 0.03  &  {\ } 6.8 & 5 & 0.24  &  {\ } 8.8 & 5 & 0.12  &  {\ } 8.1 & 5 & 0.15  &  {\ } 4.0 & 5 & 0.55 \\
$p_{\text{T}}^{t,2}$ &
{\ } 8.6 & 6 & 0.20  &  {\ } 2.6 & 6 & 0.86  &  {\ } 9.9 & 6 & 0.13  &  {\ } 12.2 & 6 & 0.06  &  {\ } 5.0 & 6 & 0.54  &  {\ } 5.0 & 6 & 0.55 \\
$           |y^{t,2}|$ &
{\ } 3.7 & 5 & 0.59  &  {\ } 4.6 & 5 & 0.46  &  {\ } 3.1 & 5 & 0.68  &  {\ } 3.5 & 5 & 0.63  &  {\ } 3.2 & 5 & 0.67  &  {\ } 2.9 & 5 & 0.72 \\
$        m^{t\bar{t}}$ &
{\ } 4.5 & 9 & 0.88  &  {\ } 4.7 & 9 & 0.86  &  {\ } 4.0 & 9 & 0.91  &  {\ } 5.3 & 9 & 0.81  &  {\ } 5.2 & 9 & 0.82  &  {\ } 10.0 & 9 & 0.35 \\
$p_{\text{T}}^{t\bar{t}}$ &
{\ } 7.8 & 5 & 0.17  &  {\ } 20.9 & 5 & $<$0.01  &  {\ } 12.6 & 5 & 0.03  &  {\ } 15.0 & 5 & 0.01  &  {\ } 1.9 & 5 & 0.86  &  {\ } 1.9 & 5 & 0.87 \\
$      |y^{t\bar{t}}|$ &
{\ } 1.1 & 5 & 0.95  &  {\ } 2.2 & 5 & 0.83  &  {\ } 0.9 & 5 & 0.97  &  {\ } 0.8 & 5 & 0.98  &  {\ } 1.8 & 5 & 0.88  &  {\ } 1.7 & 5 & 0.89 \\
$   {\chi}^{t\bar{t}}$ &
{\ } 14.2 & 6 & 0.03  &  {\ } 12.7 & 6 & 0.05  &  {\ } 13.6 & 6 & 0.03  &  {\ } 16.9 & 6 & $<$0.01  &  {\ } 10.1 & 6 & 0.12  &  {\ } 18.5 & 6 & $<$0.01 \\
$  y_\mathrm{B}^{t\bar{t}}$ &
{\ } 2.5 & 6 & 0.87  &  {\ } 3.3 & 6 & 0.77  &  {\ } 2.2 & 6 & 0.90  &  {\ } 2.6 & 6 & 0.86  &  {\ } 2.8 & 6 & 0.84  &  {\ } 3.0 & 6 & 0.81 \\
$\absPoutttbar$ &
{\ } 1.9 & 6 & 0.93  &  {\ } 53.1 & 6 & $<$0.01  &  {\ } 3.1 & 6 & 0.80  &  {\ } 4.2 & 6 & 0.64  &  {\ } 4.8 & 6 & 0.57  &  {\ } 5.9 & 6 & 0.44 \\
${\Delta\phi}^{t\bar{t}}$ &
{\ } 0.9 & 3 & 0.84  &  {\ } 16.3 & 3 & $<$0.01  &  {\ } 2.0 & 3 & 0.58  &  {\ } 3.0 & 3 & 0.40  &  {\ } 0.6 & 3 & 0.89  &  {\ } 3.4 & 3 & 0.33 \\
$  H_\mathrm{T}^{t\bar{t}}$ &
{\ } 4.8 & 6 & 0.57  &  {\ } 5.2 & 6 & 0.52  &  {\ } 4.5 & 6 & 0.61  &  {\ } 5.0 & 6 & 0.54  &  {\ } 5.0 & 6 & 0.55  &  {\ } 3.1 & 6 & 0.80 \\
${\cos{\theta}}^{\star}$ &
{\ } 9.9 & 5 & 0.08  &  {\ } 10.5 & 5 & 0.06  &  {\ } 9.3 & 5 & 0.10  &  {\ } 12.8 & 5 & 0.03  &  {\ } 6.5 & 5 & 0.26  &  {\ } 18.7 & 5 & $<$0.01 \\
\hline
\end{tabular}

}
\label{tab:chi2:particle:rel}
\end{table*}

\begin{table*} [t!]
\footnotesize
\caption{\small{
Comparison between the measured normalized parton-level differential cross-sections and the predictions from several MC event generators. 
For each variable and prediction, a $\chi^2$ and a $p$-value are calculated using the covariance matrix described in the text, which includes
all sources of uncertainty.
The number of degrees of freedom (NDF) is equal to $N_{\rm b}-1$, where $N_{\rm b}$ is the number of bins in the distribution.
}}
\resizebox{\textwidth}{!}{
\centering
\footnotesize
\centering
\begin{tabular}{|c | r @{/} l r<{\hspace{4pt}}  | r @{/} l r<{\hspace{4pt}}  | r @{/} l r<{\hspace{4pt}}  | r @{/} l r<{\hspace{4pt}}  | r @{/} l r<{\hspace{4pt}}  | r @{/} l r<{\hspace{4pt}} | }
\hline
Observable  & \multicolumn{3}{|c|}{\textsc{PWG+PY8}} & \multicolumn{3}{|c|}{\textsc{\AMCatNLO+PY8}} & 
\multicolumn{3}{|c|}{\textsc{PWG+H7}} & 
\multicolumn{3}{|c|}{$\begin{array}{c} \textsc{PWG+PY8} \\ \textrm{(more IFSR)} \end{array}$} & 
\multicolumn{3}{|c|}{$\begin{array}{c} \textsc{PWG+PY8} \\ \textrm{(less IFSR)} \end{array}$} & 
\multicolumn{3}{|c|}{\textsc{Sherpa~2.2.1}} \\
& \multicolumn{2}{|c}{$\chi^{2}$/NDF} & $p$-value & \multicolumn{2}{|c}{$\chi^{2}$/NDF} &  $p$-value  & \multicolumn{2}{|c}{$\chi^{2}$/NDF} &  $p$-value  & \multicolumn{2}{|c}{$\chi^{2}$/NDF} &  $p$-value  & \multicolumn{2}{|c}{$\chi^{2}$/NDF} &  $p$-value  & \multicolumn{2}{|c}{$\chi^{2}$/NDF} & $p$-value  \\
\hline
\hline
$  p_{\mathrm{T}}^{t}$ &
{\ } 3.7 & 6 & 0.72  &  {\ } 4.5 & 6 & 0.61  &  {\ } 4.0 & 6 & 0.67  &  {\ } 3.9 & 6 & 0.69  &  {\ } 4.0 & 6 & 0.68  &  {\ } 4.3 & 6 & 0.64 \\
$             |y^{t}|$ &
{\ } 4.3 & 7 & 0.75  &  {\ } 4.1 & 7 & 0.77  &  {\ } 4.0 & 7 & 0.78  &  {\ } 4.4 & 7 & 0.73  &  {\ } 4.3 & 7 & 0.74  &  {\ } 5.3 & 7 & 0.62 \\
$p_{\mathrm{T}}^{t,1}$ &
{\ } 5.9 & 7 & 0.55  &  {\ } 7.0 & 7 & 0.43  &  {\ } 5.9 & 7 & 0.55  &  {\ } 6.4 & 7 & 0.50  &  {\ } 6.2 & 7 & 0.52  &  {\ } 7.6 & 7 & 0.37 \\
$           |y^{t,1}|$ &
{\ } 5.5 & 7 & 0.60  &  {\ } 8.3 & 7 & 0.31  &  {\ } 5.1 & 7 & 0.65  &  {\ } 5.9 & 7 & 0.55  &  {\ } 5.5 & 7 & 0.60  &  {\ } 4.7 & 7 & 0.70 \\
$p_{\mathrm{T}}^{t,2}$ &
{\ } 5.7 & 6 & 0.46  &  {\ } 2.8 & 6 & 0.83  &  {\ } 6.1 & 6 & 0.41  &  {\ } 4.6 & 6 & 0.60  &  {\ } 7.4 & 6 & 0.29  &  {\ } 7.0 & 6 & 0.32 \\
$         |{y}^{t,2}|$ &
{\ } 4.4 & 7 & 0.73  &  {\ } 5.1 & 7 & 0.65  &  {\ } 4.2 & 7 & 0.76  &  {\ } 4.4 & 7 & 0.73  &  {\ } 4.3 & 7 & 0.74  &  {\ } 5.9 & 7 & 0.55 \\
$        m^{t\bar{t}}$ &
{\ } 4.0 & 9 & 0.91  &  {\ } 3.7 & 9 & 0.93  &  {\ } 3.9 & 9 & 0.92  &  {\ } 3.9 & 9 & 0.92  &  {\ } 4.3 & 9 & 0.89  &  {\ } 4.6 & 9 & 0.86 \\
$p_{\mathrm{T}}^{t\bar{t}}$ &
{\ } 5.1 & 7 & 0.65  &  {\ } 7.0 & 7 & 0.42  &  {\ } 6.2 & 7 & 0.52  &  {\ } 3.7 & 7 & 0.81  &  {\ } 6.8 & 7 & 0.45  &  {\ } 30.1 & 7 & $<$0.01 \\
$      {y}^{t\bar{t}}$ &
{\ } 1.8 & 7 & 0.97  &  {\ } 2.9 & 7 & 0.90  &  {\ } 2.0 & 7 & 0.96  &  {\ } 2.0 & 7 & 0.96  &  {\ } 1.9 & 7 & 0.97  &  {\ } 4.2 & 7 & 0.76 \\
$   {\chi}^{t\bar{t}}$ &
{\ } 7.9 & 6 & 0.24  &  {\ } 5.0 & 6 & 0.55  &  {\ } 7.3 & 6 & 0.29  &  {\ } 6.4 & 6 & 0.38  &  {\ } 9.0 & 6 & 0.17  &  {\ } 7.6 & 6 & 0.27 \\
$  y_\mathrm{B}^{t\bar{t}}$ &
{\ } 1.0 & 6 & 0.99  &  {\ } 1.4 & 6 & 0.96  &  {\ } 1.0 & 6 & 0.98  &  {\ } 1.1 & 6 & 0.98  &  {\ } 1.0 & 6 & 0.99  &  {\ } 1.0 & 6 & 0.99 \\
\absPoutttbar  &
{\ } 1.7 & 6 & 0.94  &  {\ } 16.9 & 6 & $<$0.01  &  {\ } 1.2 & 6 & 0.98  &  {\ } 1.9 & 6 & 0.93  &  {\ } 2.7 & 6 & 0.84  &  {\ } 3.9 & 6 & 0.69 \\
${\Delta\phi}^{t\bar{t}}$ &
{\ } 0.5 & 3 & 0.93  &  {\ } 13.1 & 3 & $<$0.01  &  {\ } 0.7 & 3 & 0.87  &  {\ } 0.1 & 3 & 1.00  &  {\ } 1.1 & 3 & 0.78  &  {\ } 0.2 & 3 & 0.98 \\
$  H_\mathrm{T}^{t\bar{t}}$ &
{\ } 5.2 & 9 & 0.81  &  {\ } 5.7 & 9 & 0.77  &  {\ } 7.4 & 9 & 0.60  &  {\ } 6.9 & 9 & 0.64  &  {\ } 5.6 & 9 & 0.78  &  {\ } 5.9 & 9 & 0.75 \\
${\cos{\theta}^{\star}}$ &
{\ } 5.5 & 5 & 0.35  &  {\ } 3.2 & 5 & 0.66  &  {\ } 5.3 & 5 & 0.38  &  {\ } 5.0 & 5 & 0.42  &  {\ } 6.2 & 5 & 0.29  &  {\ } 7.8 & 5 & 0.17 \\
\hline
\end{tabular}

}
\label{tab:chi2:parton:rel}
\end{table*}

These results are in agreement with earlier differential cross-section measurements in the \ttbar\ final
states involving at least one lepton  
\cite{atlasDiff1,atlasDiff2,atlasDiff3,atlasDiff4,atlasDiff5,atlasDiff6,atlasDiff7,atlasDiff8,cmsDiff2,cmsDiff3,cmsDiff4,cmsDiff5}. 
Those studies observed a ``softer'' \pT\ spectrum for the top-quark final states, although the
statistical and systematic uncertainties for top quarks with $\pT > 500$~\GeV\ are larger than the 
measurements reported here.  
Together, the previous measurements and these results provide a coherent picture
that the current NLO Monte Carlo models for \ttbar\ production and decay overestimate 
the production of highly boosted top quarks.

\FloatBarrier

\section{Conclusion}
\label{sec:conclusion}

Measurements of differential cross-sections of highly boosted pair-produced top quarks in 13~\TeV\ $pp$\ collisions 
are presented in a data sample of \lumitot~collected by the ATLAS detector at the LHC.
The top-quark pairs are observed in their all-hadronic decay modes. 
With a combination of top-tagging and $b$-tagging techniques,
an event sample with a \ttbar\ signal-to-background ratio of approximately 3-to-1 is selected.  
Because most of the decay products of the top quarks are observed in a \largeR\ jet, the kinematics of the top quarks and 
the \ttbar\ system are well-measured compared with final states involving energetic neutrinos.
The measurements are corrected to a fiducial phase space and normalized to 
the total cross-section for events with leading top quarks with $\pt >$ 500 \GeV~ and second-leading top quarks 
with $\pt>350$~\GeV.
Parton-level differential cross-sections are also determined. 

The leading and second-leading top-quark \pT\ differential cross-sections fall by two orders of magnitude over 
the \pT\ range from 500~\GeV\ to 1~\TeV.  
The top-quark rapidity distributions show a plateau out to $|\yt|\sim 0.6$\ and then fall rapidly, reflecting the central production of these top-quark pairs.
The measurements show that the \ttbar\ system is produced centrally with limited transverse momentum, though events are 
observed up to a $\pT^{\ttbar}$\ of 500~\GeV.  

The normalized differential cross-sections are compared with several Standard Model predictions for highly boosted 
pair-produced top quarks,
and there is generally good agreement of the predictions with the particle-level and parton-level differential results.
In particular, the \POWHEG{}+\PYTHIA{8}, \POWHEG{}+\HERWIG{7}\ and \SHERPA\ predictions are consistent with the observed 
differential cross-sections at particle level and parton level.
The most significant discrepancy is in the \AMCatNLO{}+\PYTHIA{8}\ predictions for the kinematics of the \ttbar\ system.  
Qualitatively, both particle- and parton-level rapidity distributions of the leading top quark and 
of the \ttbar{}~system are broader in the data compared with the Monte Carlo generator predictions.
Also, there are more modest differences between predicted and observed differential cross-sections as a 
function of the production angle $\cos \theta^{*}$ and the variable \chittbar.  

The cross-section for \ttbar\ production in the particle-level fiducial phase space is
$292 \pm 7\ \rm{(stat)} \pm 76\ \rm{(syst)}$~fb, which can be compared with the \POWHEG{}+\PYTHIA{8}\ prediction of
$384\pm36$~fb, where the total cross-section has been calculated up to NNLO+NNLL corrections. 
Improvements in this measurement will come from a better understanding of 
the models of \ttbar\ production that are the source of the modelling uncertainties.

This analysis shows that studies of boosted top-quark jets can be done with good efficiency and signal-to-background ratios
in the all-hadronic channel. 
This creates opportunities for more detailed studies of high-\pt\ Standard Model processes, and provides
data to test and improve models of \ttbar\ production.

\section*{Acknowledgments}
We thank CERN for the very successful operation of the LHC, as well as the
support staff from our institutions without whom ATLAS could not be
operated efficiently.

We acknowledge the support of ANPCyT, Argentina; YerPhI, Armenia; ARC, Australia; BMWFW and FWF, Austria; ANAS, Azerbaijan; SSTC, Belarus; CNPq and FAPESP, Brazil; NSERC, NRC and CFI, Canada; CERN; CONICYT, Chile; CAS, MOST and NSFC, China; COLCIENCIAS, Colombia; MSMT CR, MPO CR and VSC CR, Czech Republic; DNRF and DNSRC, Denmark; IN2P3-CNRS, CEA-DRF/IRFU, France; SRNSF, Georgia; BMBF, HGF, and MPG, Germany; GSRT, Greece; RGC, Hong Kong SAR, China; ISF, I-CORE and Benoziyo Center, Israel; INFN, Italy; MEXT and JSPS, Japan; CNRST, Morocco; NWO, Netherlands; RCN, Norway; MNiSW and NCN, Poland; FCT, Portugal; MNE/IFA, Romania; MES of Russia and NRC KI, Russian Federation; JINR; MESTD, Serbia; MSSR, Slovakia; ARRS and MIZ\v{S}, Slovenia; DST/NRF, South Africa; MINECO, Spain; SRC and Wallenberg Foundation, Sweden; SERI, SNSF and Cantons of Bern and Geneva, Switzerland; MOST, Taiwan; TAEK, Turkey; STFC, United Kingdom; DOE and NSF, United States of America. In addition, individual groups and members have received support from BCKDF, the Canada Council, CANARIE, CRC, Compute Canada, FQRNT, and the Ontario Innovation Trust, Canada; EPLANET, ERC, ERDF, FP7, Horizon 2020 and Marie Sk{\l}odowska-Curie Actions, European Union; Investissements d'Avenir Labex and Idex, ANR, R{\'e}gion Auvergne and Fondation Partager le Savoir, France; DFG and AvH Foundation, Germany; Herakleitos, Thales and Aristeia programmes co-financed by EU-ESF and the Greek NSRF; BSF, GIF and Minerva, Israel; BRF, Norway; CERCA Programme Generalitat de Catalunya, Generalitat Valenciana, Spain; the Royal Society and Leverhulme Trust, United Kingdom.

The crucial computing support from all WLCG partners is acknowledged gratefully, in particular from CERN, the ATLAS Tier-1 facilities at TRIUMF (Canada), NDGF (Denmark, Norway, Sweden), CC-IN2P3 (France), KIT/GridKA (Germany), INFN-CNAF (Italy), NL-T1 (Netherlands), PIC (Spain), ASGC (Taiwan), RAL (UK) and BNL (USA), the Tier-2 facilities worldwide and large non-WLCG resource providers. Major contributors of computing resources are listed in Ref.~\cite{ATL-GEN-PUB-2016-002}.

\clearpage
\label{app:References}
\printbibliography

\clearpage 
\begin{flushleft}
{\Large The ATLAS Collaboration}

\bigskip

M.~Aaboud$^\textrm{\scriptsize 137d}$,
G.~Aad$^\textrm{\scriptsize 88}$,
B.~Abbott$^\textrm{\scriptsize 115}$,
O.~Abdinov$^\textrm{\scriptsize 12}$$^{,*}$,
B.~Abeloos$^\textrm{\scriptsize 119}$,
S.H.~Abidi$^\textrm{\scriptsize 161}$,
O.S.~AbouZeid$^\textrm{\scriptsize 139}$,
N.L.~Abraham$^\textrm{\scriptsize 151}$,
H.~Abramowicz$^\textrm{\scriptsize 155}$,
H.~Abreu$^\textrm{\scriptsize 154}$
% ...
\bigskip
\\
% ...
$^{12}$ Institute of Physics, Azerbaijan Academy of Sciences, Baku, Azerbaijan\\
% ...
$^{88}$ CPPM, Aix-Marseille Universit{\'e} and CNRS/IN2P3, Marseille, France\\
% ...
$^{115}$ Homer L. Dodge Department of Physics and Astronomy, University of Oklahoma, Norman OK, United States of America\\
% ...
$^{119}$ LAL, Univ. Paris-Sud, CNRS/IN2P3, Universit{\'e} Paris-Saclay, Orsay, France\\
% ...
$^{137}$ $^{(a)}$ Facult{\'e} des Sciences Ain Chock, R{\'e}seau Universitaire de Physique des Hautes Energies - Universit{\'e} Hassan II, Casablanca; $^{(b)}$ Centre National de l'Energie des Sciences Techniques Nucleaires, Rabat; $^{(c)}$ Facult{\'e} des Sciences Semlalia, Universit{\'e} Cadi Ayyad, LPHEA-Marrakech; $^{(d)}$ Facult{\'e} des Sciences, Universit{\'e} Mohamed Premier and LPTPM, Oujda; $^{(e)}$ Facult{\'e} des sciences, Universit{\'e} Mohammed V, Rabat, Morocco\\
% ...
$^{139}$ Santa Cruz Institute for Particle Physics, University of California Santa Cruz, Santa Cruz CA, United States of America\\
% ...
$^{151}$ Department of Physics and Astronomy, University of Sussex, Brighton, United Kingdom\\
% ...
$^{154}$ Department of Physics, Technion: Israel Institute of Technology, Haifa, Israel\\
$^{155}$ Raymond and Beverly Sackler School of Physics and Astronomy, Tel Aviv University, Tel Aviv, Israel\\
% ...
$^{161}$ Department of Physics, University of Toronto, Toronto ON, Canada\\
% ...
$^{a}$ Also at Department of Physics, King's College London, London, United Kingdom\\
$^{b}$ Also at Institute of Physics, Azerbaijan Academy of Sciences, Baku, Azerbaijan\\
$^{c}$ Also at Novosibirsk State University, Novosibirsk, Russia\\
$^{d}$ Also at TRIUMF, Vancouver BC, Canada\\
$^{e}$ Also at Department of Physics {\&} Astronomy, University of Louisville, Louisville, KY, United States of America\\
% ...
$^{*}$ Deceased
\end{flushleft}

\end{document}